\pdfoutput=1

\documentclass[11pt,twoside,a4paper,cmspaper,final,collab]{cms-tdr}
\def\svnVersion{281f9ec-D}\def\svnDate{2019/09/04}\def\cmsCernNoTag{CERN-EP-2019-119}\def\cmsCernDate{\today}\def\cmsMessage{"Published in Physical Review D as \href{http://dx.doi.org/10.1103/PhysRevD.100.072007}{\doi{10.1103/PhysRevD.100.072007}.}"}

\begin{document}\cmsNoteHeader{TOP-17-004}

\hyphenation{had-ron-i-za-tion}
\hyphenation{cal-or-i-me-ter}
\hyphenation{de-vices}
\RCS$Revision$
\RCS$HeadURL$
\RCS$Id$

\newlength\cmsFigWidth
\newlength\cmsTabSkip\setlength{\cmsTabSkip}{1ex}
\ifthenelse{\boolean{cms@external}}{\setlength\cmsFigWidth{0.48\textwidth}}{\setlength\cmsFigWidth{0.32\textwidth}}
\ifthenelse{\boolean{cms@external}}{\providecommand{\cmsLeft}{top\xspace}}{\providecommand{\cmsLeft}{left\xspace}}
\ifthenelse{\boolean{cms@external}}{\providecommand{\cmsRight}{bottom\xspace}}{\providecommand{\cmsRight}{right\xspace}}
\ifthenelse{\boolean{cms@external}}{\providecommand{\NA}{\ensuremath{\cdots}\xspace}}{\providecommand{\NA}{\ensuremath{\text{---}}\xspace}}
\ifthenelse{\boolean{cms@external}}{\providecommand{\CL}{C.L.\xspace}}{\providecommand{\CL}{CL\xspace}}

\newcommand{\Mtop}{\ensuremath{m_{\PQt}}\xspace}
\newcommand{\Dn}{\ensuremath{D_{\nu,\mathrm{min}}}\xspace}
\newcommand{\Hathor}{\textsc{hathor}\xspace}
\newcommand{\yukawa}{\ensuremath{Y_{\PQt}}\xspace}
\newcommand{\deltaY}{\ensuremath{\Delta y_{\ttbar}}\xspace}
\newcommand{\Mttb}{\ensuremath{M_{\ttbar}}\xspace}
\newcommand{\MWhad}{\ensuremath{M_{\PW_{\mathrm{h}}}}\xspace}
\newcommand{\Mthad}{\ensuremath{M_{\PQt_{\mathrm{h}}}}\xspace}
\newcommand{\blep}{\ensuremath{\cPqb_\ell}\xspace}
\newcommand{\bhad}{\ensuremath{\cPqb_\mathrm{h}}\xspace}
\newcommand{\alpW}{\ensuremath{\alpha_\text{weak}}\xspace}

\cmsNoteHeader{TOP-17-004}
\title{Measurement of the top quark Yukawa coupling from \texorpdfstring{\ttbar}{ttbar} kinematic distributions in the lepton+jets final state in proton-proton collisions at \texorpdfstring{$\sqrt{s} = 13\TeV$}{sqrt(s) = 13 TeV}}

\date{\today}

\abstract{
   Results are presented for an extraction of the top quark Yukawa
   coupling from top quark-antiquark (\ttbar) kinematic distributions
   in the lepton plus jets final state in proton-proton collisions, based on
   data collected by the CMS experiment at the LHC at $\sqrt{s} = 13\TeV$,
   corresponding to an integrated luminosity of 35.8\fbinv. Corrections
   from weak boson exchange, including Higgs bosons, between the
   top quarks can produce large distortions of differential
   distributions near the energy threshold of \ttbar production.
   Therefore, precise measurements of these distributions are sensitive
   to the Yukawa coupling. Top quark events are reconstructed with at
   least three jets in the final state, and a novel technique is introduced
   to reconstruct the \ttbar system for events with one missing jet.
   This technique enhances the experimental sensitivity in the low
   invariant mass region, \Mttb. The data yields in
   \Mttb, the rapidity difference
   $\abs{y_{\PQt}-y_{\PAQt}}$, and the number of reconstructed
   jets are compared with distributions representing different Yukawa
   couplings. These comparisons are used to measure the ratio of the
   top quark Yukawa coupling to its standard model predicted value to
   be $1.07^{+0.34}_{-0.43}$ with an upper limit of 1.67 at the 95\%
   confidence level.
}

\hypersetup{%
pdfauthor={CMS Collaboration},%
pdftitle={Measurement of the top quark Yukawa coupling from ttbar differential cross sections in the lepton+jets final state at 13 TeV},%
pdfsubject={CMS},%
pdfkeywords={CMS, physics, top quark, top Yukawa coupling, differential cross section}}

\maketitle

\section{Introduction}
\label{sec:intro}
The study of the properties of the Higgs boson, which is responsible
for electroweak symmetry breaking, is one of the main goals of the LHC
program. The standard model (SM) relates the mass of a fermion to its
Yukawa coupling, \ie, the strength of its interaction with the Higgs
boson, as $g_\mathrm{f} = \sqrt{2} m_\mathrm{f}/v$, where
$m_\mathrm{f}$ is the fermion mass and $v = 246.22\GeV$ is the vacuum
expectation value of the Higgs potential~\cite{Weinberg:1967tq},
obtained from a measurement of the $\mu^+$ lifetime~\cite{Webber:2010zf}.
Since fermionic masses are not predicted by the SM, their values are
only constrained by experimental observations. Given the measured value
of the top quark mass of $\Mtop = 172.4 \pm
0.5\GeV$~\cite{Khachatryan:2015hba}, the top quark is the heaviest
fermion and therefore provides access to the largest Yukawa coupling,
which is expected to be close to unity in the SM\@.
It is important to verify this prediction experimentally.
We define \yukawa as the ratio of the
top quark Yukawa coupling to its SM value. In this definition, \yukawa is
equal to $\kappa_\PQt$ as defined in the ``$\kappa$
framework''~\cite{Heinemeyer:2013tqa}, which introduces coupling
modifiers to test for deviations in the SM couplings of the
Higgs boson to other particles. Several Higgs boson production
processes are sensitive to \yukawa, in particular Higgs boson
production via gluon fusion~\cite{Sirunyan:2017exp,Sirunyan:2018ouh}
and Higgs boson production in association with top quark pairs,
$\ttbar\PH$~\cite{Sirunyan:2018hoz}. In both cases, in addition to
\yukawa, the rate depends on the Higgs boson coupling to the decay
products, \eg, bottom quarks or $\tau$ leptons. The only Higgs boson
production process that is sensitive exclusively to \yukawa is
$\ttbar\PH$ production with the Higgs boson decaying to a \ttbar pair,
leading to a four top quark final state~\cite{Sirunyan:2017roi}. In
this paper, we explore a complementary approach to measure \yukawa
independently of the Higgs coupling to other particles by utilizing a
precise measurement of the top quark pair production cross section,
which is affected by a virtual Higgs boson exchange. It has been shown
that in the top quark pair production threshold region, which
corresponds to  a small relative velocity between the top quark and
antiquark, the \ttbar cross section is sensitive to the top quark
Yukawa coupling through weak force mediated
corrections~\cite{Kuhn:2013zoa}. For example, doubling the Yukawa
coupling would lead to a change in the observed differential cross
section comparable to the current experimental precision of
around 6\%~\cite{Khachatryan:2016mnb}. A detailed study of the
differential \ttbar kinematic properties close to the production
threshold could, therefore, determine the value of the top quark Yukawa
coupling. This approach is similar to the threshold scan methods
proposed for \Pe{}$^+$\Pe{}$^-$
colliders~\cite{Strassler:1990nw,Beneke:2015lwa}.

We calculate the weak interaction correction factors for different
values of \yukawa using \Hathor (v2.1)~\cite{Aliev:2010zk} and apply
them at the parton level to existing \ttbar simulated samples. From
these modified simulations, we obtain distributions at detector level
that can be directly compared to data. The Yukawa coupling is extracted
from the distributions of the invariant mass of the top quark pair,
\Mttb, and the rapidity difference between the top quark and antiquark,
$\deltaY = y_{\PQt} - y_{\PAQt}$, for different jet
multiplicities. The low \Mttb and small $\abs{\deltaY}$ regions are
the most sensitive to \yukawa.

Top quarks decay almost exclusively via $\PQt\to\PW\PQb$ and the
final topology depends on the \PW~boson decays. When one \PW~boson
decays leptonically and the other decays hadronically, $\PQt\PAQt
\to \PW^+\PQb\,\PW^-\PAQb\to \ell^+\nu \PQb \, \PQq\PAQq^\prime
\PAQb$ + charge conjugate, the final state at leading order (LO) consists of an
isolated lepton (electron or muon in this analysis), missing transverse
momentum (from the neutrino), and four jets (from two {\cPqb} quarks and two
light quarks). This final state has a sizable branching fraction of
34\%, low backgrounds, and allows for the kinematic reconstruction of
the original top quark candidates. This analysis follows the methodology
employed in Ref.~\cite{Sirunyan:2018wem} and introduces a novel
algorithm to reconstruct the \ttbar pair when only three jets are
detected.

The outline of this paper is as follows. Section~\ref{sec:reweight} introduces
the method of implementing the weak force corrections in simulated events
as well as the variables sensitive to the top quark Yukawa coupling.
Section~\ref{sec:detector} describes the CMS detector. The
data and simulated samples used in the analysis are described in
Section~\ref{sec:dataset}. The event selection criteria are discussed in
Section~\ref{sec:selection}. The algorithm used to reconstruct \ttbar
events is described in Section~\ref{sec:reco}. Details on background
estimation and event yields are covered in Sections~\ref{sec:bck}
and~\ref{sec:controlplots}. The statistical methodologies and the
systematic uncertainties are described in Sections~\ref{sec:stat}
and~\ref{sec:sys}, respectively. Section~\ref{sec:limit} presents the
results of the fit to data. Section~\ref{sec:summary} summarizes the results.

\section{Weak interaction corrections to \texorpdfstring{\ttbar}{ttbar} production}
\label{sec:reweight}
\begin{figure}[htbp]
\centering
\includegraphics[width=0.35\textwidth]{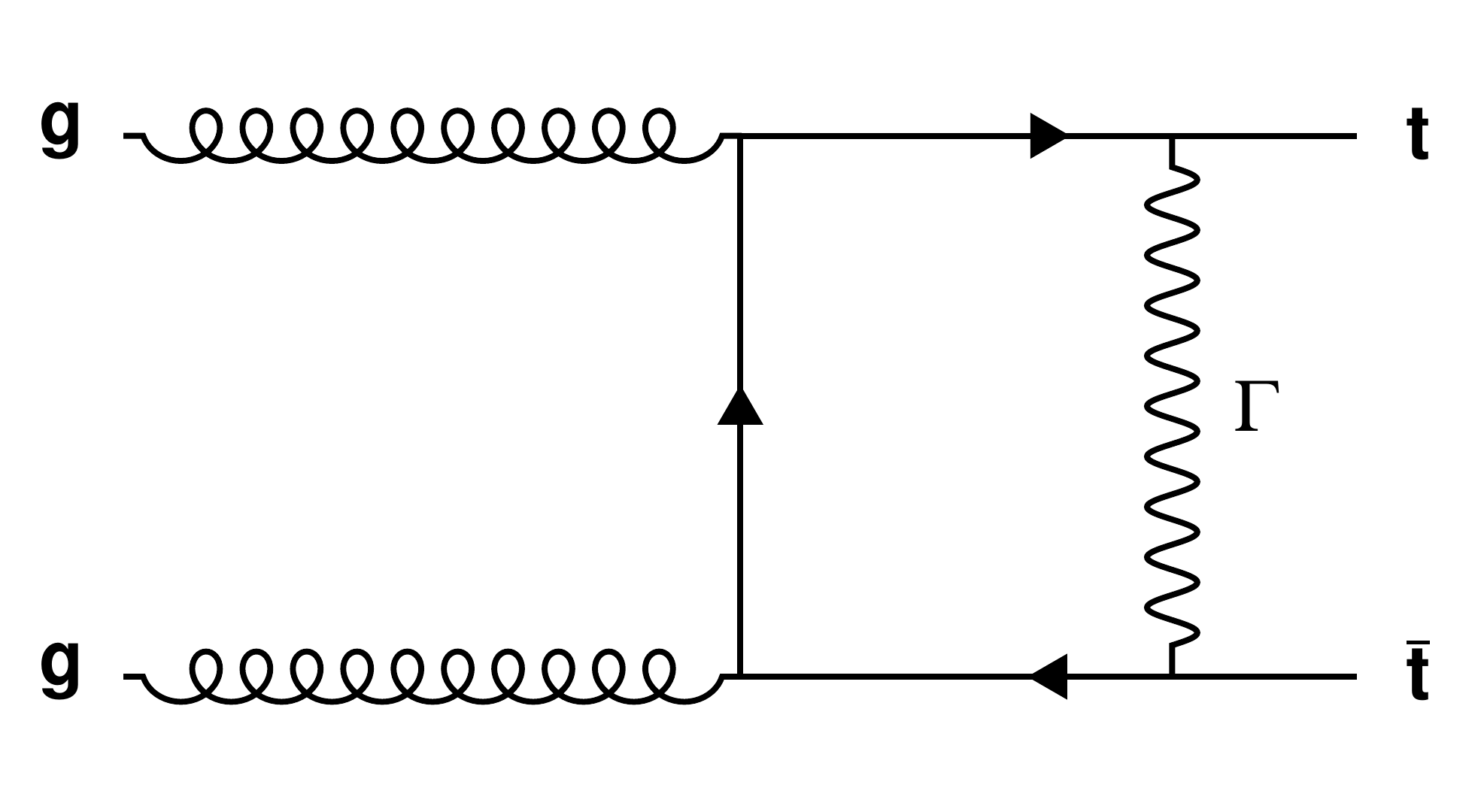}
\hspace{1cm}
\includegraphics[width=0.35\textwidth]{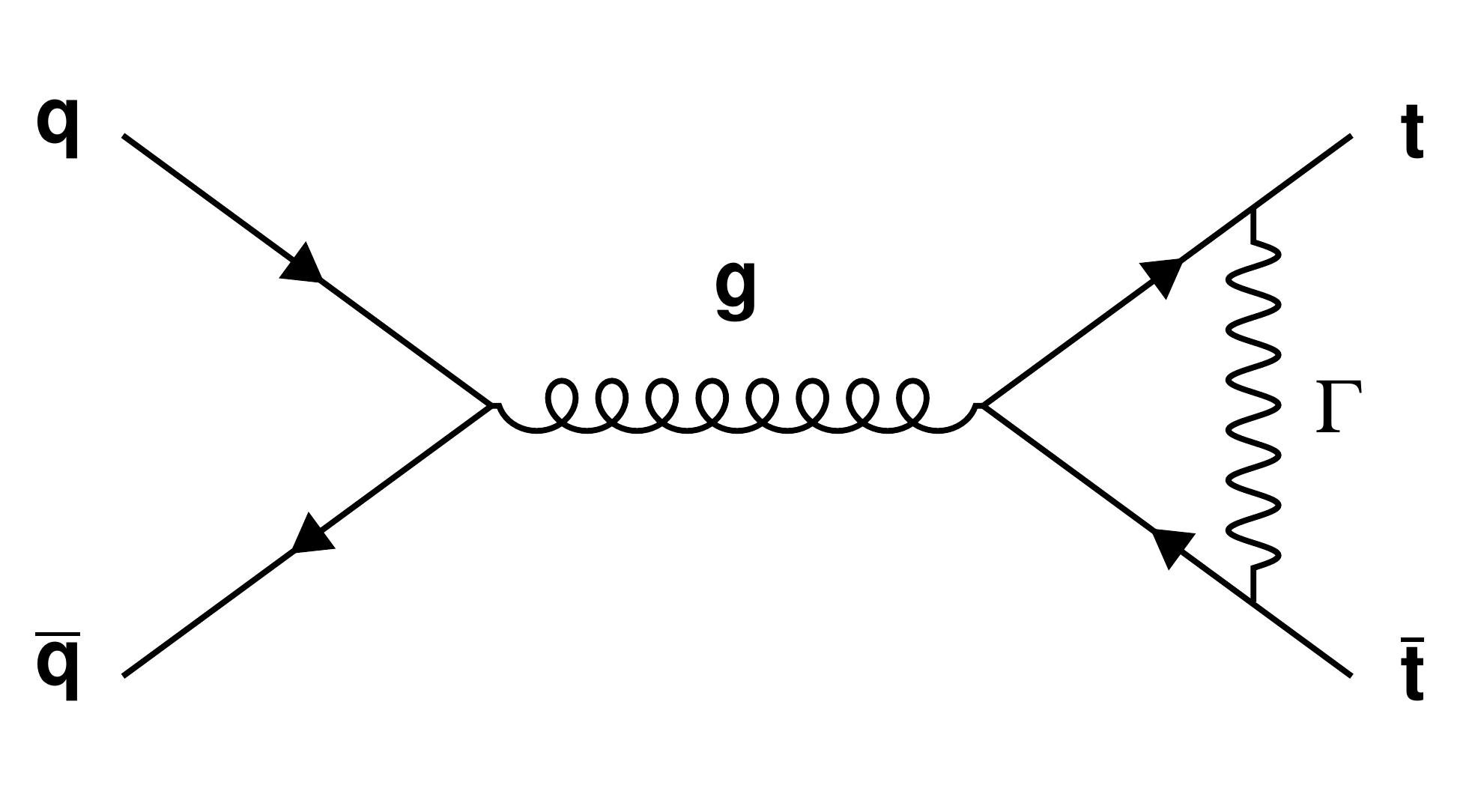}
\caption{Example of Feynman diagrams for gluon- and \qqbar-induced processes of \ttbar production and the virtual corrections. The symbol $\Gamma$ stands for all contributions from gauge and Higgs boson exchanges.}
\label{p:intro:feynman}
\end{figure}

Recent calculations provide next-to-next-to-leading-order (NNLO) predictions
within the framework of perturbative quantum chromodynamics (QCD) for
the \ttbar production cross
section~\cite{Czakon:2017wor,Czakon:2019txp}. Photon-mediated
corrections have been determined to be small~\cite{Hollik:2007sw}. The weak
force corrections to the \ttbar production cross section were
originally calculated~\cite{Beenakker:1993yr} before the top quark
discovery and were found to have a very small effect on the total cross
section, so they are typically not implemented in Monte Carlo (MC)
event generators. Nevertheless, they can have a sizable impact on differential
distributions and on the top quark charge asymmetry. There is no
interference term of order $\alpS\alpW$ between
the lowest-order strong force mediated and neutral current amplitudes
in the quark-induced processes. The weak force corrections start entering
the cross section at loop-induced order $\alpS^2\alpW$
(as shown in Fig.~\ref{p:intro:feynman}). A majority of weak
corrections do not depend on the top quark Yukawa coupling. Amplitudes
linear in \yukawa, which arise from the production of an intermediate
$s$-channel Higgs boson through a closed {\cPqb} quark loop, can be ignored
because of the small {\cPqb} quark mass. However, the amplitude of the Higgs
boson contribution to the loop ($\Gamma=\PH$ in
Fig.~\ref{p:intro:feynman}) is proportional to $\yukawa^2$. The
interference of this process with the Born-level \ttbar production has a
cross section proportional to $\alpS^2\yukawa^2$. Thus, in
some kinematic regions, the weak corrections become large and may lead
to significant distortions of differential distributions.

\begin{figure}[htbp]
\centering
\includegraphics[width=0.45\textwidth]{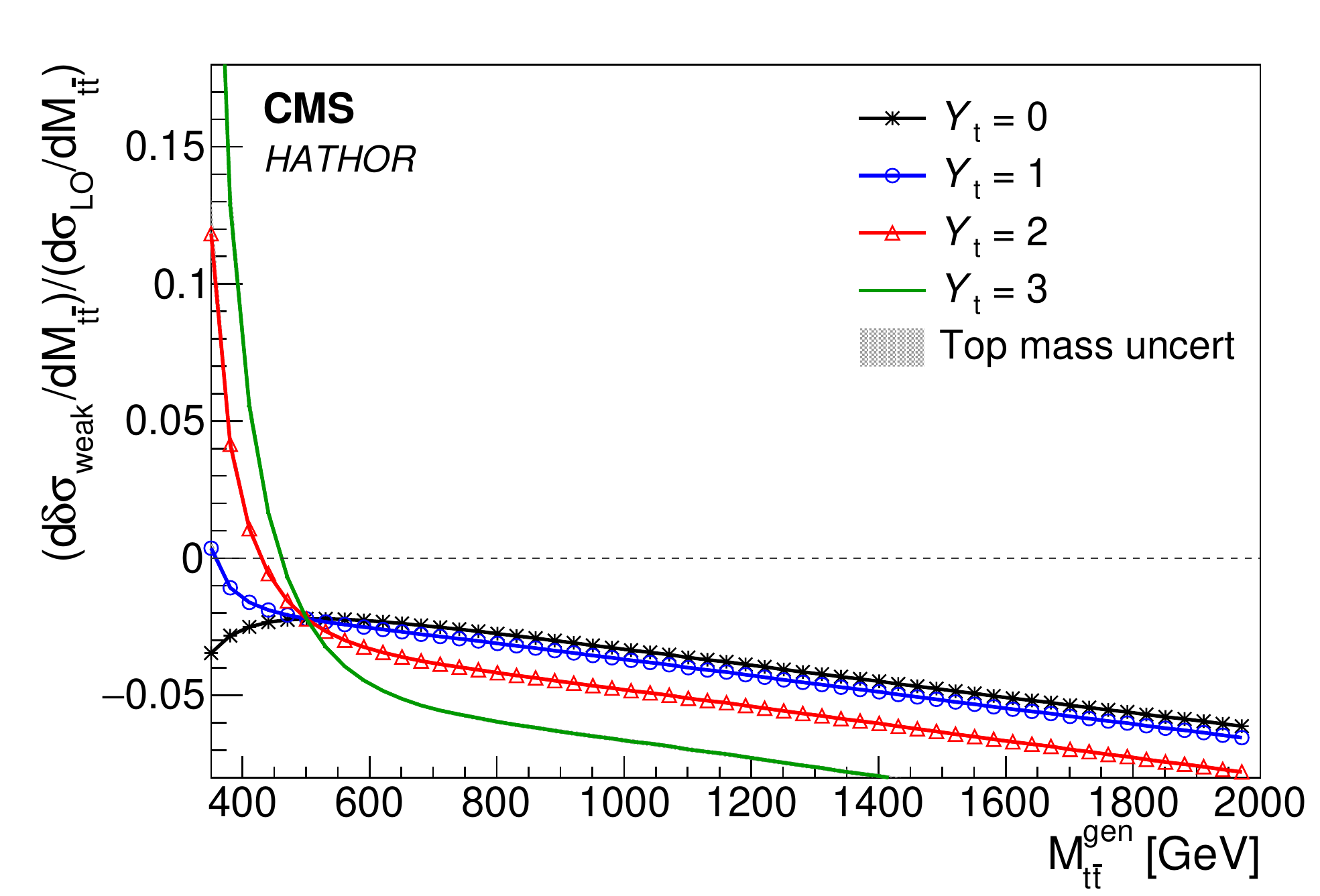}
\includegraphics[width=0.45\textwidth]{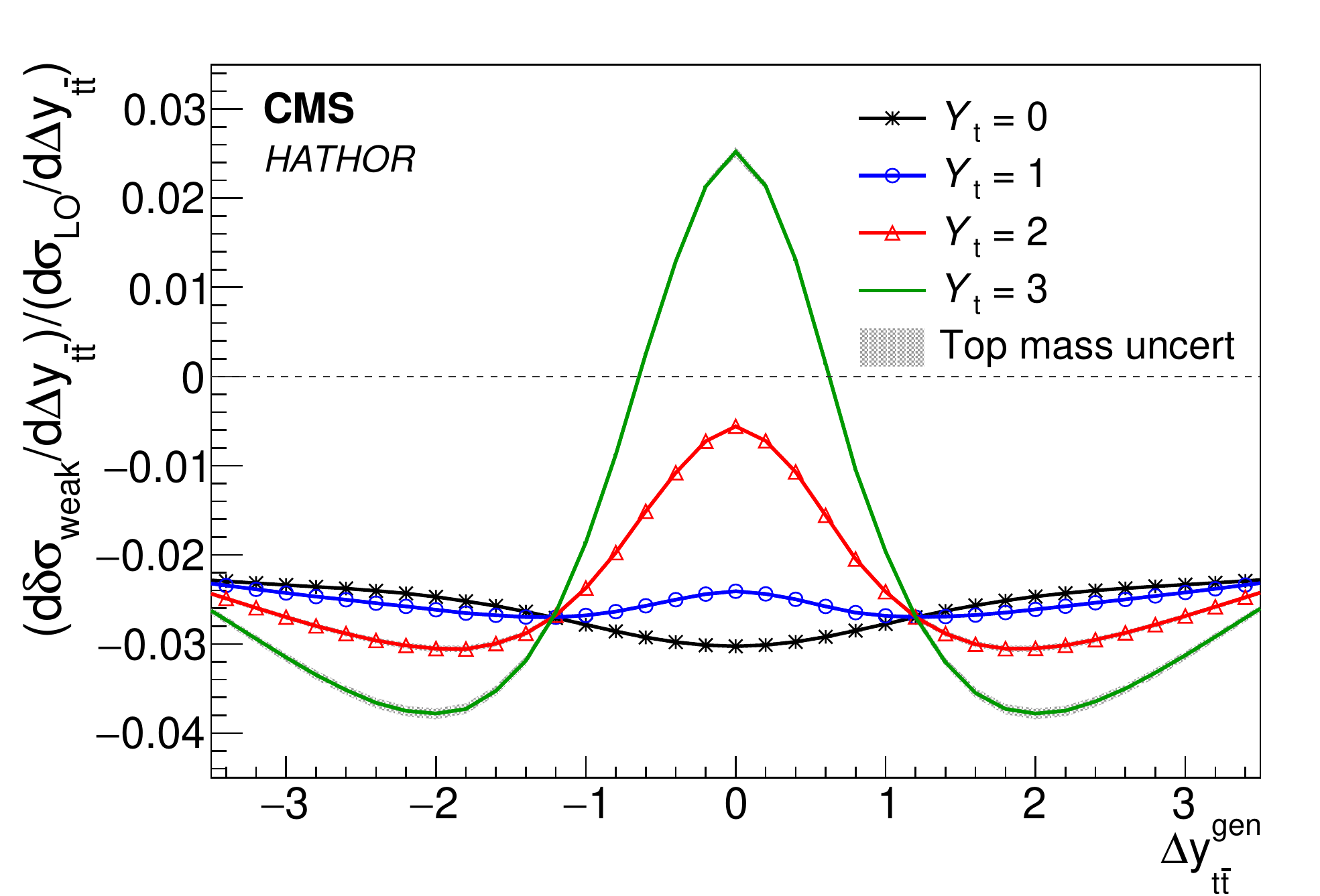}
\caption{The dependence of the ratio of weak force corrections over the LO QCD production cross section as calculated by \Hathor on the sensitive kinematic variables \Mttb and \deltaY at the generator level for different values of \yukawa. The lines contain an uncertainty band (generally not visible) derived from the dependence of the weak correction on the top quark mass varied by $\pm$1\GeV.
}
\label{p:reweight:1d}
\end{figure}

The \Hathor generator calculates the partonic cross section value,
including the next-to-leading-order (NLO) weak corrections at order
$\mathcal{O}(\alpS^2\alpW)$ for given \Mttb and
$\abs{\deltaY}$. The mass of the top quark is fixed at $\Mtop = 172.5\GeV$,
and its uncertainty is treated as a source of systematic uncertainty.
We use \Hathor to extract a two-dimensional correction factor that
contains the ratio of the \ttbar production cross section with weak
corrections over the LO QCD production cross section in bins of \Mttb
and $\abs{\deltaY}$. This is done for different hypothesized values of
\yukawa, as shown in projections in Fig.~\ref{p:reweight:1d}.
The largest effects arise near the \ttbar production threshold region and can be as high as 12\% for \yukawa = 2.
We then apply this correction factor at the parton level as a weight to each \ttbar event
simulated with \POWHEG
(v2)~\cite{Nason:2004rx,Frixione:2007vw,Alioli:2010xd,Campbell:2014kua}.
In the distributions at the detector level,
the experimental resolutions and the systematic uncertainties, which
are especially significant in the low-\Mttb region, will reduce the
sensitivity to this effect.

\section{The CMS detector}
\label{sec:detector}

The central feature of the CMS detector is a superconducting solenoid
of 6\unit{m} internal diameter, providing a magnetic field of 3.8\unit{T}. Within
the solenoid volume are a silicon pixel and strip tracker, a lead
tungstate crystal electromagnetic calorimeter (ECAL), and a brass and
scintillator hadron calorimeter (HCAL), each composed of a barrel and
two endcap sections. Forward calorimeters extend the coverage provided
by the barrel and endcap detectors. Muons are measured in
gas-ionization detectors embedded in the steel flux-return yoke outside
the solenoid. A more detailed description of the CMS detector, together
with a definition of the coordinate system and relevant kinematical
variables, can be found in Ref.~\cite{Chatrchyan:2008aa}.

The particle-flow (PF) algorithm~\cite{ref:particleflow} reconstructs
and identifies each individual particle with an optimized combination
of information from the various elements of the detector systems. The
energy of photons is directly obtained from the ECAL measurements,
corrected for zero-suppression effects. The energy of electrons is
determined from a combination of the electron momentum at the primary
interaction vertex as determined by the tracker, the energy of the
corresponding ECAL cluster, and the energy sum of all bremsstrahlung
photons spatially compatible with originating from the electron track.
The momentum of muons is obtained from the curvature of the
corresponding track, combining information from the silicon tracker and
the muon system. The energy of charged hadrons is determined from a
combination of their momentum measured in the tracker and the matching
ECAL and HCAL energy deposits, corrected for zero-suppression effects
and for the response function of the calorimeters to hadronic showers.
Finally, the energy of neutral hadrons is obtained from the
corresponding corrected ECAL and HCAL energy. The reconstructed vertex
with the largest value of the sum of the physics objects transverse
momentum squared, $\pt^2$, is taken to be
the primary proton-proton (\Pp{}\Pp{}) interaction vertex.

\section{Data set and modeling}
\label{sec:dataset}

The data used for this analysis corresponds to an integrated luminosity
of 35.8\fbinv at a center-of-mass energy of 13\TeV. Events are selected
if they pass single-lepton triggers~\cite{Khachatryan:2016bia}. These
require a transverse momentum $\pt > 27\GeV$ for electrons and $\pt >
24\GeV$ for muons, each within pseudorapidity $\abs{\eta} < 2.4$, as
well as various quality and isolation criteria.

The MC event generator \POWHEG is used to simulate \ttbar events. It calculates up to NLO QCD matrix
elements and uses \PYTHIA (v8.205)~\cite{Sjostrand:2014zea} with the
CUETP8M2T4 tune~\cite{ISR_FSR} for the parton shower simulations. The
default parametrization of the parton distribution functions (PDFs)
used in all simulations is NNPDF3.0~\cite{Ball:2014uwa}. A top quark
mass of 172.5\GeV is used. When compared to the data, the simulation
is normalized to an inclusive \ttbar production cross section of
$832^{+40}_{-46}$\unit{pb}~\cite{Czakon:2011xx}. This value is calculated at
NNLO accuracy, including the resummation of
next-to-next-to-leading-logarithmic soft gluon terms. The quoted
uncertainty is from the choice of hadronization, factorization, and
renormalization scales and the PDF uncertainties.

The background processes are modeled using the same techniques. The
\MGvATNLO generator~\cite{Alwall:2014hca} is used to simulate \PW~boson
and Drell--Yan (DY) production in association with jets and $t$-channel
single top quark production. The \POWHEG generator is used to simulate
a single top quark produced in association with a \PW~boson
($\PW\cPqt$), and \PYTHIA is used for QCD multijet production. In all
cases, the parton shower and the hadronization are simulated by
\PYTHIA. The \PW~boson and DY backgrounds are normalized to their NNLO
cross sections calculated with \FEWZ~\cite{Li:2012wna}. The cross
sections of single top quark processes are normalized to NLO
calculations~\cite{Kant:2014oha,Kidonakis:2012rm}, and the QCD multijet
simulation is normalized to the LO cross section from \PYTHIA. As
explained in Section~\ref{sec:bck}, the shape and the overall
normalization of the QCD multijet contribution to the background are
derived using data in a control region. The QCD multijet simulation is
only used to determine relative contributions from different regions.

The detector response is simulated using
\GEANTfour{}~\cite{Agostinelli:2002hh}. The same algorithms that are
applied to the collider data are used to reconstruct the simulated
data. Multiple proton-proton interactions per bunch crossing (pileup)
are included in the simulation. To correct the simulation to be in
agreement with the pileup conditions observed during the data taking,
the average number of pileup events is calculated
for the measured instantaneous luminosity. The simulated events are
weighted, depending on their number of pileup interactions, to
reproduce the measured pileup distribution.

\section{Event reconstruction and selection}
\label{sec:selection}

Jets are reconstructed from the PF candidates and are clustered by the anti-\kt
algorithm~\cite{Cacciari:2008gp,Cacciari:2011ma} with a distance
parameter $R = 0.4$.
The jet momentum is determined as the vectorial sum of the momenta of all
PF candidates in the jet. An offset correction is applied to jet
energies to take into account the contribution from pileup within the
same or nearby bunch crossings. Jet energy corrections are derived from
simulation and are improved with in situ measurements of the energy
balance in dijet, QCD multijet, photon+jet, and leptonically decaying \PZ+jet
events~\cite{Chatrchyan:2011ds,Khachatryan:2016kdb}. Additional
selection criteria are applied to each event to remove spurious
jet-like features originating from isolated noise patterns in certain
HCAL and ECAL regions~\cite{CMS-PAS-JME-10-003}.

Jets are identified as originating from {\cPqb} quarks using the combined
secondary vertex algorithm (CSV) v2~\cite{Sirunyan:2017ezt}. Data
samples are used to measure the probability of correctly identifying
jets as originating from {\cPqb} quarks ({\cPqb} tagging efficiency), and the
probability of misidentifying jets originating from light-flavor
partons (\cPqu, \cPqd, \cPqs quarks or gluons) or a charm quark as a {\cPqb}-tagged jet
(the light-flavor and charm mistag
probabilities)~\cite{Sirunyan:2017ezt}. To identify a jet as a {\cPqb} jet, its
CSV discriminant is required to be greater than 0.85. This working
point yields a {\cPqb} tagging efficiency of 63\% for jets with \pt typical
of \ttbar events, and charm and light-flavor mistag probabilities of
approximately 12 and 2\%, respectively (around 3\% in total).

The missing transverse momentum, \ptvecmiss, is calculated as the
negative vector sum of the transverse momenta of all PF candidates in
an event. The energy scale corrections applied to jets are propagated
to \ptvecmiss. Its magnitude is referred to as \ptmiss.

Candidate signal events are defined by the presence of a muon or an
electron that is isolated from other activity in the event,
specifically jets, and \ptvecmiss associated with a neutrino. The
isolation variables exclude the contributions from the physics object
itself and from pileup events. The efficiencies of lepton
identification and selection criteria are derived using a tag-and-probe
method in \pt and $\eta$ regions~\cite{TNPREF}. The same lepton isolation
criteria described in Ref.~\cite{Sirunyan:2018wem} are followed here.

To reduce the background contributions and to optimize the \ttbar
reconstruction, additional requirements on the events
are imposed. Only events with exactly one isolated
muon~\cite{Chatrchyan:2012xi} or electron~\cite{Khachatryan:2015hwa}
with $\pt > 30\GeV$ and $\abs{\eta} < 2.4$ are selected; no additional
isolated muons or electrons with $\pt > 15\GeV$ and $\abs{\eta} < 2.4$
are allowed; at least three jets with $\pt > 30\GeV$ and $\abs{\eta}
< 2.4$ are required, and at least two of them must be {\cPqb} tagged.
The \PW~boson transverse mass, defined as $M_\mathrm{T}(\PW) =
\sqrt{\smash[b]{2\pt^\ell\ptmiss[1-\cos(\Delta\phi_{\ell,\ptvecmiss})]}}$, is required to be less than 140\GeV, where
$\pt^\ell$ is the transverse momentum of the lepton. For \ttbar
events with only three jets in the final state, the \pt of the leading
\cPqb-tagged jet is required to be greater than 50\GeV.

\section{Reconstruction of the top quark-antiquark system}
\label{sec:reco}

The goal of reconstructing \ttbar events is to determine the top quark
and antiquark four-momenta. For this, it is necessary to correctly
match the final-state objects to the top quark and antiquark decay products.
We always assume that the two \cPqb-tagged jets with the
highest CSV discriminant values are associated with the two \cPqb quarks
from \ttbar decays. For each event, we test the possible assignments of
jets as \ttbar decay products and select the one with the highest value
of a likelihood discriminant constructed based on the available
information.

The first step in building the likelihood discriminant is to
reconstruct the neutrino four-momentum $p_\nu$ based on the measured
\ptvecmiss, the lepton momentum $p_\ell$, and the momentum
$p_{\blep}$ of the jet associated with the \cPqb quark from the top
quark decay. The neutrino solver algorithm~\cite{Betchart:2013nba} uses
a geometric approach to find all possible solutions for the neutrino
momentum based on the two mass constraints $(p_\nu + p_\ell)^2 =
m_{\PW}^2 = (80.4\GeV)^2$ and $(p_\nu + p_\ell + p_{\blep})^2 = m_\cPqt^2$. Each
equation describes an ellipsoid in the three-dimensional neutrino
momentum space. The intersection of these two ellipsoids is usually an
ellipse. We select $p_\nu$ as the point on the ellipse for which the
distance $\Dn$ between the ellipse projection onto the
transverse plane ($p_{\nu x}$,$p_{\nu y}$) and the measured \ptvecmiss
is minimal. The algorithm leads to a unique solution for the
longitudinal component of the neutrino momentum and an improved
resolution for its transverse component. When the invariant mass of
the lepton and the \blep candidate is above \Mtop, no solution
can be found and this jet assignment is discarded. If both
\blep candidates fail this requirement, then the event is
rejected. The algorithm is applied for each of the two $\cPqb$ jet possibilities
 and the minimum distance $\Dn$ is used to identify the correct
 $\cPqb$ jet in the leptonic top quark decay, \blep, as described below.

\subsection{Reconstruction of events with at least four jets}
\label{sec:reco4j}

The likelihood discriminant for events with at least four reconstructed
jets is built to minimize the calculated \Dn, and to simultaneously
ensure that the invariant mass of the two jets hypothesized to
originate from the \PW~boson decay (\MWhad) is
consistent with the \PW~boson mass, and that the invariant mass of the
three jets hypothesized to originate from the hadronically decaying top
quark (\Mthad) is consistent with $\Mtop$. The
likelihood discriminant for events with at least four jets, $\lambda_4$, is constructed as
\begin{equation}
{-}\ln [ \lambda_4 ] = {-}\ln \left [ P_\mathrm{m}(\MWhad, \Mthad ) \right ] {-}\ln \left [ P_{\nu}(\Dn) \right ],
\label{TTRECEQ1}
\end{equation}
where $P_\mathrm{m}$ is the two-dimensional probability density to
correctly reconstruct the \PW~boson and top quark invariant masses, and
$P_{\nu}$ is the probability density describing the distribution of \Dn for a
correctly selected \blep. On average, the distance \Dn for a
correctly selected \blep is smaller and has a lower tail
compared to the distance obtained for other jets. Jet assignments with
values of $\Dn > 150\GeV$ are rejected since they are very unlikely
to originate from a correct \blep association. The
distributions from which $P_\mathrm{m}$ and $P_{\nu}$ are derived,
together with $\lambda_4$ are shown in Figs.~2 (top-left), 2 (bottom-left) and~4 (left) of Ref.~\cite{Sirunyan:2018wem}, respectively.

The efficiency of the reconstruction algorithm is defined as the
probability that the most likely assignment, as identified by  the
largest value of  $\lambda_4$, is the correct one, given that all decay
products from the \ttbar decay are reconstructed and selected. Since
the number of possible assignments increases drastically with the
number of jets, it is more likely to select a wrong assignment if there
are additional jets. The algorithm identifies the correct assignment in
around 84\% of the four-jet events, 69\% of the five-jet events, and
53\% of the six-jet events.

\subsection{Reconstruction of events with exactly three jets}
\label{sec:reco3j}

The most sensitive region of the phase space to probe the size of the
top quark Yukawa coupling is at the threshold of \ttbar production.
However, the efficiency for selecting \ttbar events in this region is
rather low, since one or more quarks from the \ttbar decay are likely
to have \pt or $\eta$ outside of the selection thresholds resulting in
a missing jet. To mitigate this effect, an algorithm was developed for
the reconstruction of
\ttbar events with one missing jet~\cite{Demina:2013wda}.

\begin{figure}[tbp]
\centering
\includegraphics[width=\cmsFigWidth]{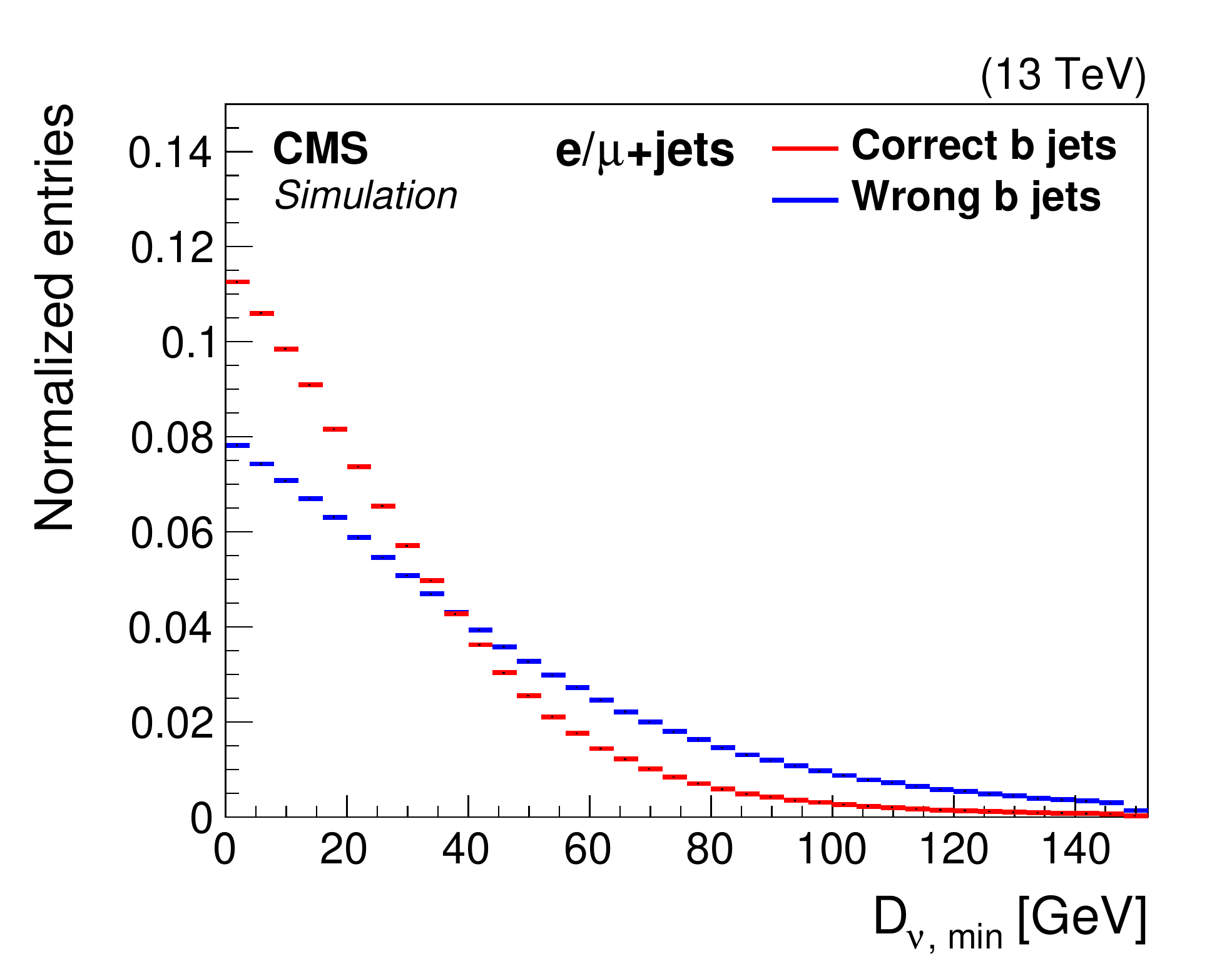}
\includegraphics[width=\cmsFigWidth]{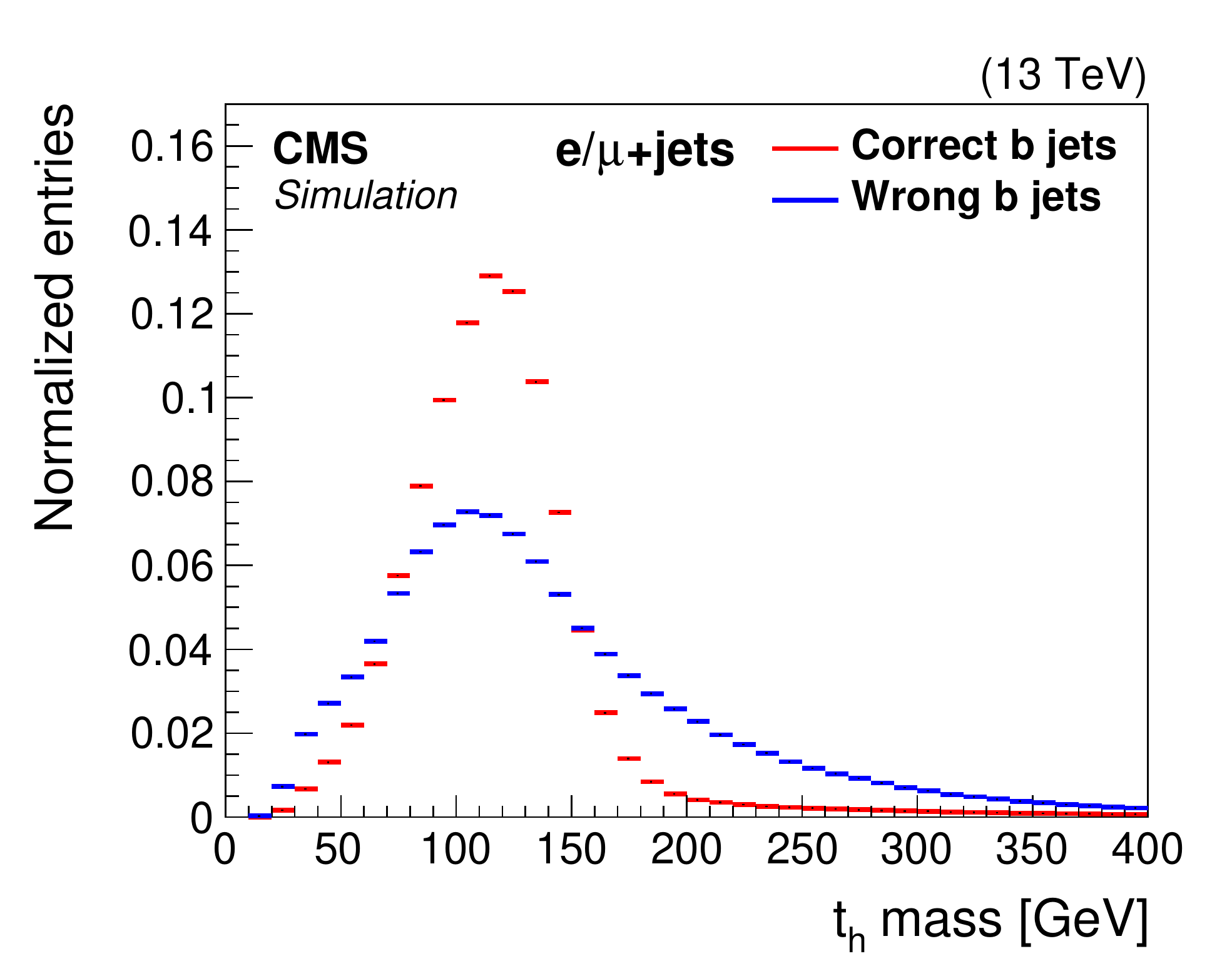}
\includegraphics[width=\cmsFigWidth]{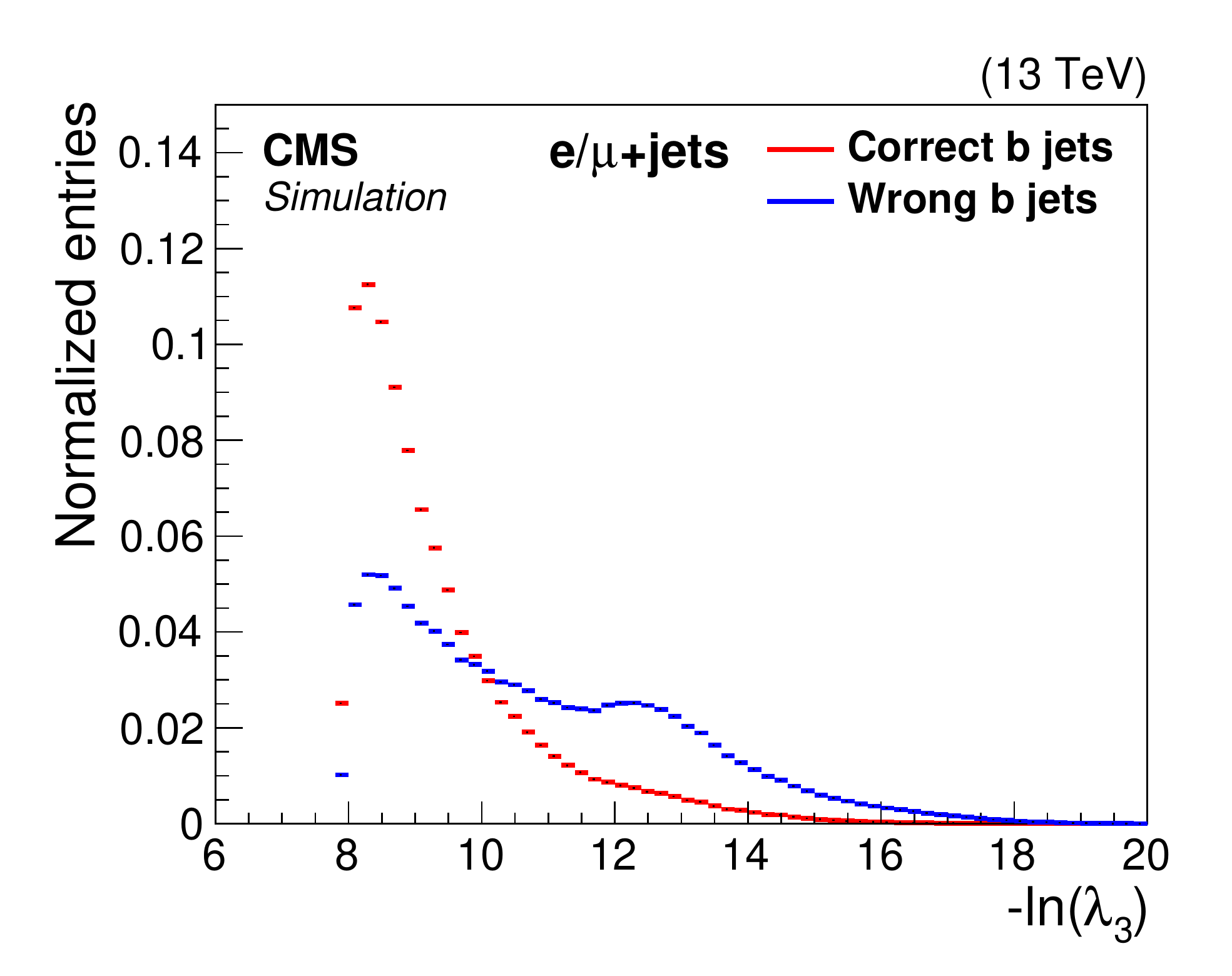}
\caption{Three-jet reconstruction. Distributions of the distance \Dn for correctly and wrongly selected \blep candidates (\cmsLeft). Mass distribution of the correctly and wrongly selected \bhad and the jet from the \PW~boson (middle). Distribution of the negative combined log-likelihood (\cmsRight). All distributions are normalized to have unit area.}
\label{p:reco3j}
\end{figure}

As the missing jet in 93\% of the selected three-jet events is associated
with a quark from the \PW~boson decay, we assume the two jets with the highest
CSV discriminant are associated with {\cPqb} quarks from the \ttbar decay.
The remaining two-fold ambiguity is in the assignment of the {\cPqb}-tagged jets:
which one originates from the hadronic and which one from the
semileptonic top quark decay. For each of the two possible {\cPqb} jet
assignments, the algorithm uses the neutrino solver to calculate the
corresponding minimum distance $\Dn$. If the
neutrino solver yields no solution, this jet assignment is discarded
and the other solution is used if available. Events with no solutions
are discarded. If both {\cPqb} jet candidates have solutions for
neutrino momentum, a likelihood discriminant is constructed using the
minimum distance $\Dn$ and the invariant mass
\Mthad of the two jets hypothesized to belong to the
hadronic top quark decay. We choose the jet assignment with the lowest
value of the negative log likelihood ${-}\ln[\lambda_3]$ defined as
\begin{equation}
{-}\ln[\lambda_3] = {-}\ln \left [ P_{\Mthad} \right ] {-}\ln \left [ P_{\nu}(\Dn) \right ],
\label{TTRECEQ2}
\end{equation}
where the label 3 refers to the requirement of three jets.
The function $P_{\nu}(\Dn)$ is the probability density of
$\Dn$ to correctly identify \blep, and
$P_{\Mthad}$ is the probability density of the invariant
mass of the hypothesized \bhad and the jet from the
\PW~boson decay. Figures~\ref{p:reco3j} (\cmsLeft) and (middle) show
the separation between correct and incorrect $\cPqb$ assignments in the
relevant variables for signal events.
The distribution of $-\ln[\lambda_3]$ is shown in the \cmsRight plot
of Fig.~\ref{p:reco3j}. Jet assignments with values of $-\ln[\lambda_3]
> 13$ are discarded to improve the signal-to-background ratio.
Overall, this algorithm identifies the correct {\cPqb} jet assignment in 80\% of three-jet events.

\begin{figure}[tbp]
\centering
\includegraphics[width=\cmsFigWidth]{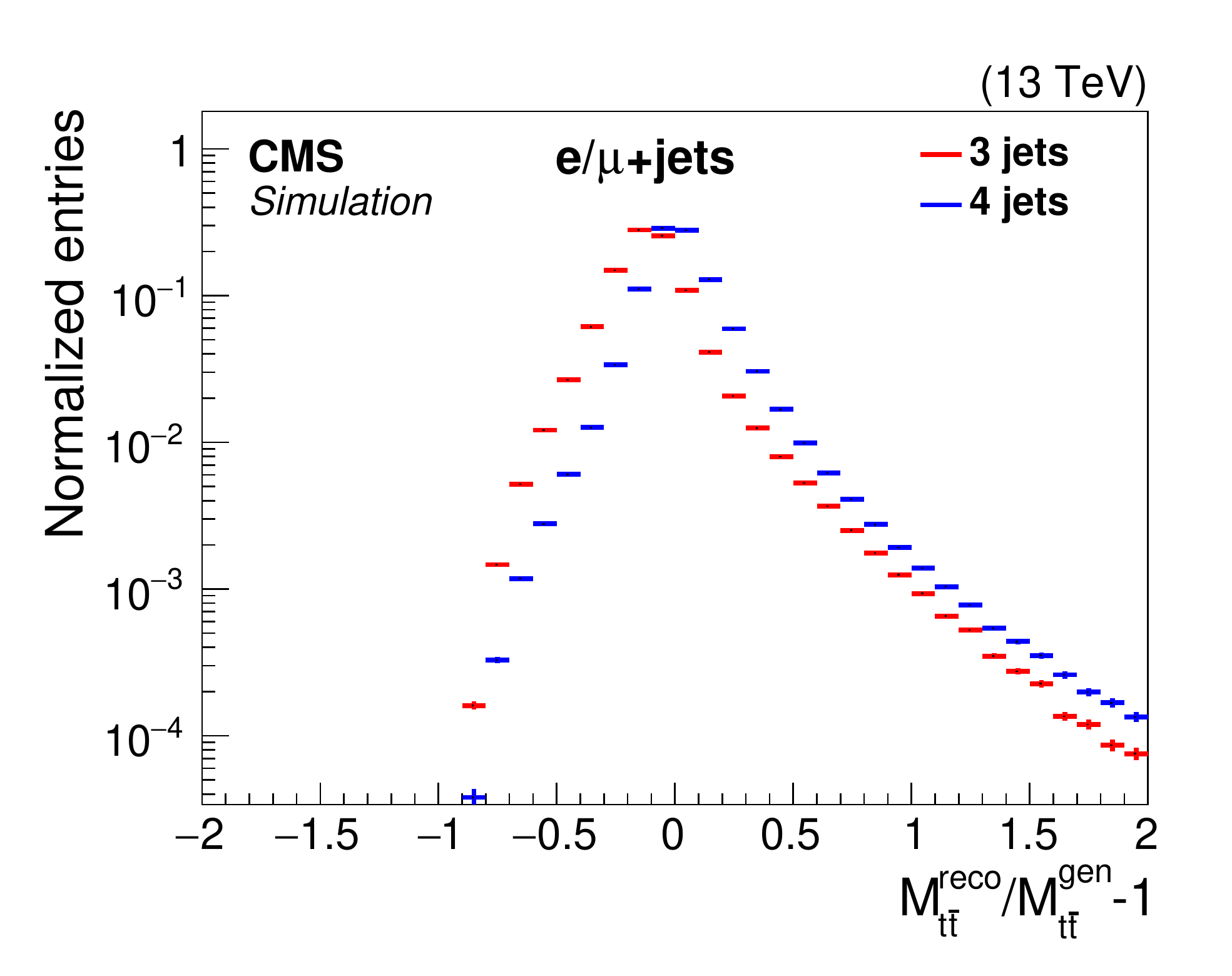}
\includegraphics[width=\cmsFigWidth]{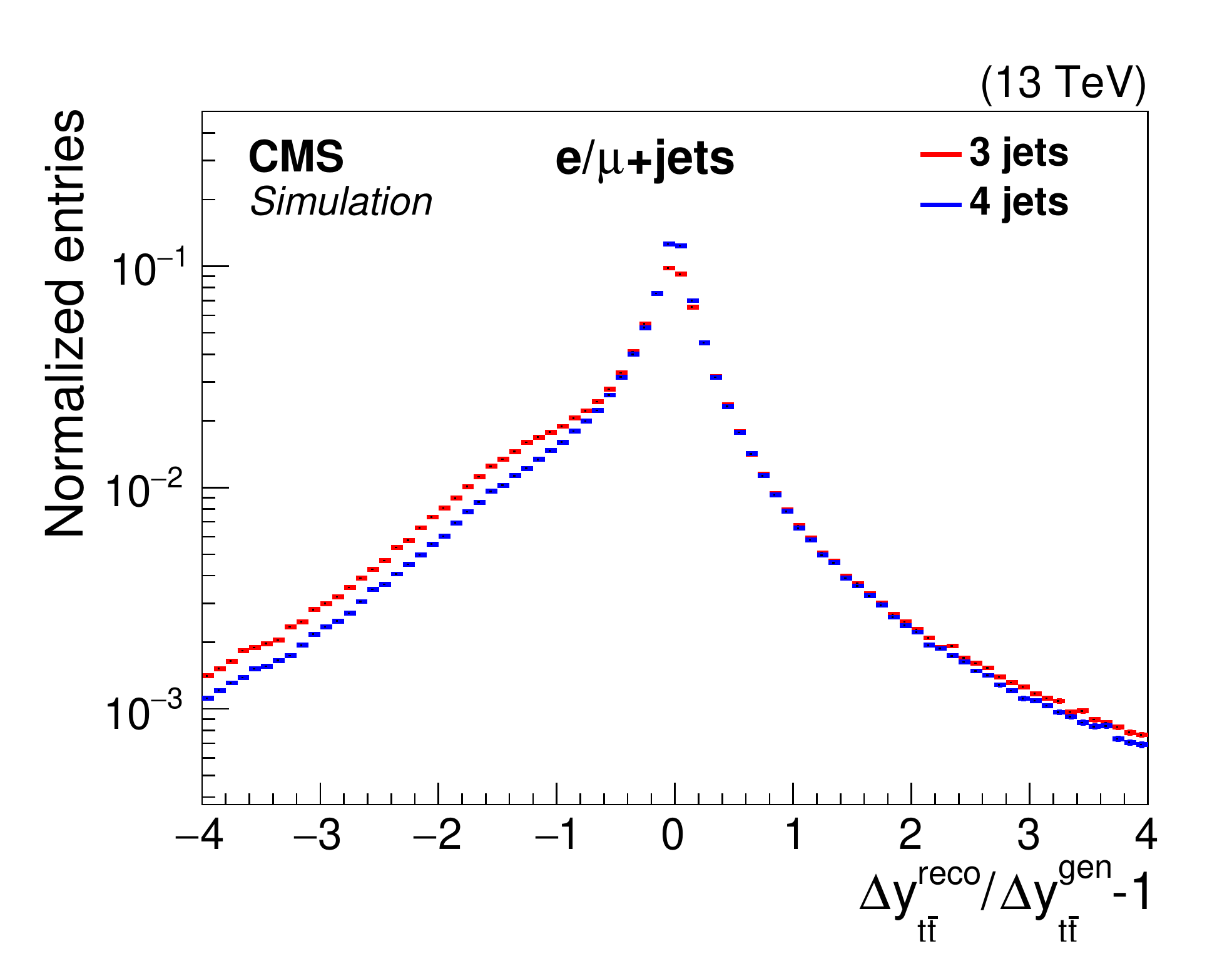}
\caption{Relative difference between the reconstructed and generated \Mttb (\cmsLeft) and \deltaY (\cmsRight) for three-jet and four-jet event categories.}
\label{p:reco3j:resol}
\end{figure}

Semileptonic top quark decays are fully reconstructible, regardless of
whether the event has three or four jets. The hadronically decaying top
quark candidate in the missing jet category is approximated by the
system of two jets identified to be associated with the hadronic top
quark decay. Figure~\ref{p:reco3j:resol} shows the relative difference between the reconstructed and generated invariant mass of the \ttbar system and of the difference in rapidity
for three-jet events, compared to those with four jets. Because of the
missing jet, the observed value of \Mttb in the three-jet category
tends to be lower than in the four-jet category. However, this shift
does not affect the \yukawa measurement since the data are compared to
the simulation in each different jet multiplicity bin: only the widths
of these distributions are important.
Figure~\ref{p:reco3j:resol} demonstrates that the three-jet
reconstruction is competitive with the one achieved in the four-jet
category.

To summarize, the newly developed three-jet reconstruction algorithm
allows us to increase the yields in the sensitive low-\Mttb region. As
will be shown in Section~\ref{sec:sys}, the addition of three-jet
events also helps to reduce the systematic uncertainty from effects
that cause migration between jet multiplicity bins, \eg, jet energy
scale variation and the hadronization model. The analysis is performed
in three independent channels based on the jet multiplicity of the
event: three, four, and five or more jets.

\section{Background estimation}
\label{sec:bck}

The backgrounds in this analysis arise from QCD multijet production, single
top quark production, and vector boson production in association with
jets (V+jets). The expected number of events from $\PW\PW$ and $\PW\PZ$
production is negligible and we ignore this contribution in the signal
region (SR).

The contributions from single top quark and V+jets production are
estimated from the simulated samples. Rather than relying on the
relatively small simulated sample of QCD multijet events, smoother
distributions in \Mttb and $\abs{\deltaY}$ are obtained from data in a
control region (CR).
Events in the CR are selected in the same way as the signal events,
except that the maximum value of the CSV discriminant of jets in each
event has to be less than 0.6. Hence, events in the CR originate
predominately from V+jets and QCD multijet processes. The simulation in
this background-enriched CR describes the data well within uncertainties. We take the
distributions in \Mttb and $\abs{\deltaY}$ from data in the CR, after
subtracting the expected contribution from the V+jets, single
top quark, \ttbar, and $\PW\PW$ and $\PW\PZ$ processes. To obtain distributions in
the SR, the distributions in the CR are then normalized by the ratio of
the number of events in the SR ($\mathrm{N_{QCD MC}^{SR}}$) and CR
($\mathrm{N_{QCD MC}^{CR}}$) determined from simulated QCD multijet events:
\begin{equation}
	\mathrm{N_{QCD}^{SR}} = \mathrm{N_{resDATA}^{CR}} \, \frac{\mathrm{N_{QCD MC}^{SR}}}{\mathrm{N_{QCD MC}^{CR}}},
\label{eq:bck:qcdnorm}
\end{equation}
where $\mathrm{N_{resDATA}^{CR}}$ is the residual yield in data (after
subtracting the background contributions not from QCD multijet).
The SR-to-CR simulated events ratio in Eq.~(\ref{eq:bck:qcdnorm}) is 0.043 $\pm$ 0.014,
0.041 $\pm$ 0.012, and 0.081 $\pm$ 0.015 for three, four, and five or
more jets, respectively. The normalization
uncertainty is estimated to be 30\%. The shape uncertainty due to the
CR definition is evaluated by selecting events for which the lepton
fails the isolation requirement. The uncertainty is defined by the
difference between the distributions of events that pass or fail the
CSV discriminant requirement and can be as large as 60\% in some
regions of phase space.

\section{Event yields and control plots}
\label{sec:controlplots}

Table~\ref{tab:expyields} shows the expected and observed event yields
after event selection and \ttbar reconstruction, including the
statistical uncertainties in the expected yields.  All of the \ttbar
components depend on the top quark Yukawa coupling from the production,
so all of them are considered as signal. Here, the signal simulation is
divided into the following categories: correctly reconstructed \ttbar
systems (\ttbar right reco); events where all required decay products
are available, but the algorithm failed to identify the correct jet
assignments (\ttbar wrong reco); $\ell$+jets~\ttbar events where at
least one required decay product is missing (\ttbar
nonreconstructible); and \ttbar events from dileptonic,
$\PW \to\tau\nu$, or fully hadronic decays (\ttbar background).

\begin{table}[htbp]
\centering
\topcaption{
Expected and observed yields after event selection and \ttbar reconstruction, with statistical uncertainties in the expected yields. The QCD multijet yield is derived from Eq.~(\ref{eq:bck:qcdnorm}) and its uncertainty is the statistical uncertainty in the control region from the data-based QCD multijet determination described in Section~\ref{sec:bck}.}
\renewcommand{\arraystretch}{1.}
\begin{scotch}{l r@{$\pm$}l r@{$\pm$}l r@{$\pm$}l}
Source & \multicolumn{2}{c}{3 jets} & \multicolumn{2}{c}{4 jets} & \multicolumn{2}{c}{$\geq$5 jets} \\
\hline
\ttbar right reco	&  130\,520 &150 &  92\,900 &130 & 71\,640 &110 \\
\ttbar wrong reco	&  29\,298  &73  &  17\,356 &57  & 43\,073 &89  \\
\ttbar nonreco		&  50\,695  &96  &  88\,760 &130 & 80\,960 &120 \\
\ttbar background	&  53\,465  &99  &  26\,085 &69  & 25\,047 &68  \\[\cmsTabSkip]
Single \PQt	        &  17\,849  & 40 &  6922    &27   & 6294 &26   \\
V+jets		        &  8990     & 100&  2824    &52   & 2478 &49   \\
QCD multijet 	    &  19\,840  & 69 &  2100    &25   & 1080 &30  \\[\cmsTabSkip]
Expected sum		& 310\,650  &250  & 236\,950 &210 & 230\,570 &210\\
Data & \multicolumn{2}{l}{308\,932} & \multicolumn{2}{l}{237\,491} & \multicolumn{2}{l}{226\,788}\\
\end{scotch}
\label{tab:expyields}
\end{table}

\begin{figure*}[h!tbp]
\centering
\includegraphics[width=0.44\textwidth]{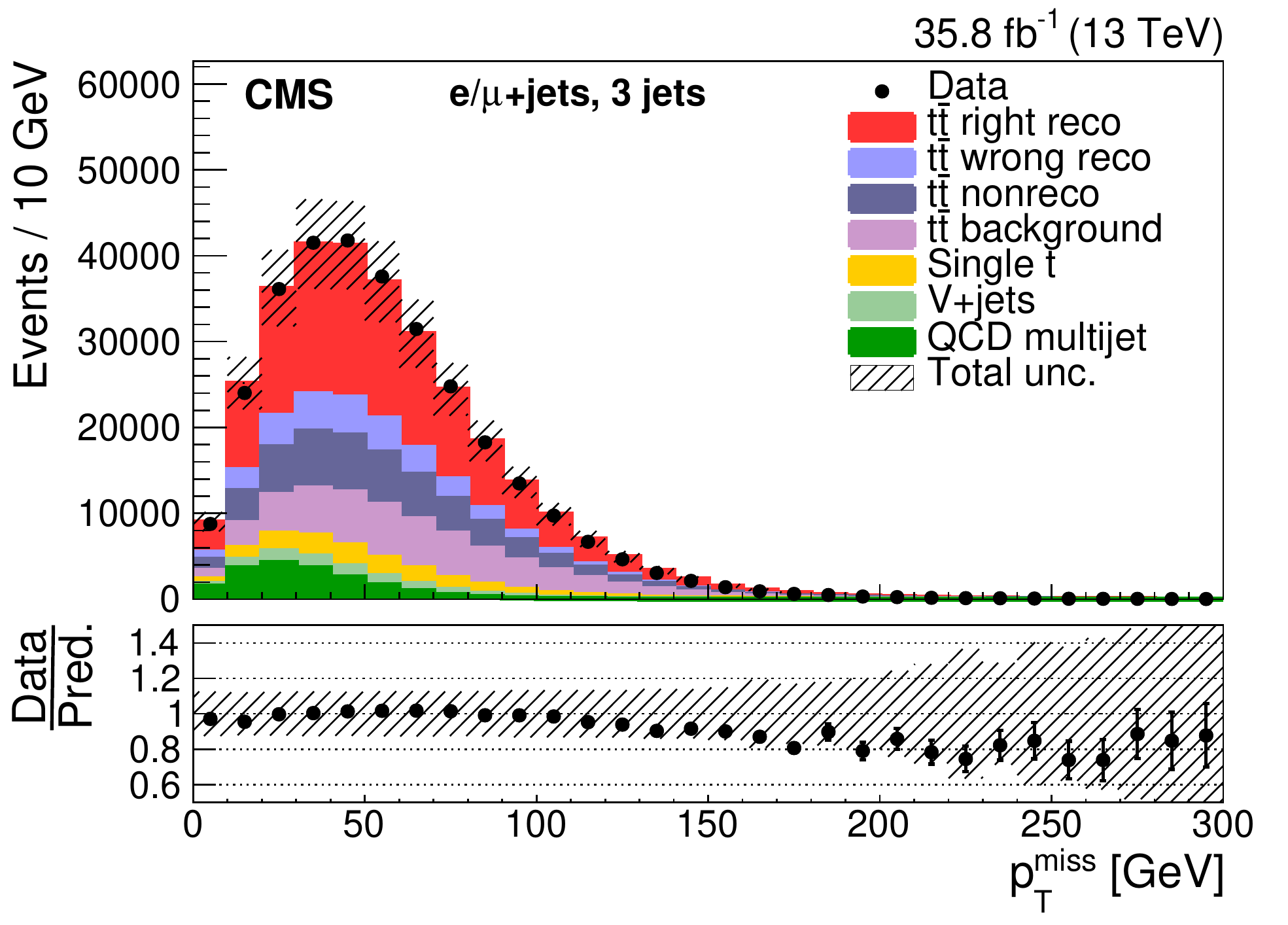}
\includegraphics[width=0.44\textwidth]{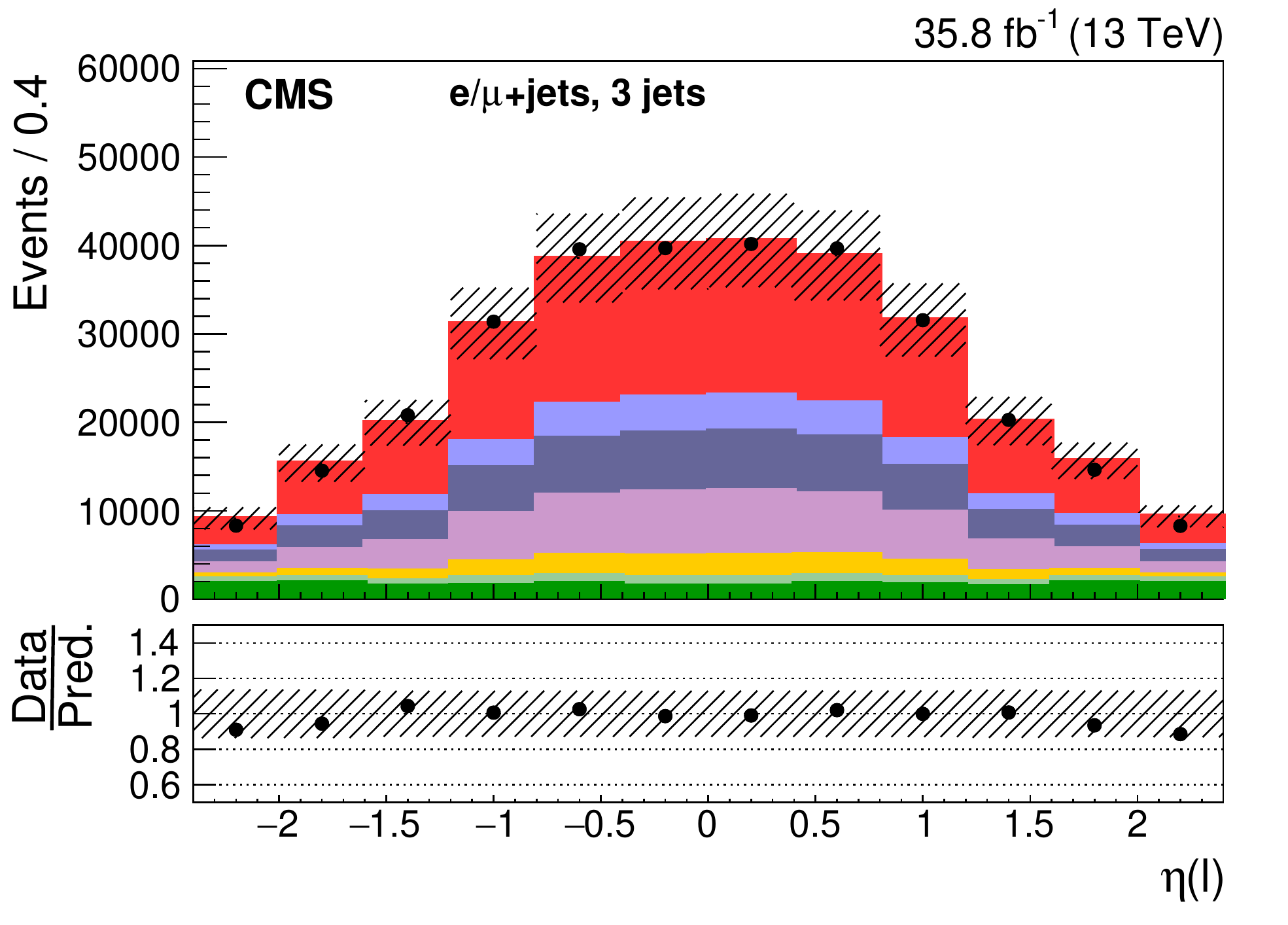}
\includegraphics[width=0.44\textwidth]{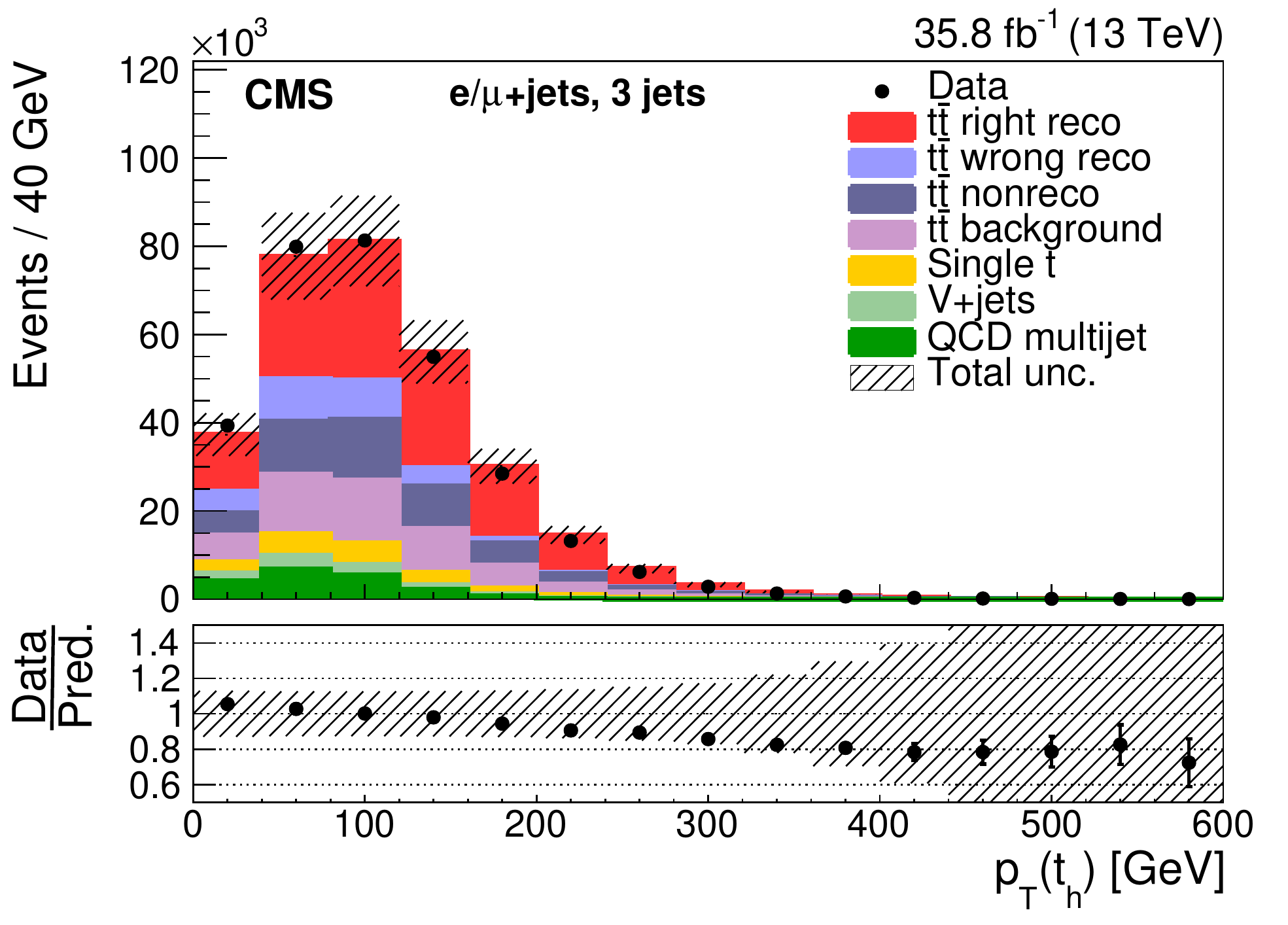}
\includegraphics[width=0.44\textwidth]{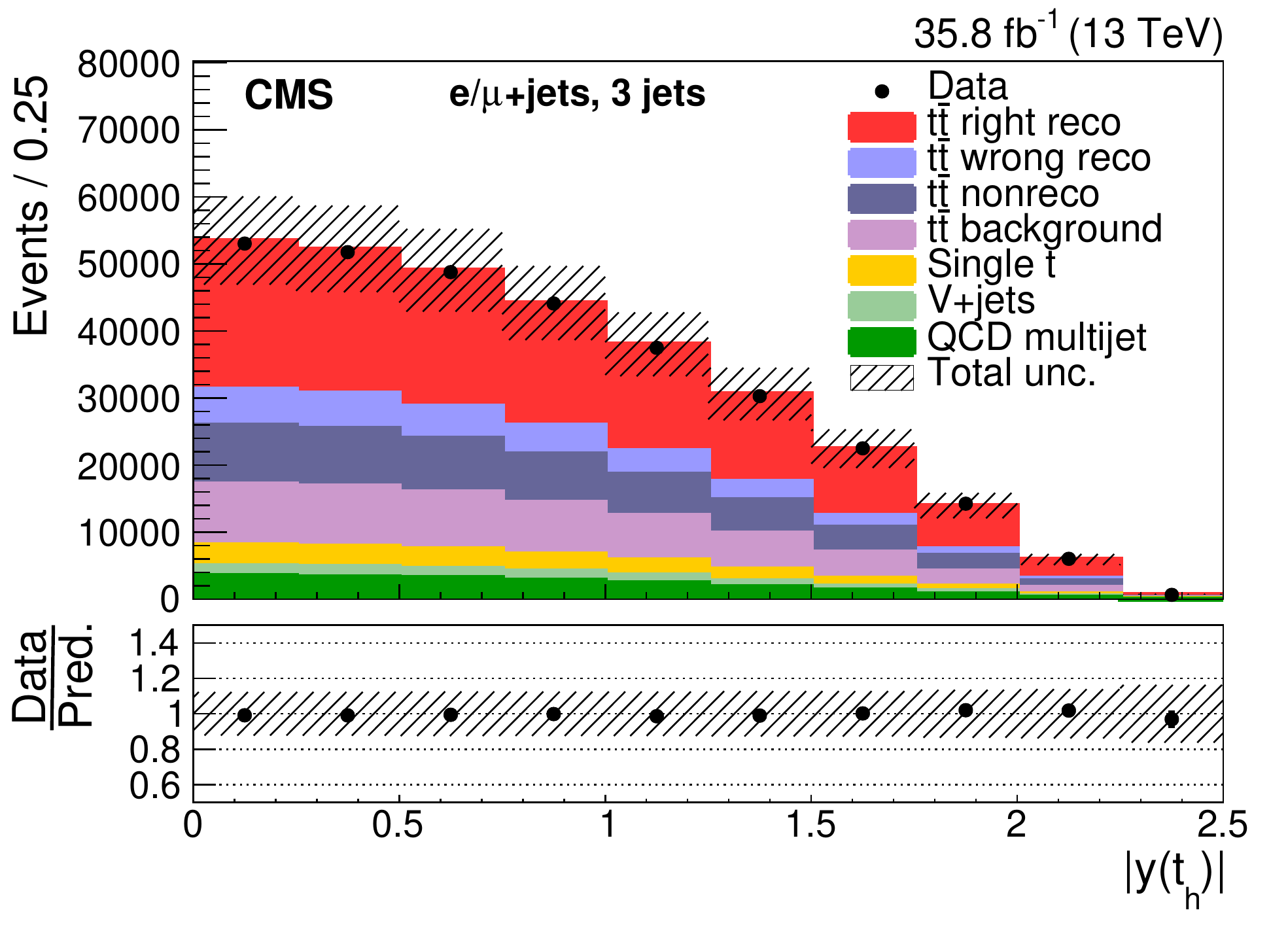}
\includegraphics[width=0.44\textwidth]{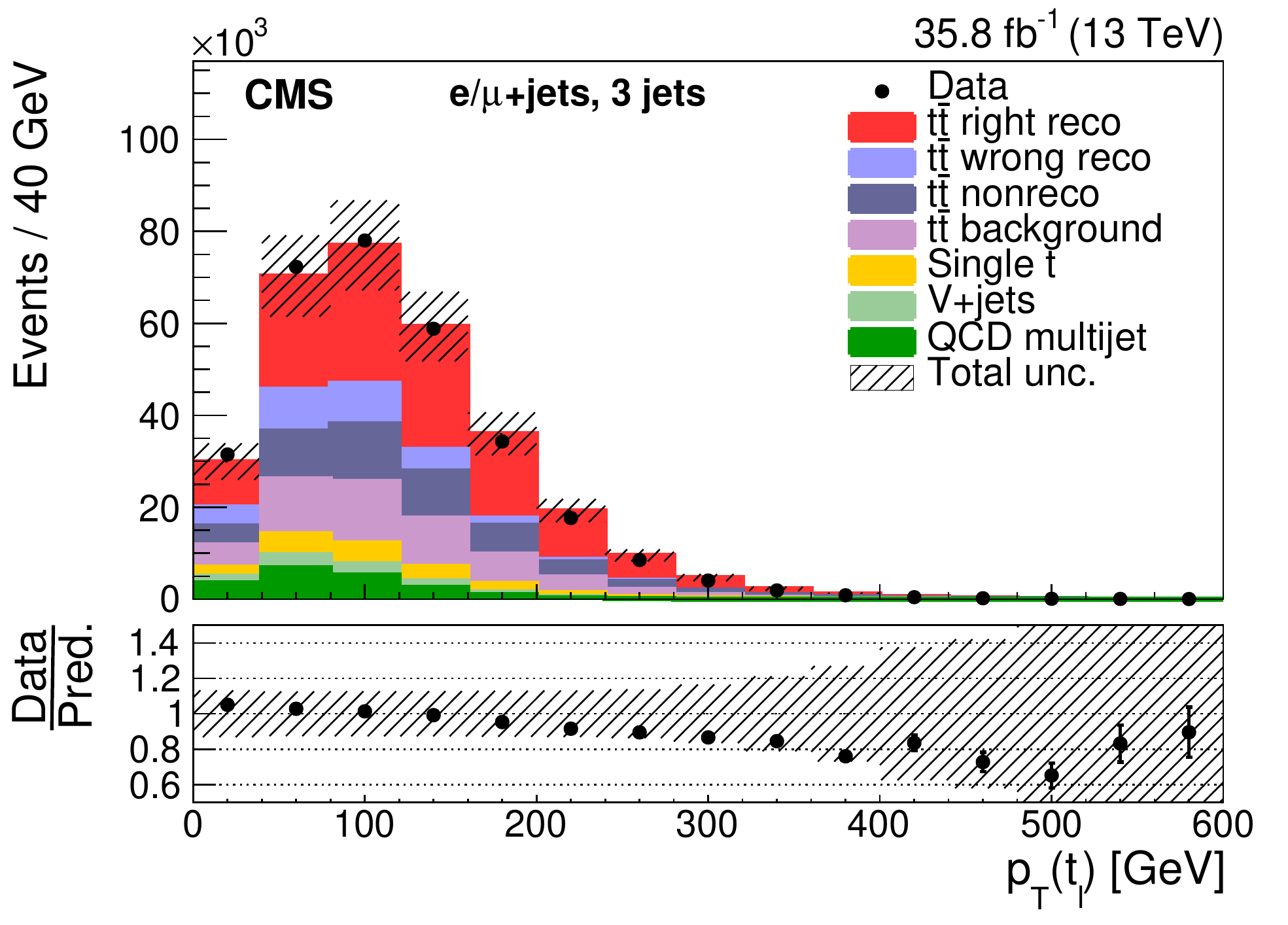}
\includegraphics[width=0.44\textwidth]{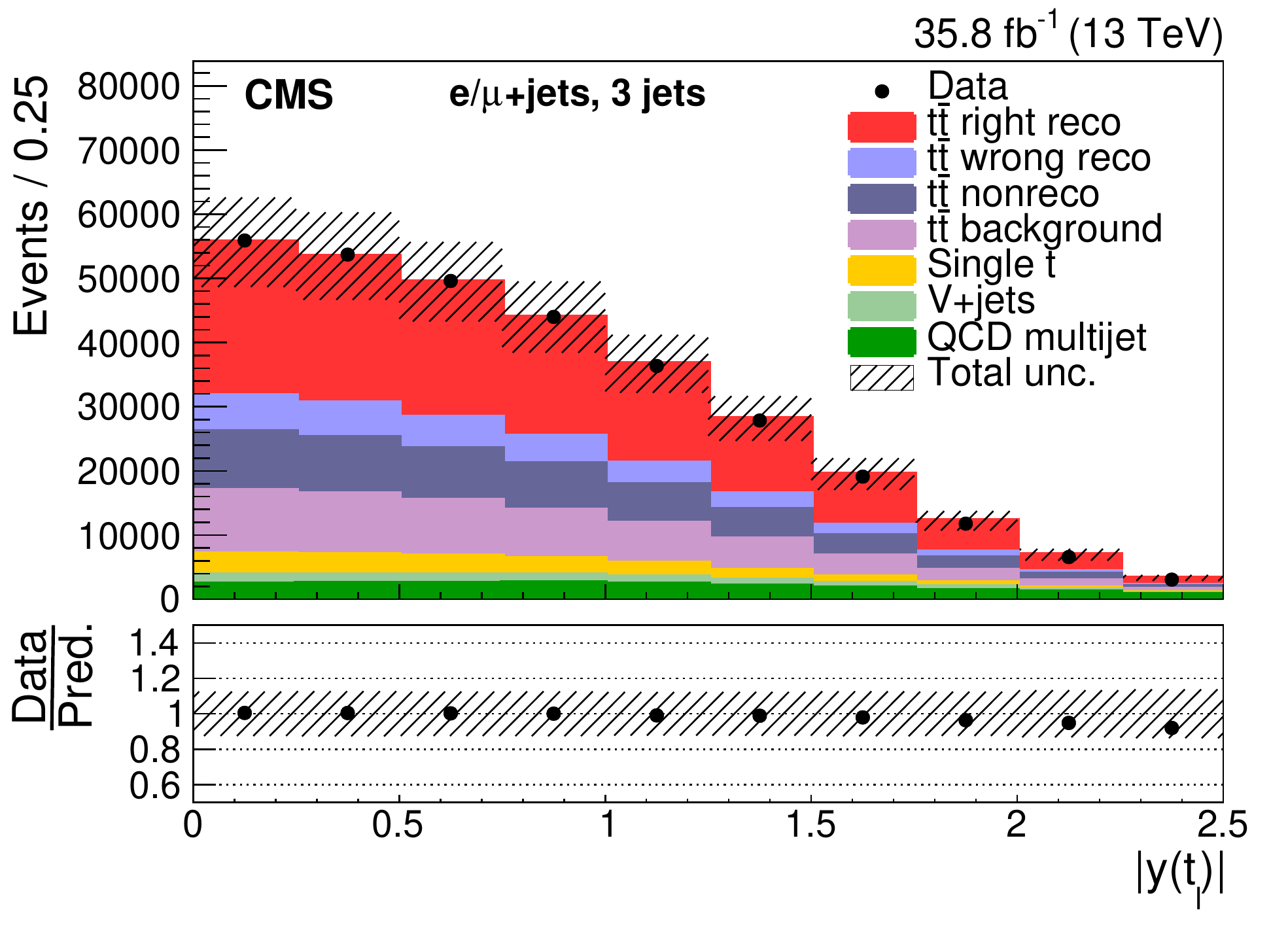}
\includegraphics[width=0.44\textwidth]{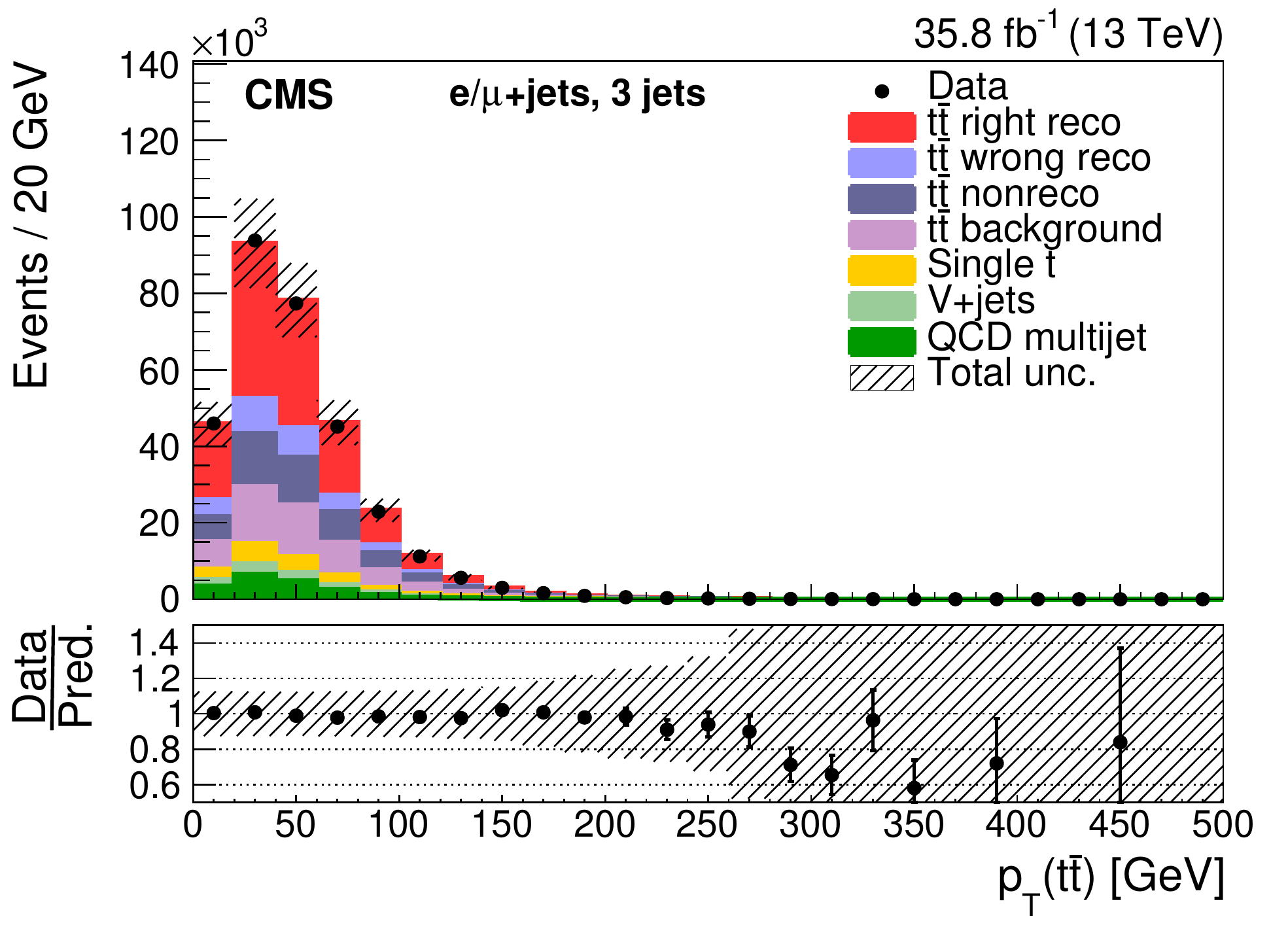}
\includegraphics[width=0.44\textwidth]{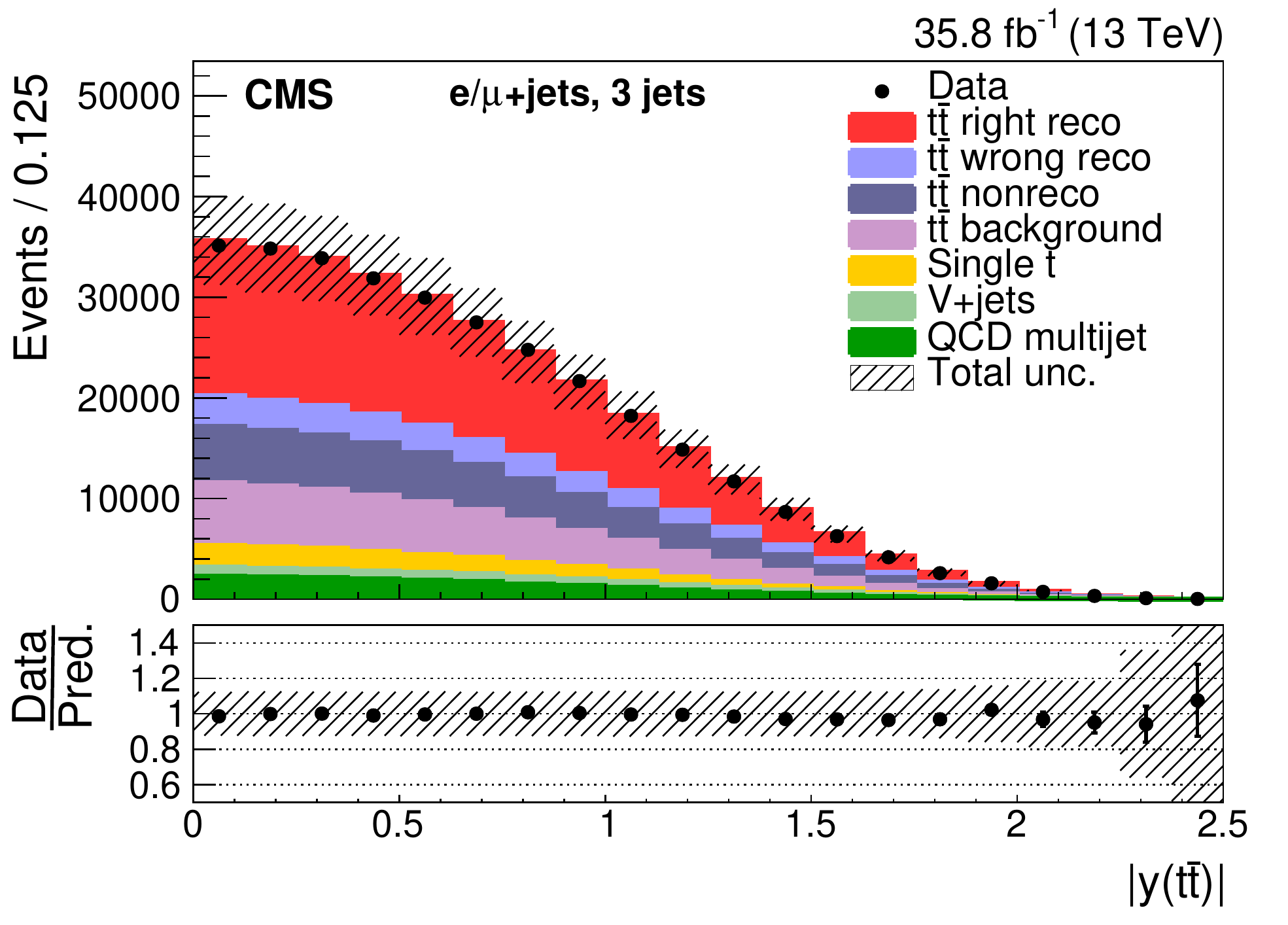}
\caption{Three-jet events after selection and \ttbar reconstruction. The plots show (left to right, upper to lower) the missing transverse momentum (\ptmiss), the lepton pseudorapidity, and $\pt$ and the absolute rapidity of the top quark decaying hadronically, semileptonically, and of the \ttbar system. The hatched band shows the total uncertainty associated with the signal and background predictions with the individual sources of uncertainty assumed to be uncorrelated. The ratios of data to the sum of the predicted yields are provided at the bottom of each panel.}
\label{p:controlplots:3j}
\end{figure*}

\begin{figure*}[h!tbp]
\centering
\includegraphics[width=0.44\textwidth]{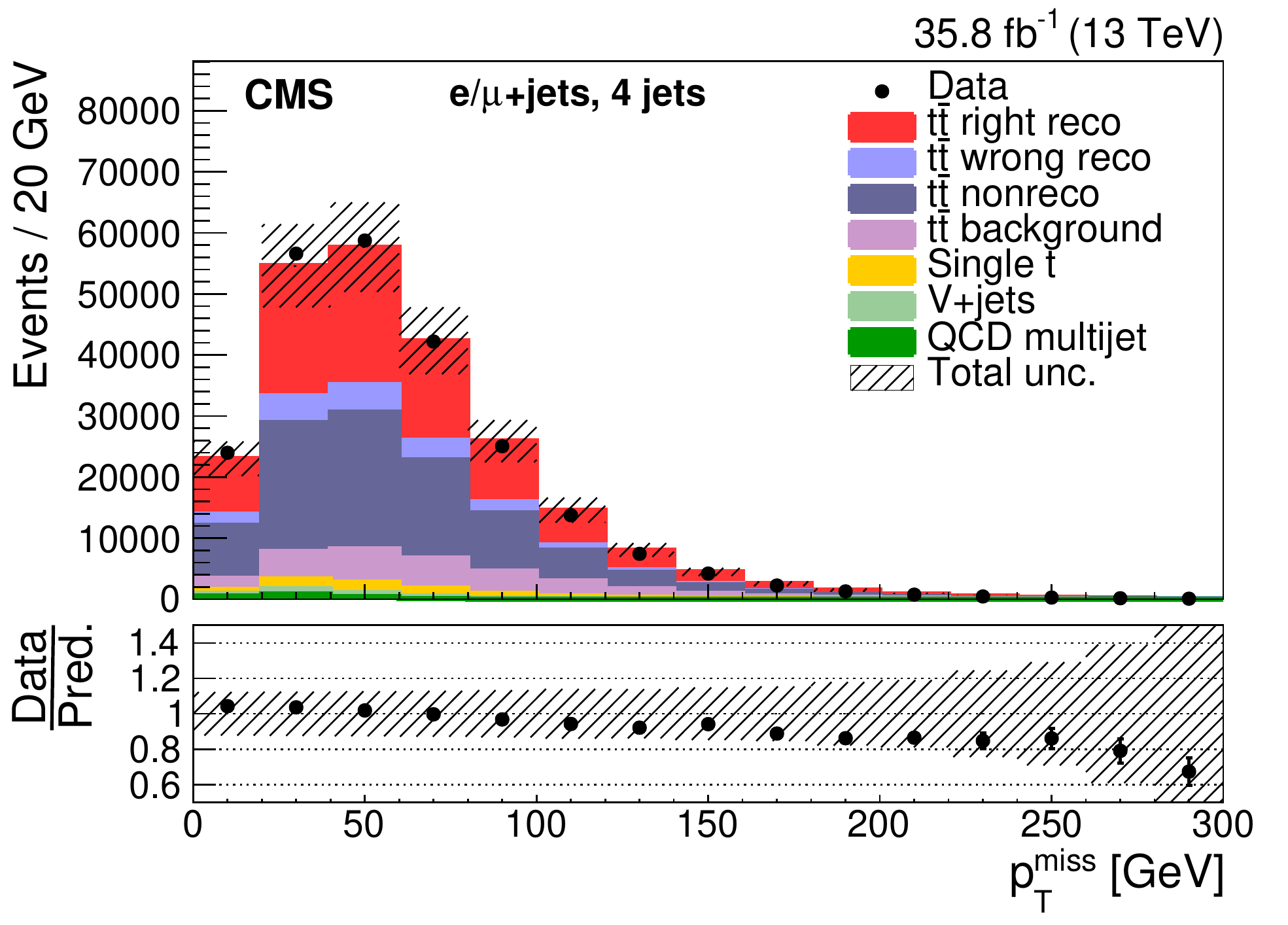}
\includegraphics[width=0.44\textwidth]{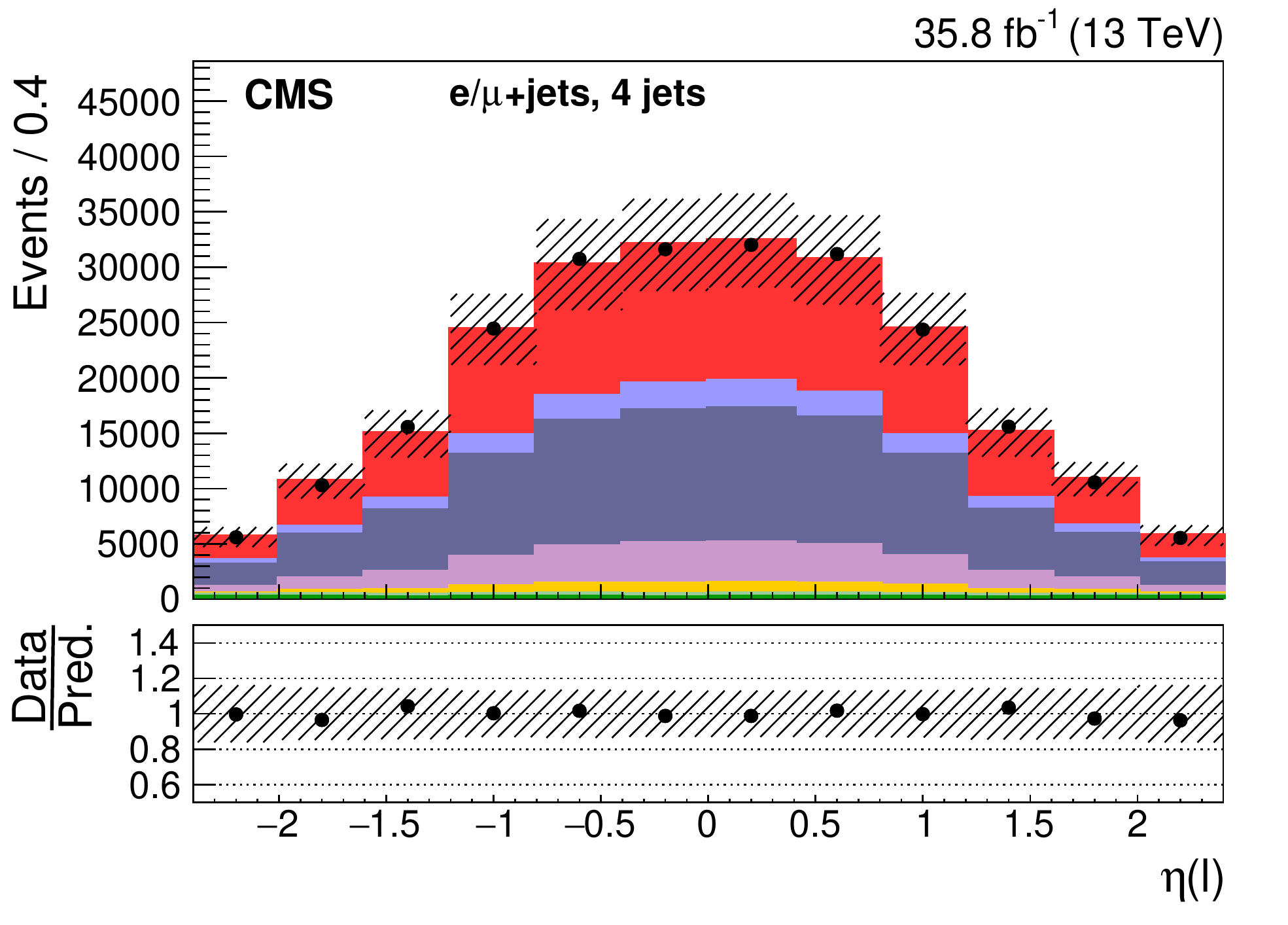}
\includegraphics[width=0.44\textwidth]{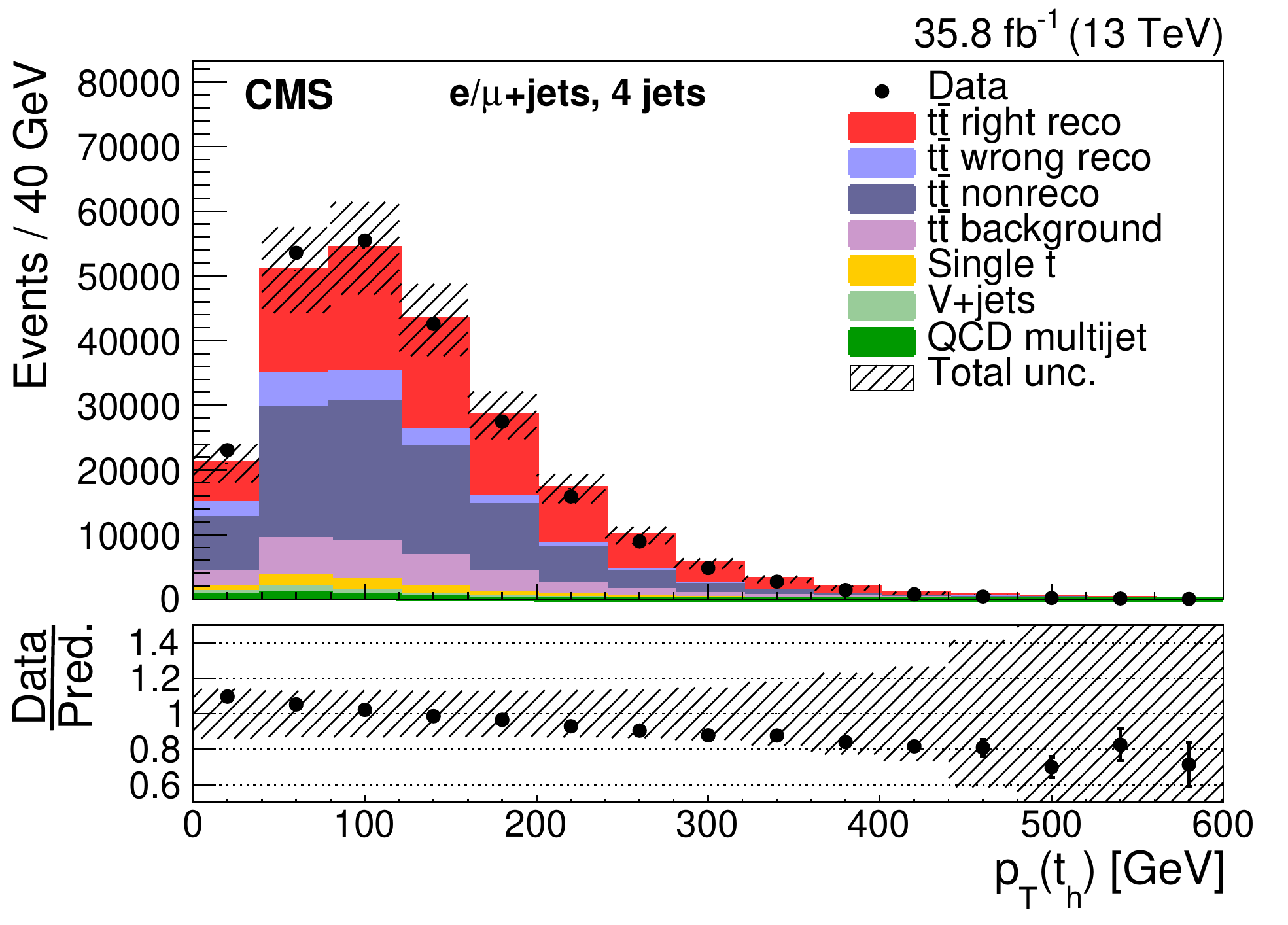}
\includegraphics[width=0.44\textwidth]{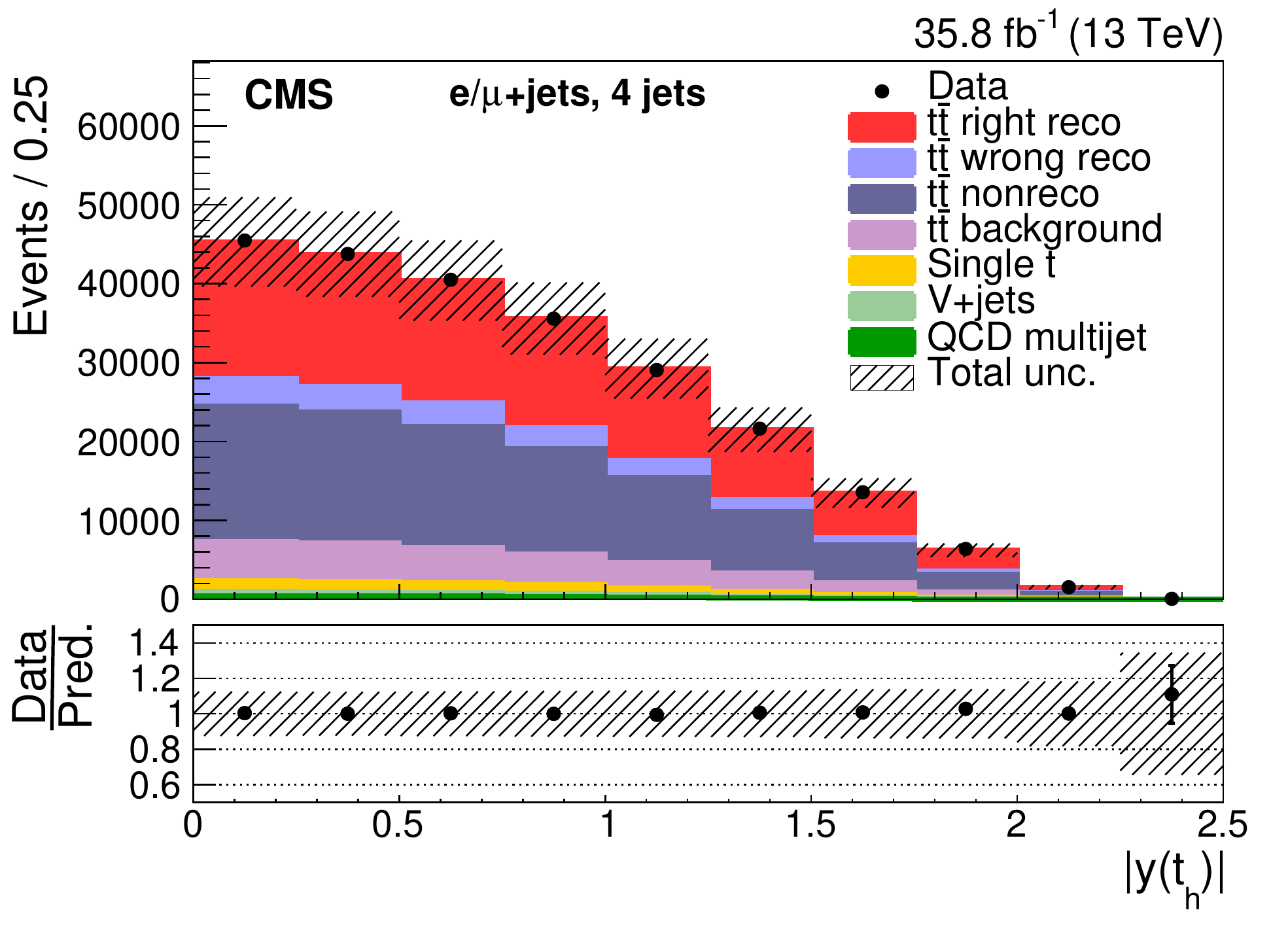}
\includegraphics[width=0.44\textwidth]{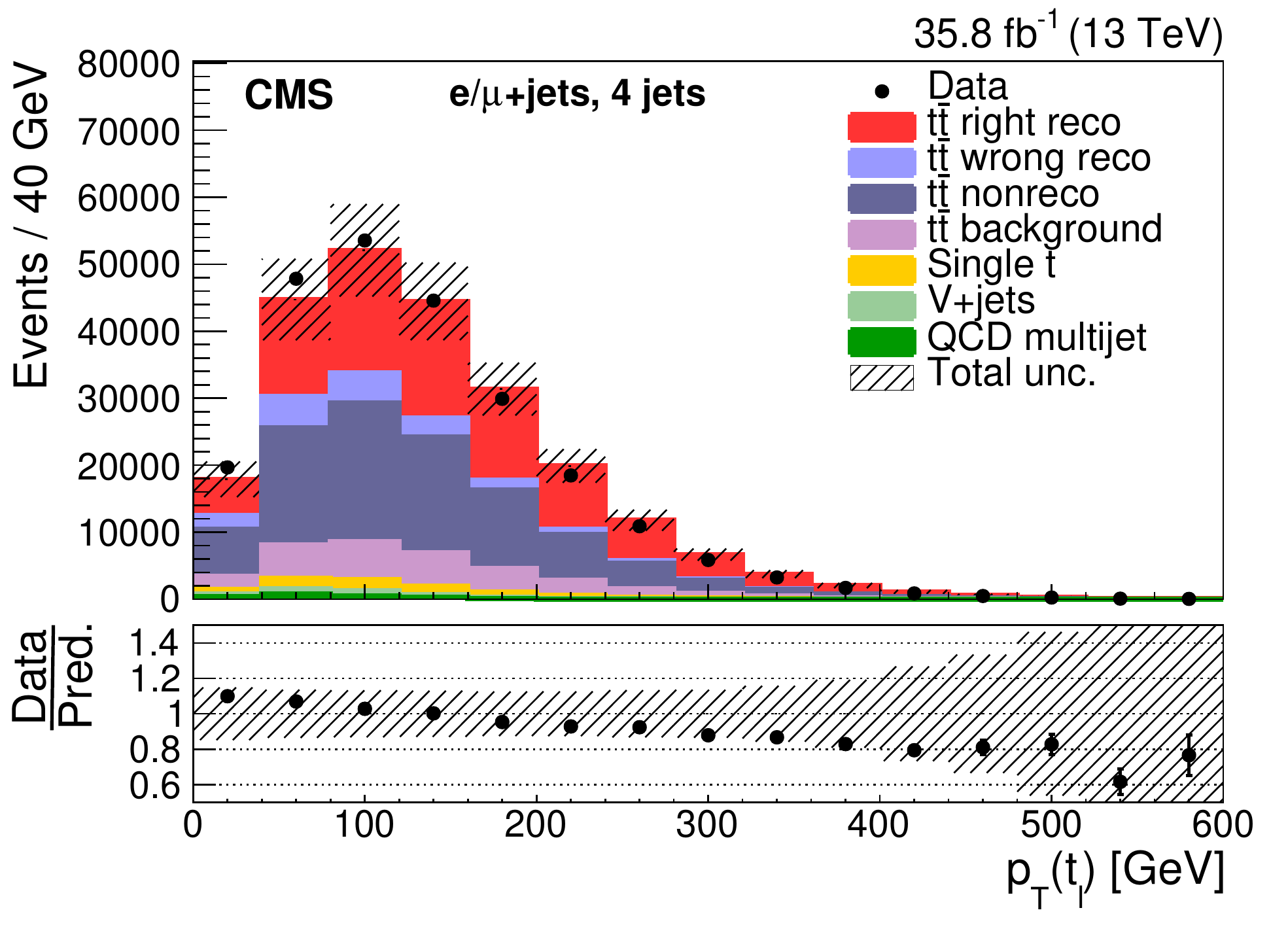}
\includegraphics[width=0.44\textwidth]{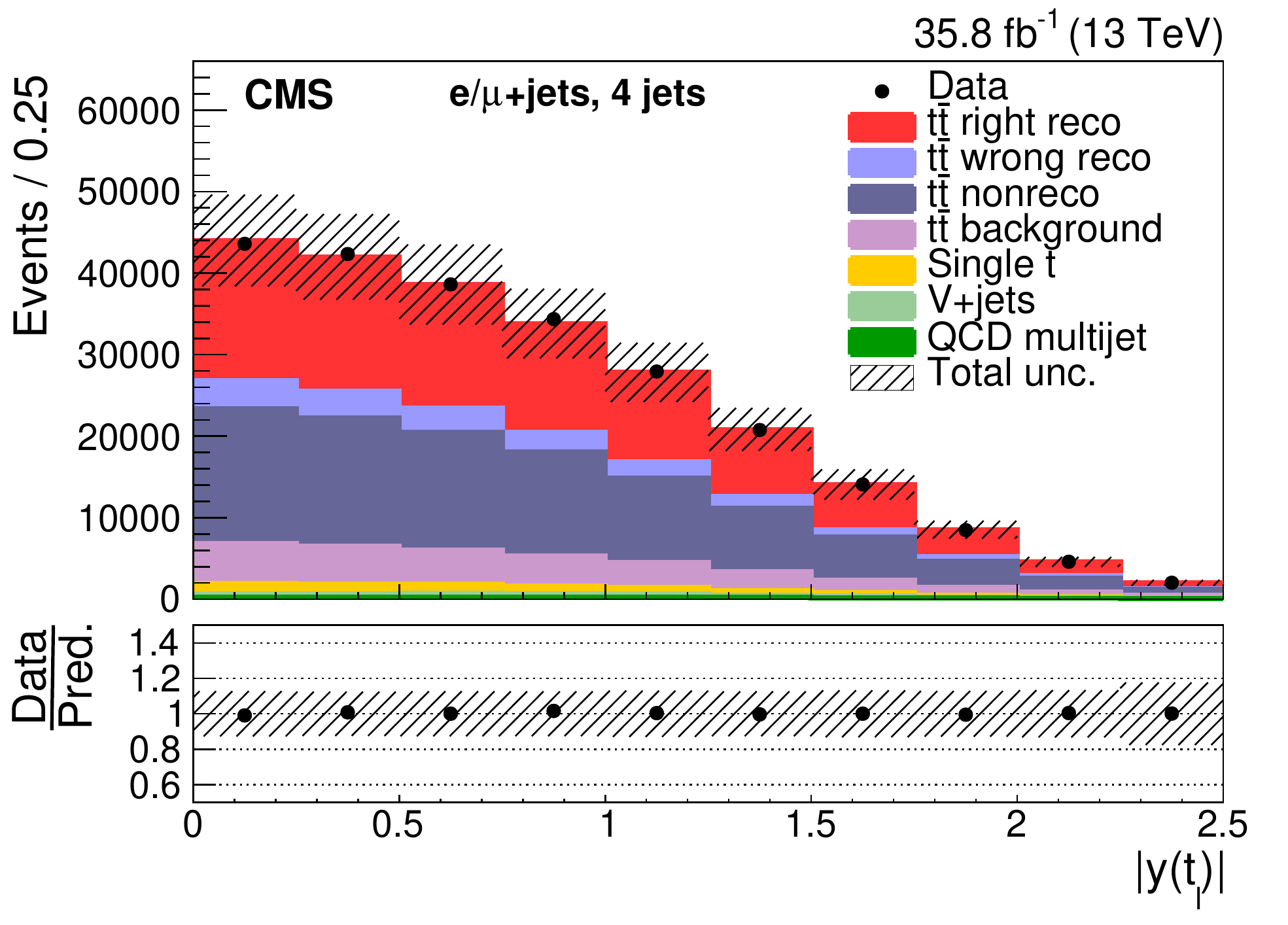}
\includegraphics[width=0.44\textwidth]{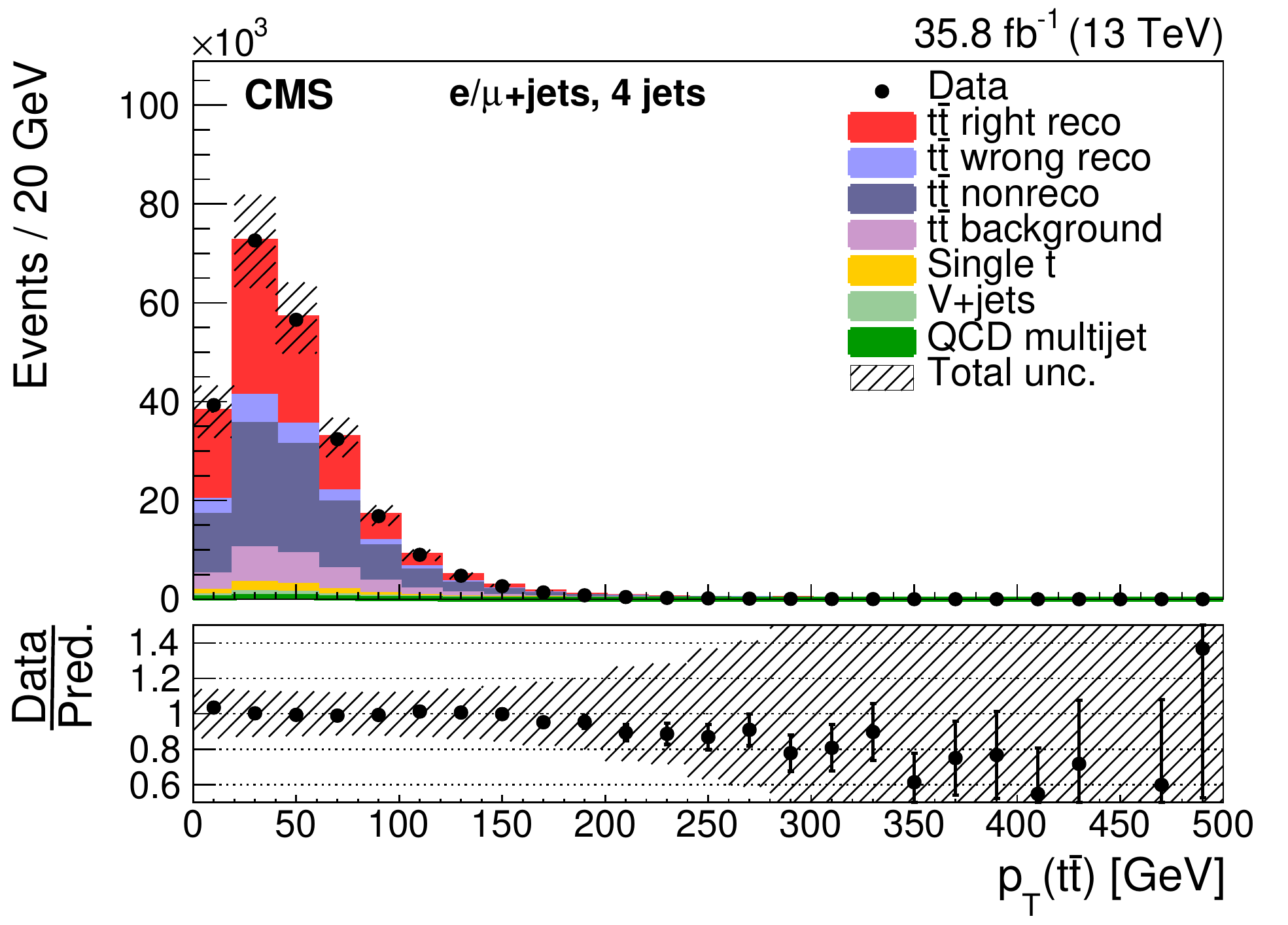}
\includegraphics[width=0.44\textwidth]{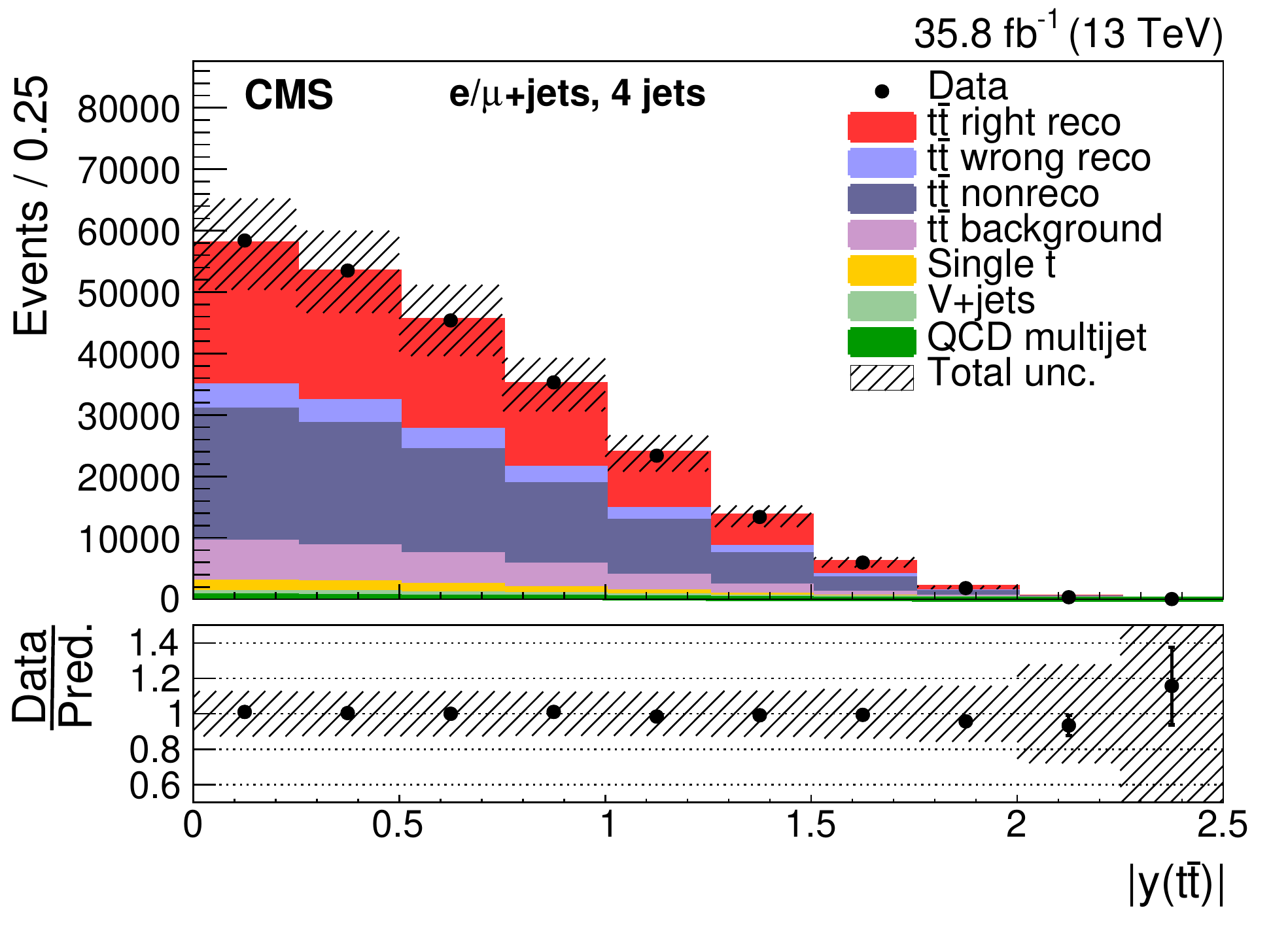}
\caption{Four-jet events after selection and \ttbar reconstruction. Same distributions as described in Fig.~\ref{p:controlplots:3j}.}
\label{p:controlplots:4j}
\end{figure*}

\begin{figure*}[h!tbp]
\centering
\includegraphics[width=0.44\textwidth]{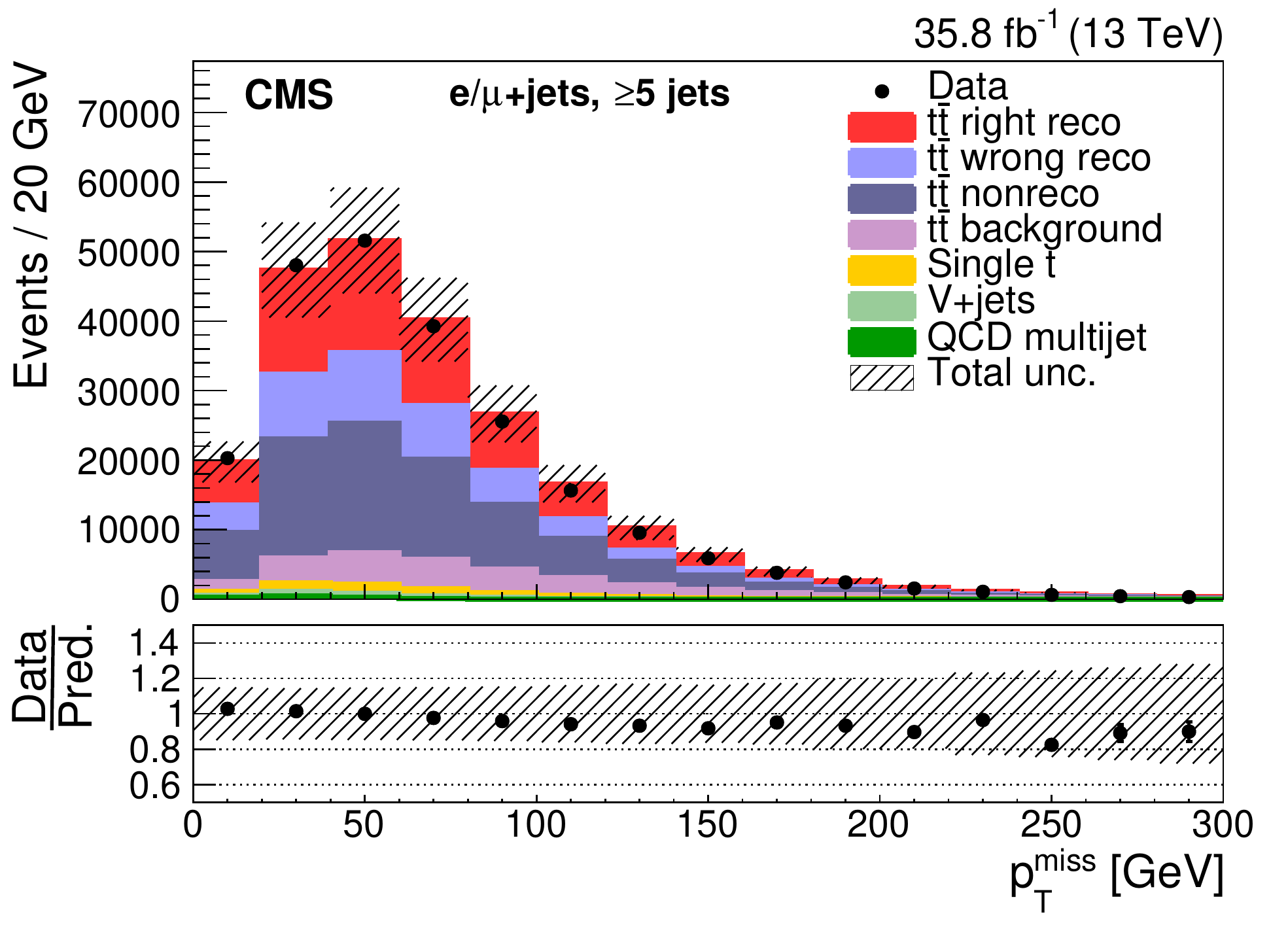}
\includegraphics[width=0.44\textwidth]{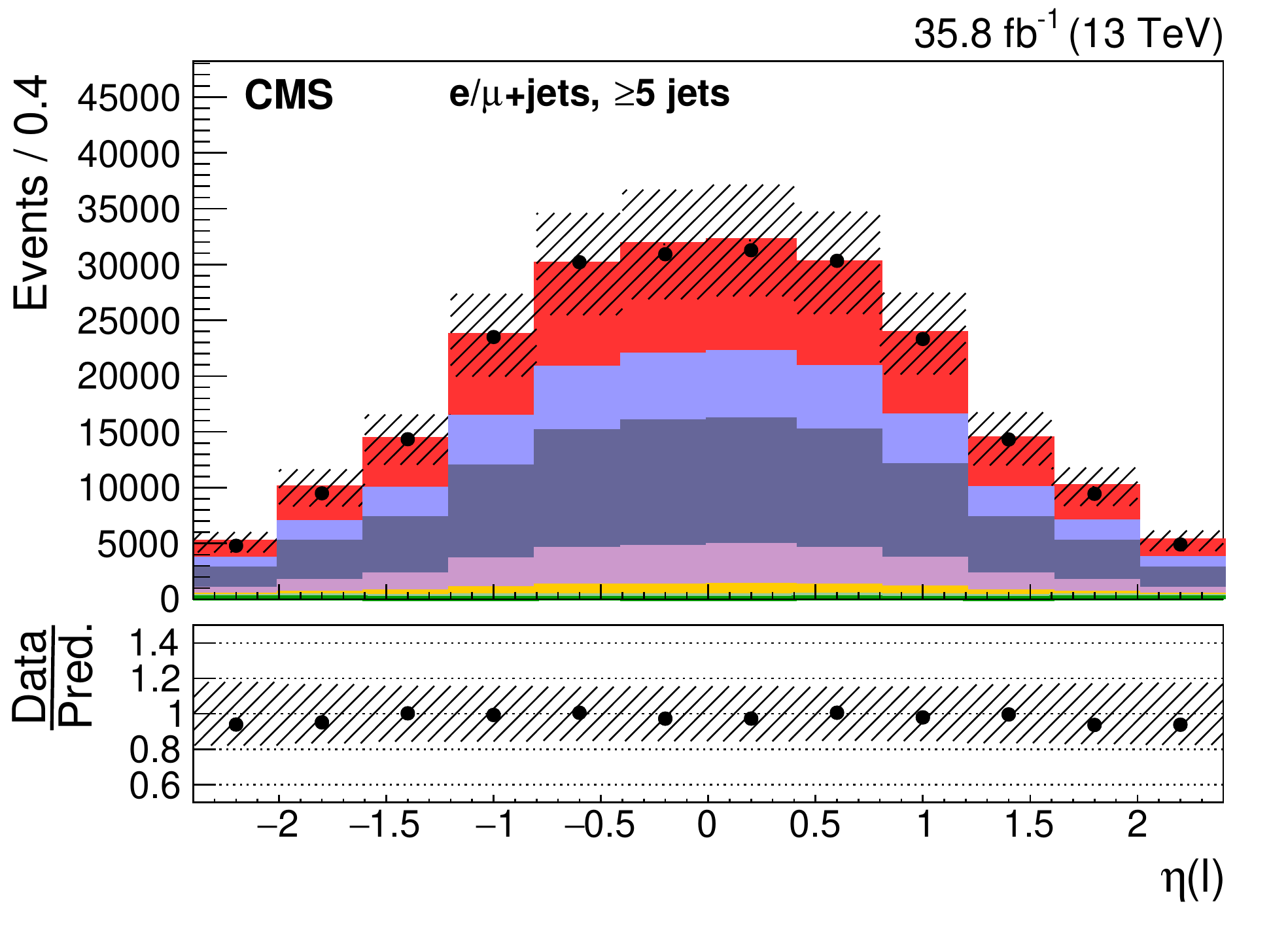}
\includegraphics[width=0.44\textwidth]{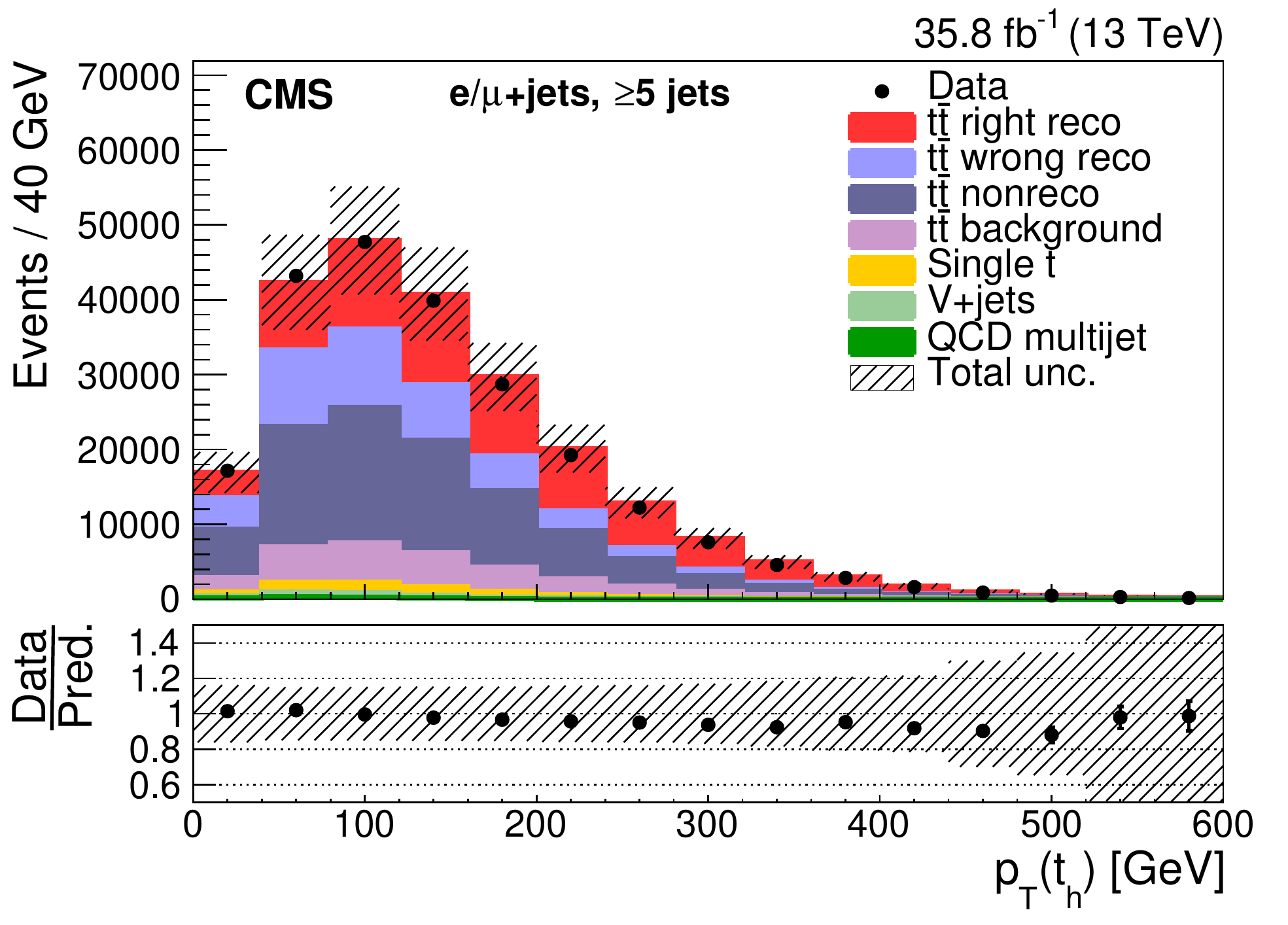}
\includegraphics[width=0.44\textwidth]{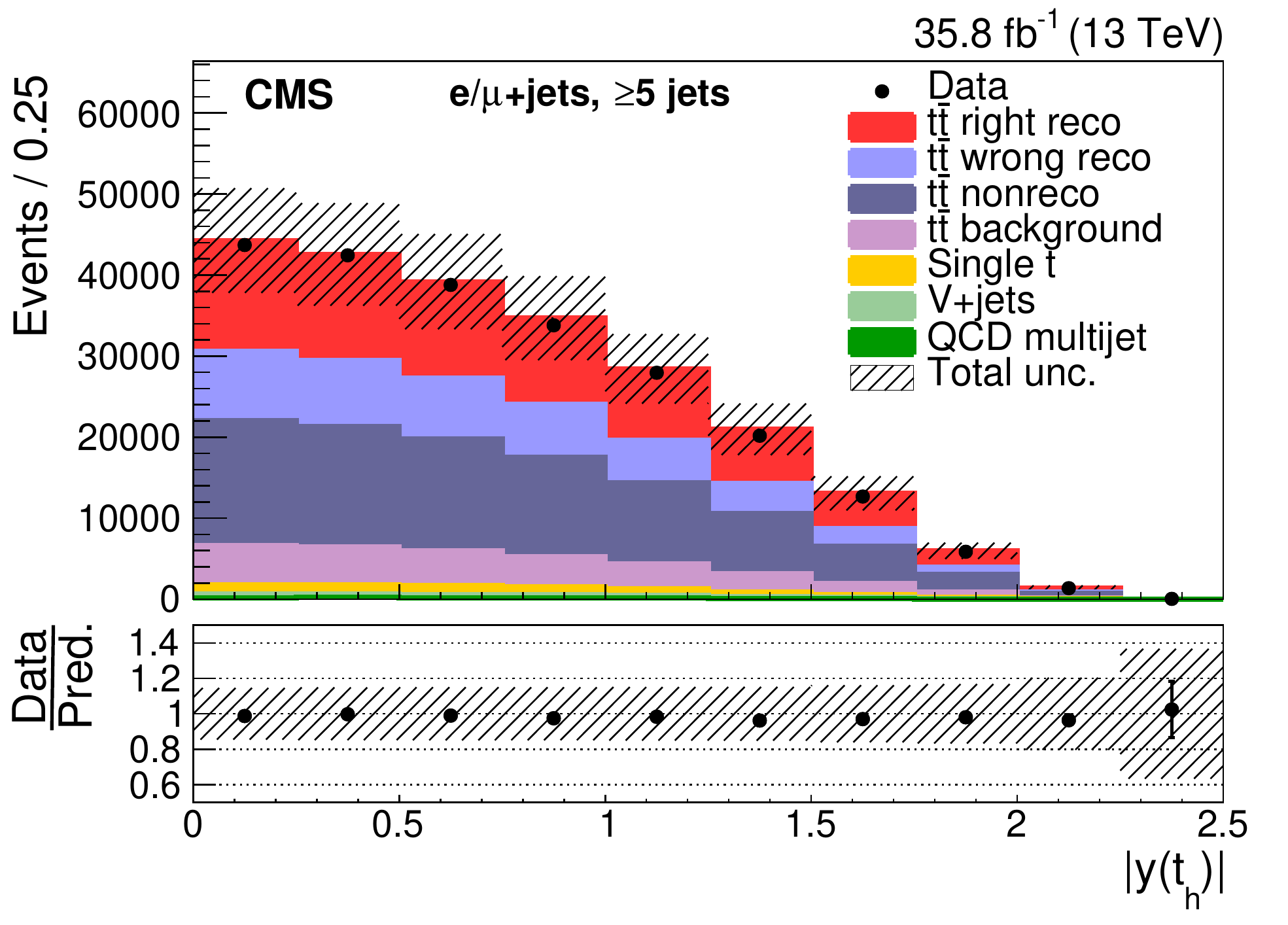}
\includegraphics[width=0.44\textwidth]{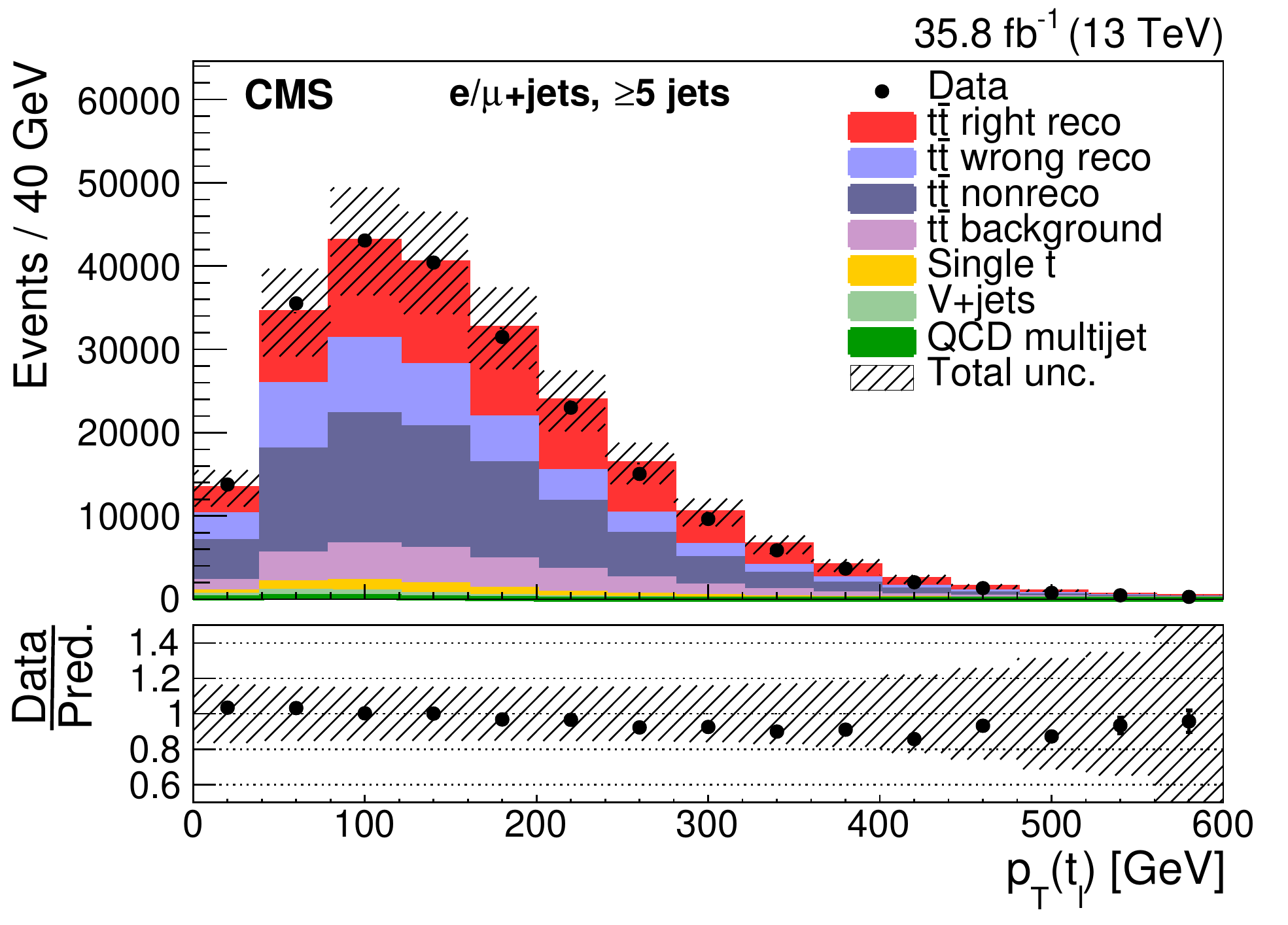}
\includegraphics[width=0.44\textwidth]{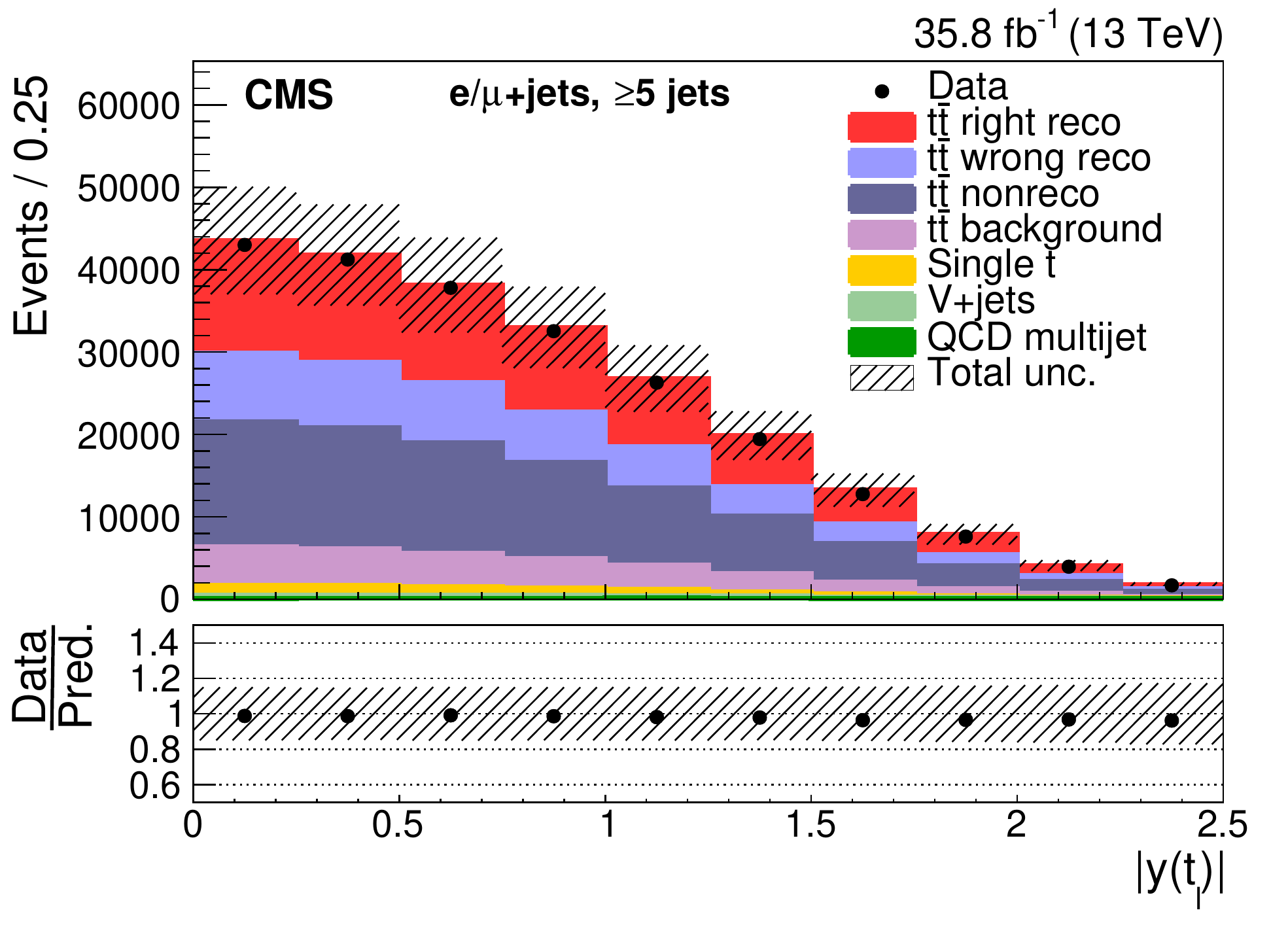}
\includegraphics[width=0.44\textwidth]{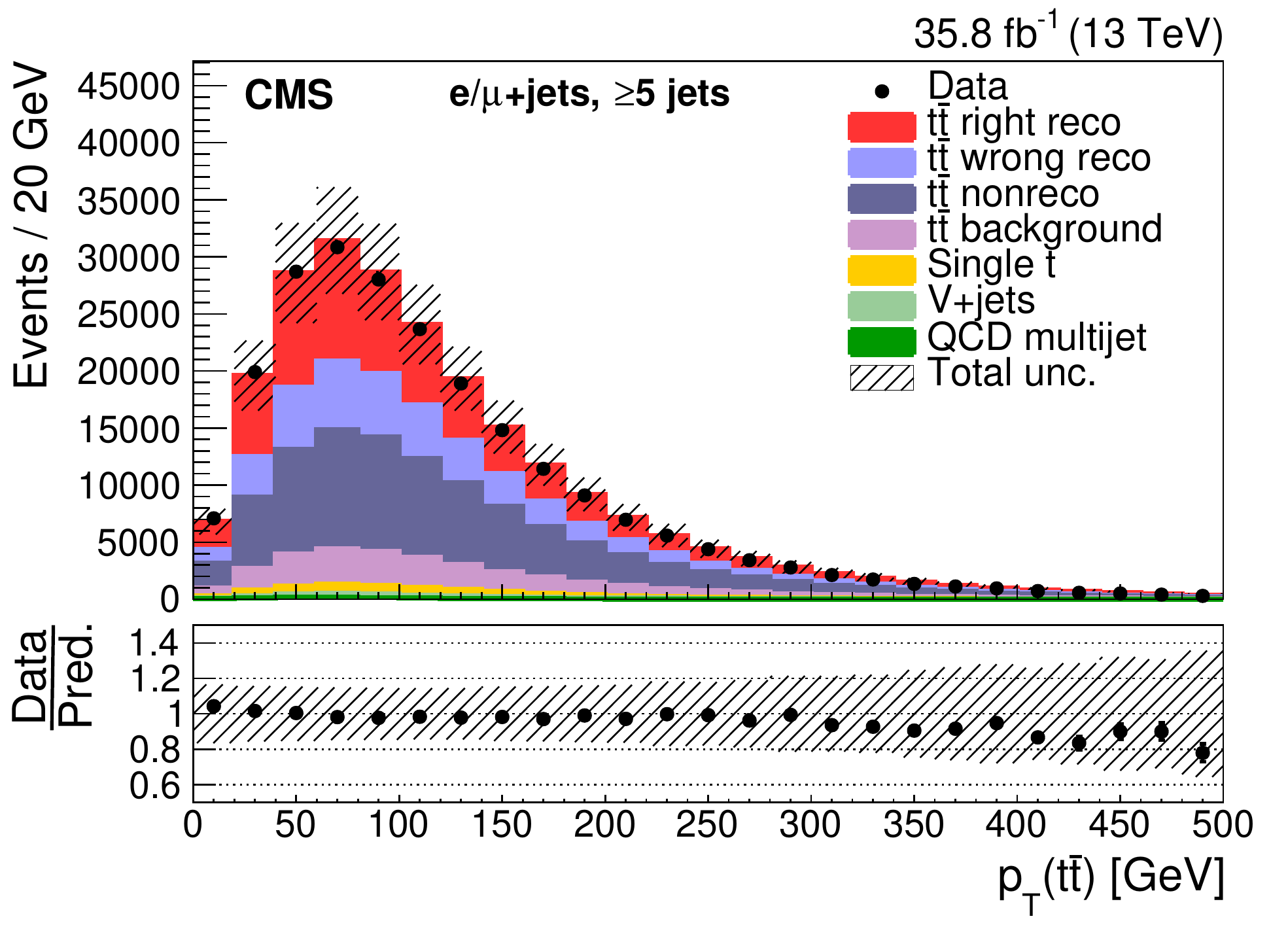}
\includegraphics[width=0.44\textwidth]{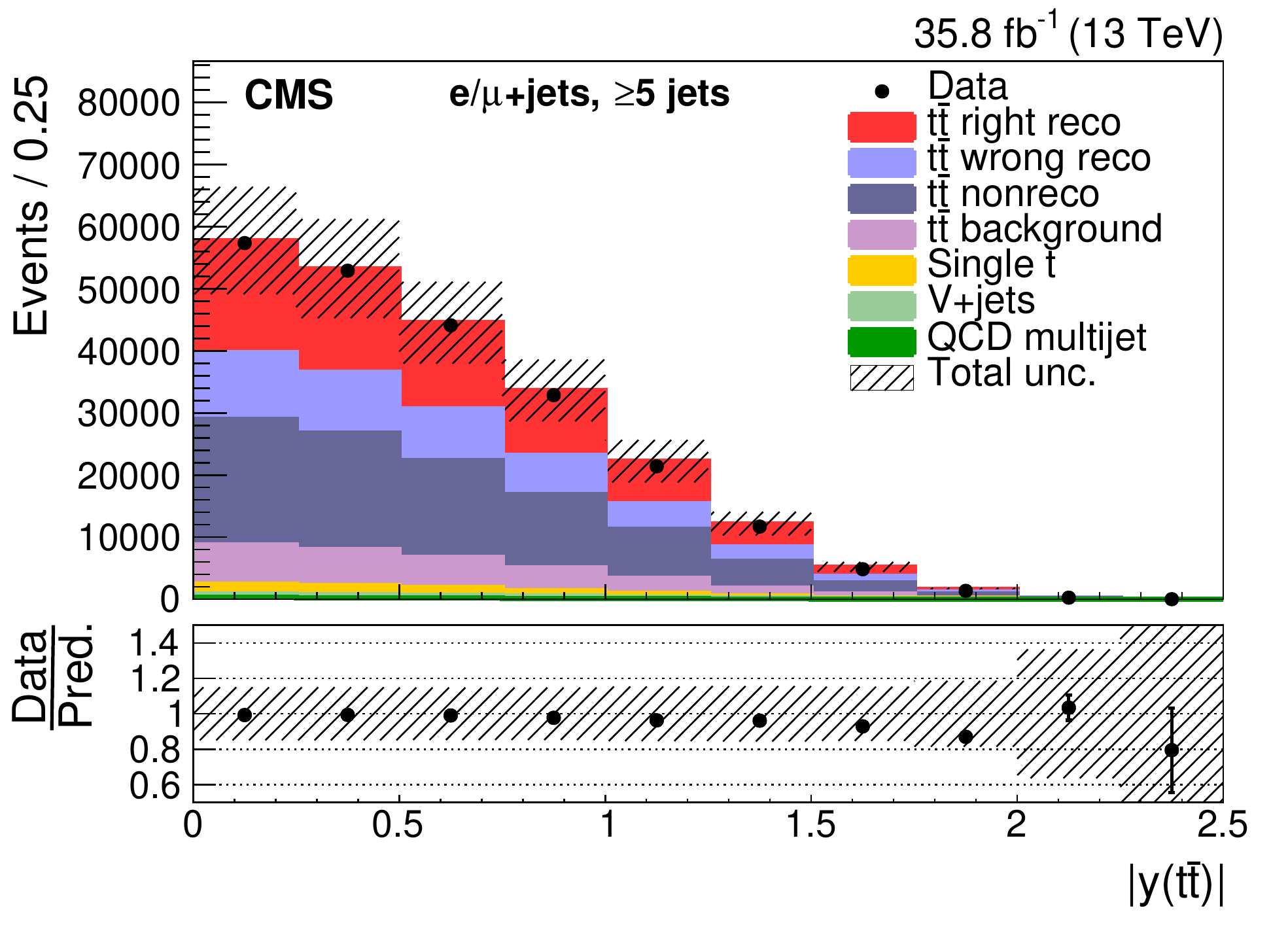}
\caption{Events with five or more jets after selection and \ttbar reconstruction. Same distributions as described in Fig.~\ref{p:controlplots:3j}.}
\label{p:controlplots:5j}
\end{figure*}

Figures~\ref{p:controlplots:3j}--\ref{p:controlplots:5j} show the
comparison of data and simulation for \ptvecmiss, the pseudorapidity
of the lepton, and several kinematic variables of the top quarks and
\ttbar system.
In general, good agreement between data and prediction is observed. 
The data appear to have a deficit for high top quark \pt with respect 
to the available MC generators. This trend has been observed before in 
Refs.~\cite{Aad:2015mbv,Khachatryan:2015oqa} and~\cite{Aaboud:2017fha,Sirunyan:2018wem}
both at 8 and 13\TeV, and recent differential 
NNLO calculations~\cite{NNLO,Catani:2019hip} reduce the discrepancy.

\section{Determination of \texorpdfstring{\yukawa}{Yukawa}}
\label{sec:stat}

The two-dimensional data distributions in (\Mttb, $\abs{\deltaY}$) are fit to
the sum of the predicted contributions to infer the value of \yukawa
for events with three, four, and five or more jets in the final state.
The bin limits are selected to capture the different behavior of the
weak interaction correction, as seen in
Fig.~\ref{p:reweight:1d}. There are three bins in $\abs{\deltaY}$: 0--0.6,
0.6--1.2, and $>$1.2. A minimum of 10\,000 simulated events are
required in each (\Mttb, $\abs{\deltaY}$) bin.
This results in 21, 17, and 17 bins for event categories with three,
four, and five or more jets, respectively.

\begin{figure*}[htbp]
\centering
\includegraphics[width=0.32\textwidth]{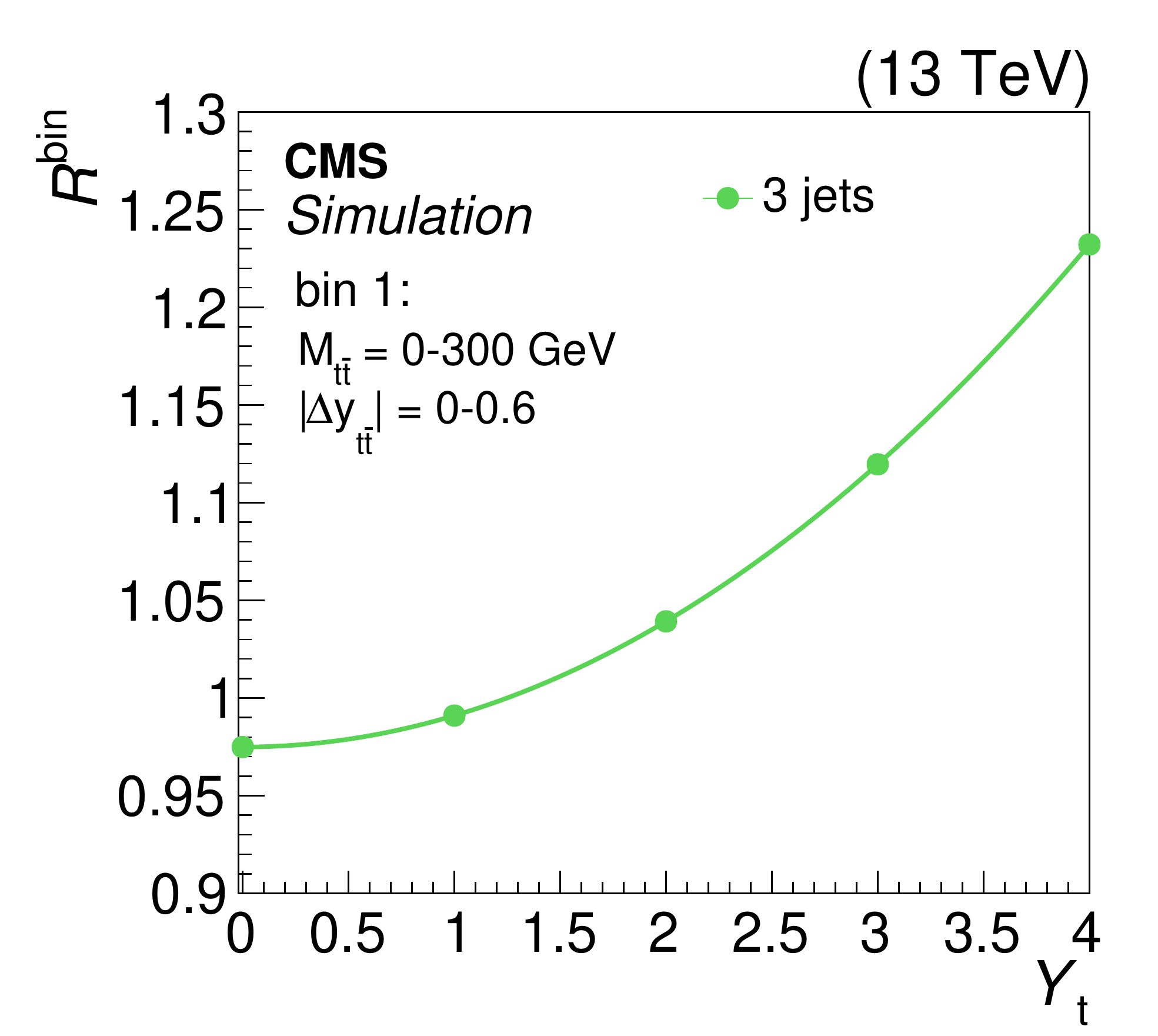}
\includegraphics[width=0.32\textwidth]{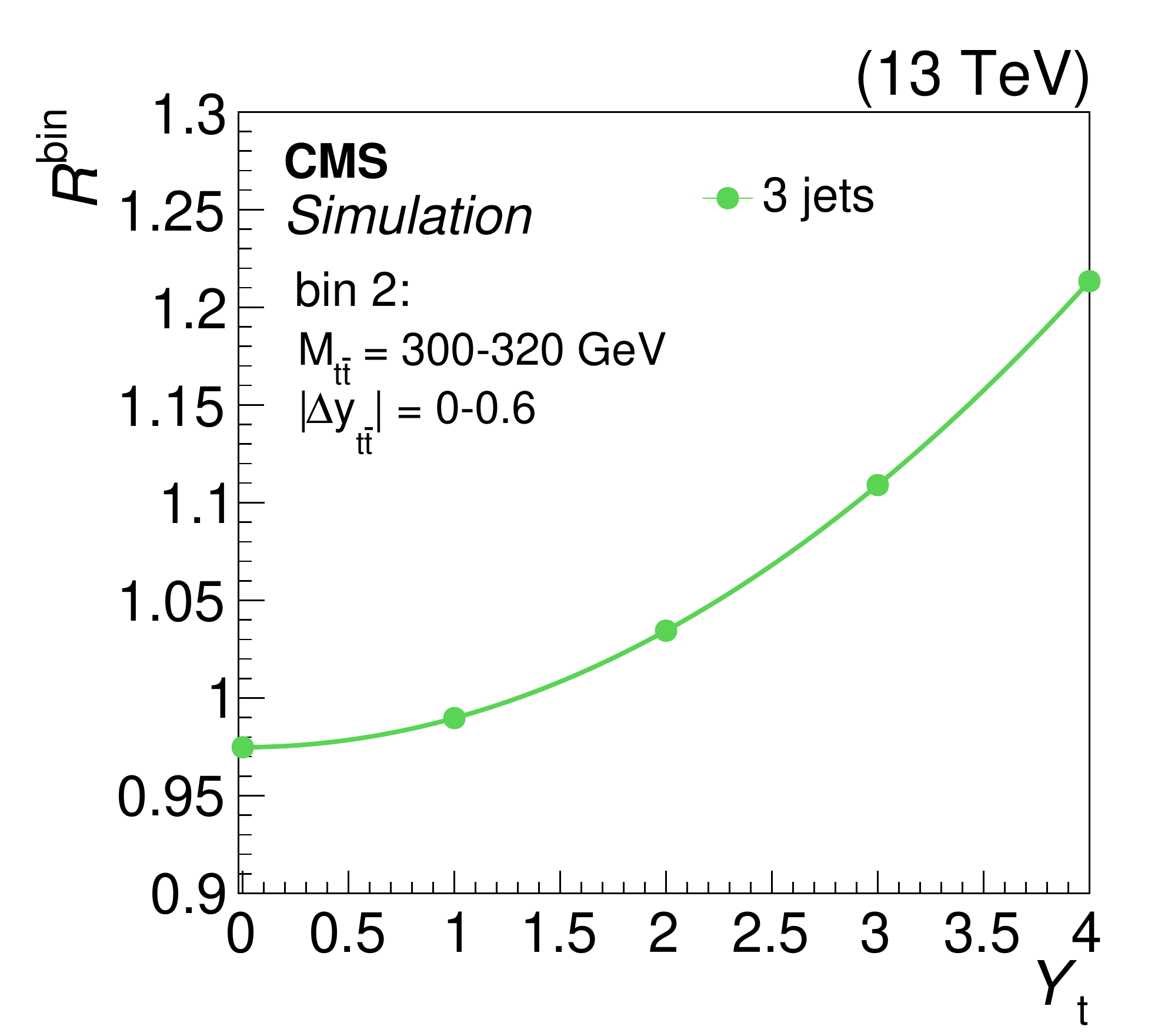}
\includegraphics[width=0.32\textwidth]{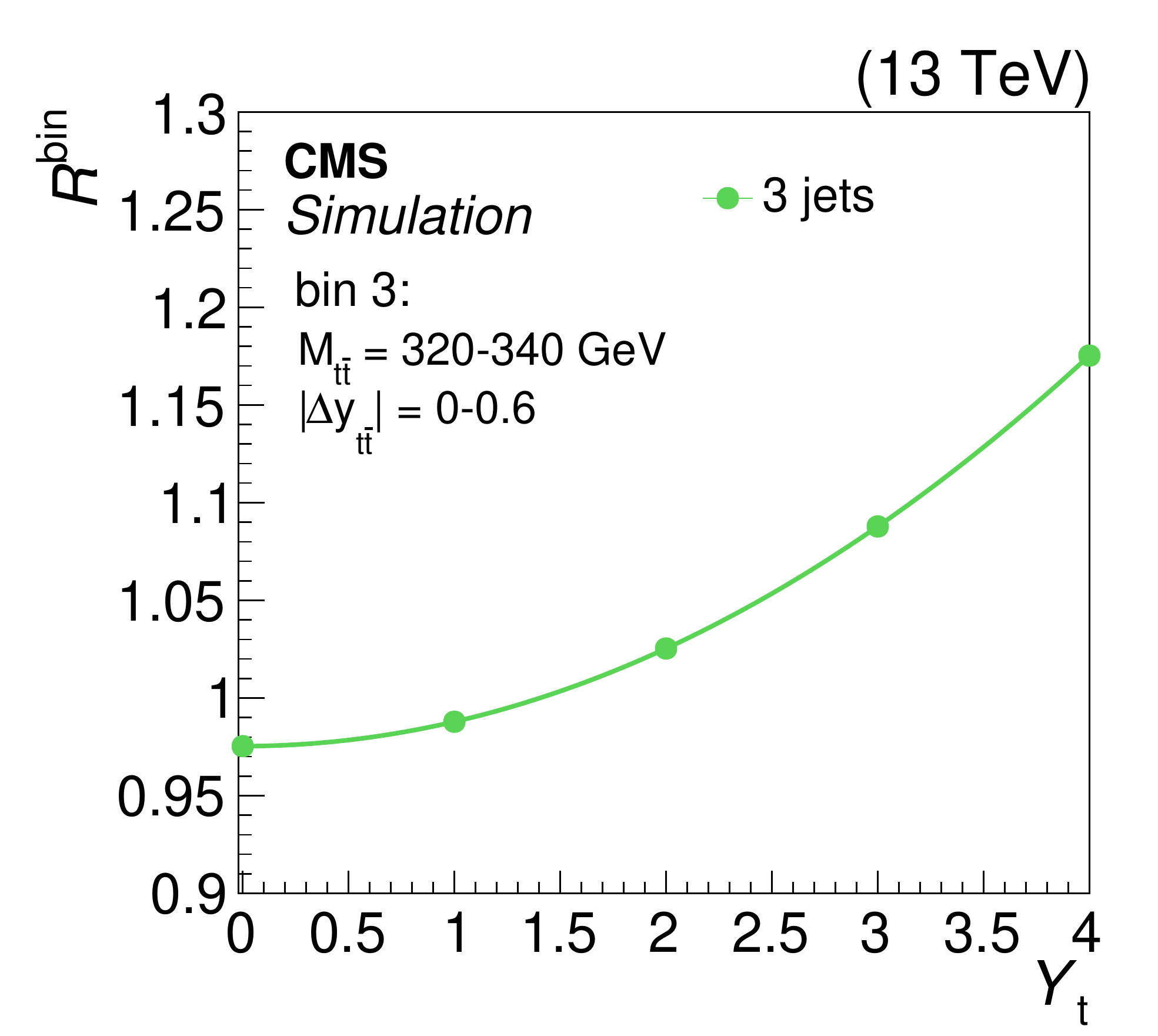}
\includegraphics[width=0.32\textwidth]{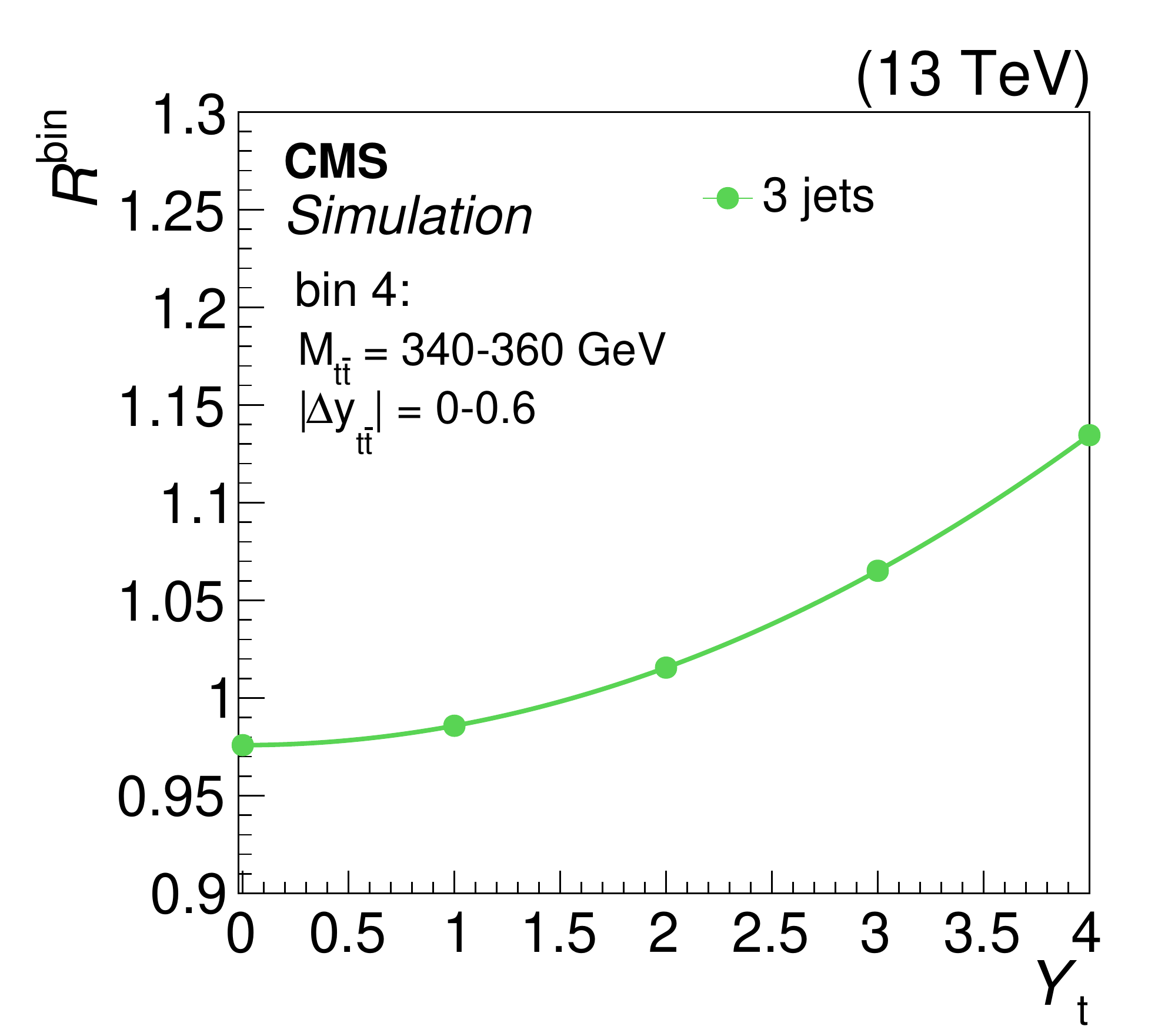}
\includegraphics[width=0.32\textwidth]{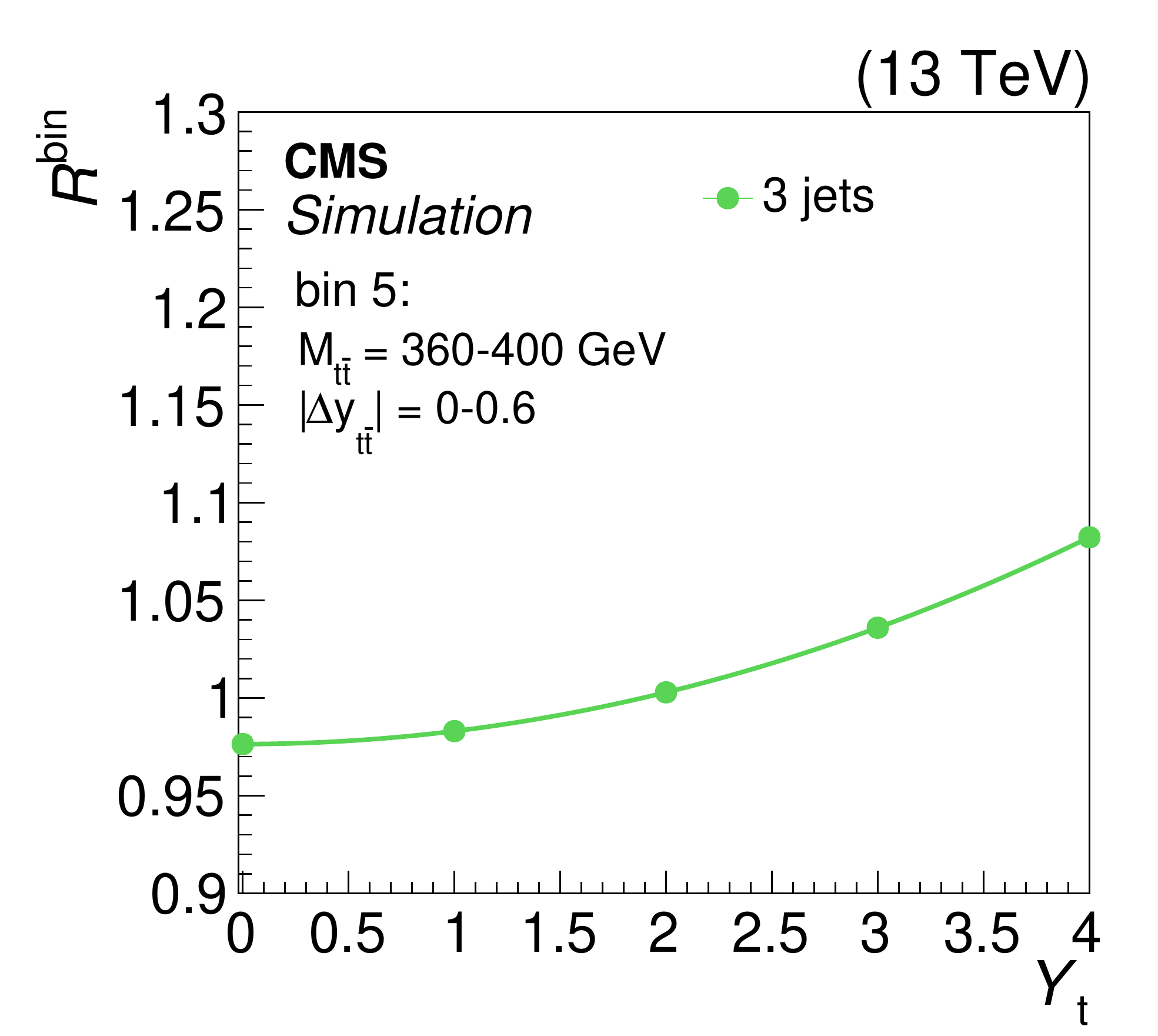}
\includegraphics[width=0.32\textwidth]{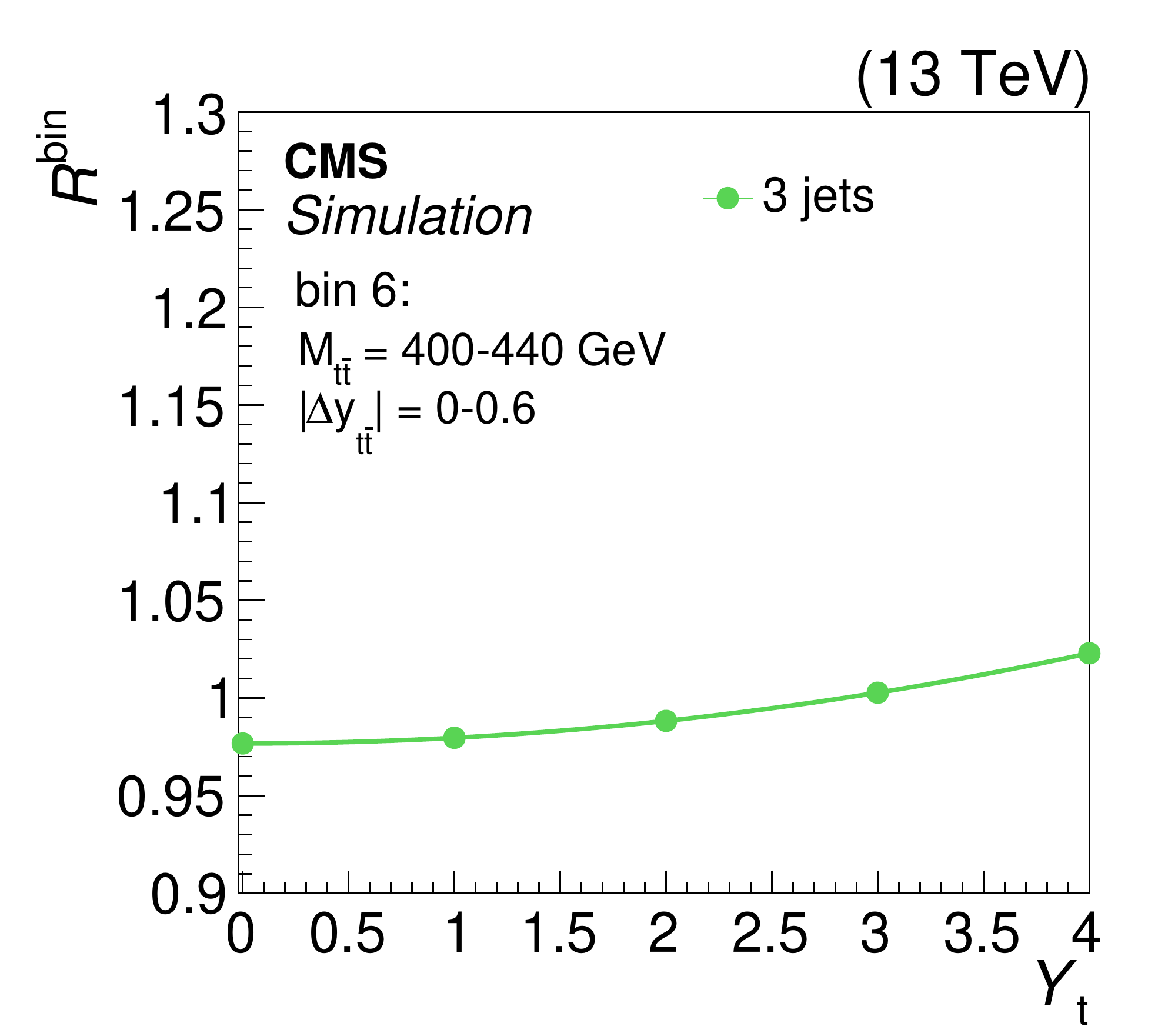}
\includegraphics[width=0.32\textwidth]{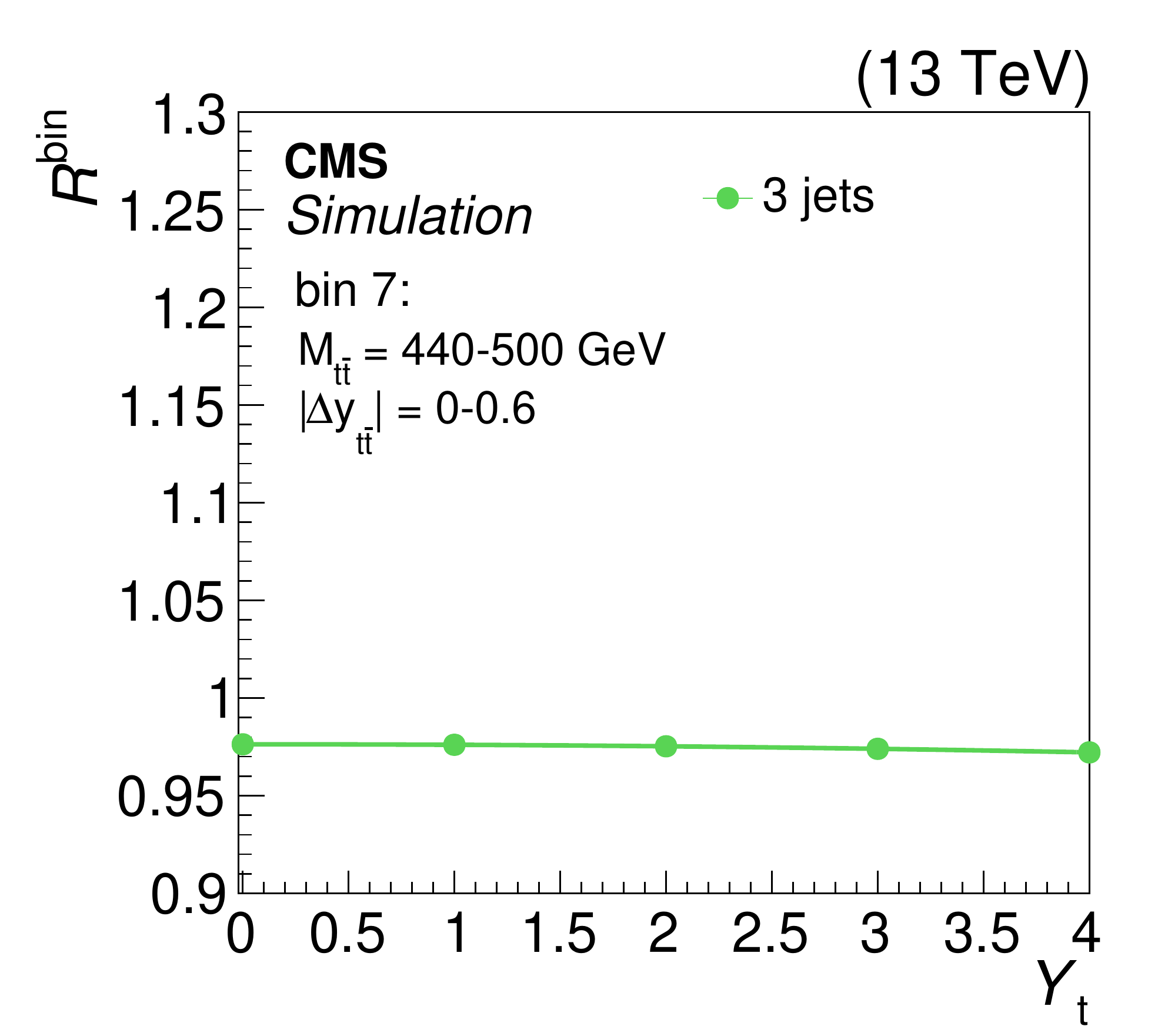}
\includegraphics[width=0.32\textwidth]{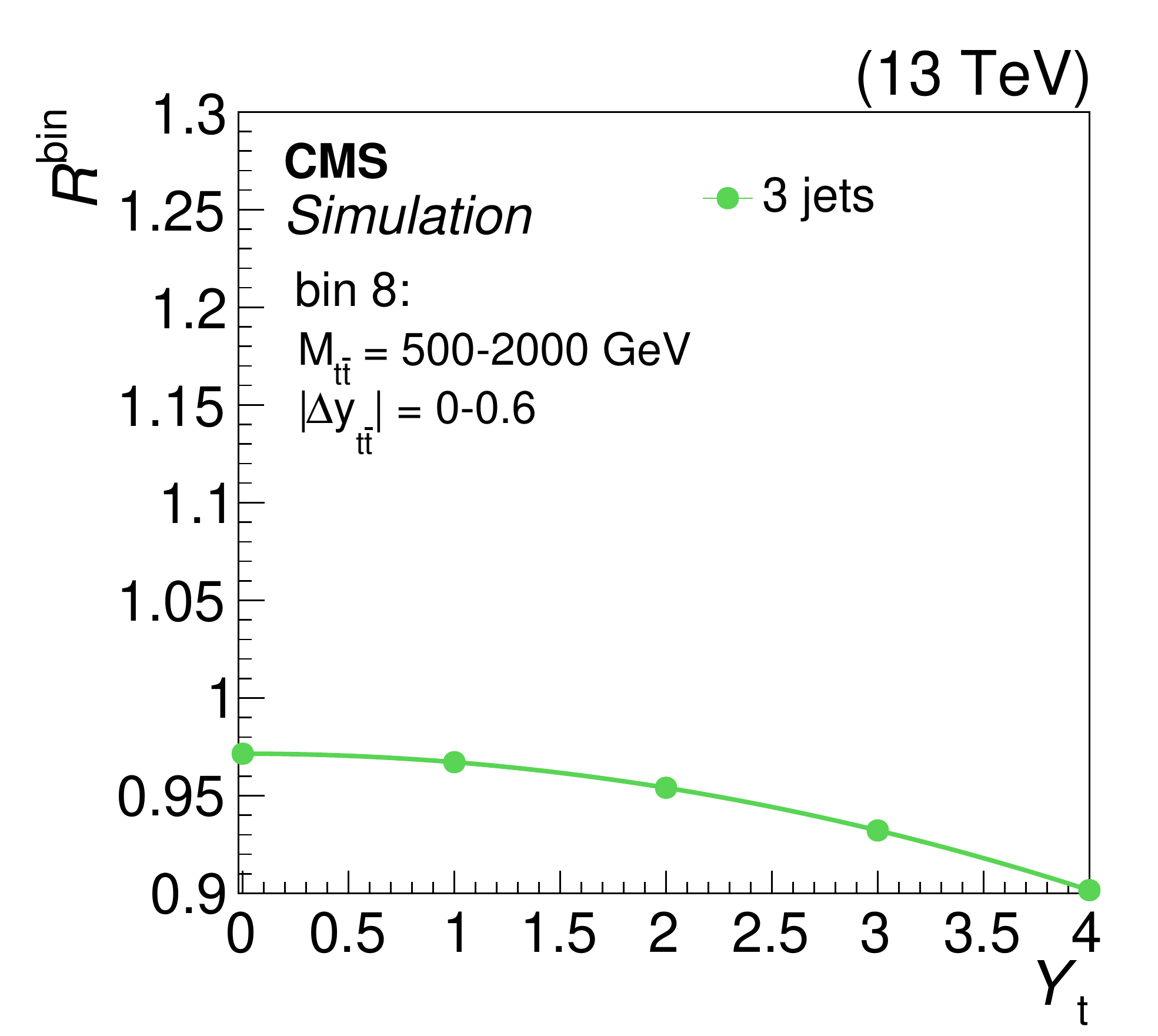}
\caption{The strength of the weak interaction correction, relative to the predicted \POWHEG signal, $R^\mathrm{bin}$, as a function of \yukawa in the three-jet category. The plots correspond to the first eight \Mttb bins for $\abs{\deltaY}<0.6$ (as shown in Fig.~\ref{p:results:2dshape:comb}). A quadratic fit is performed in each bin.}
\label{p:sys:model:3j:dely1}
\end{figure*}

\begin{figure*}[htbp]
\centering
\includegraphics[width=0.32\textwidth]{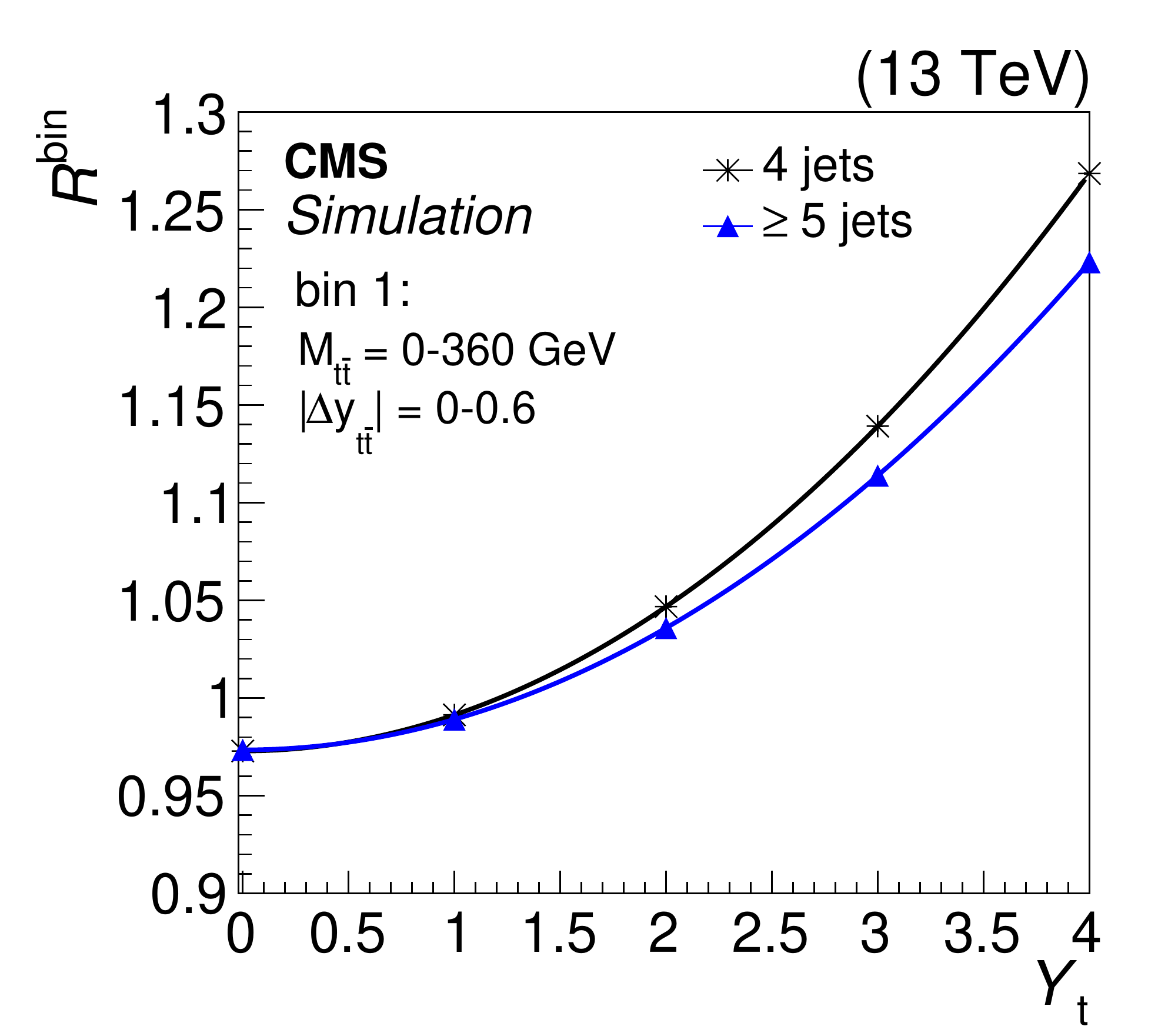}
\includegraphics[width=0.32\textwidth]{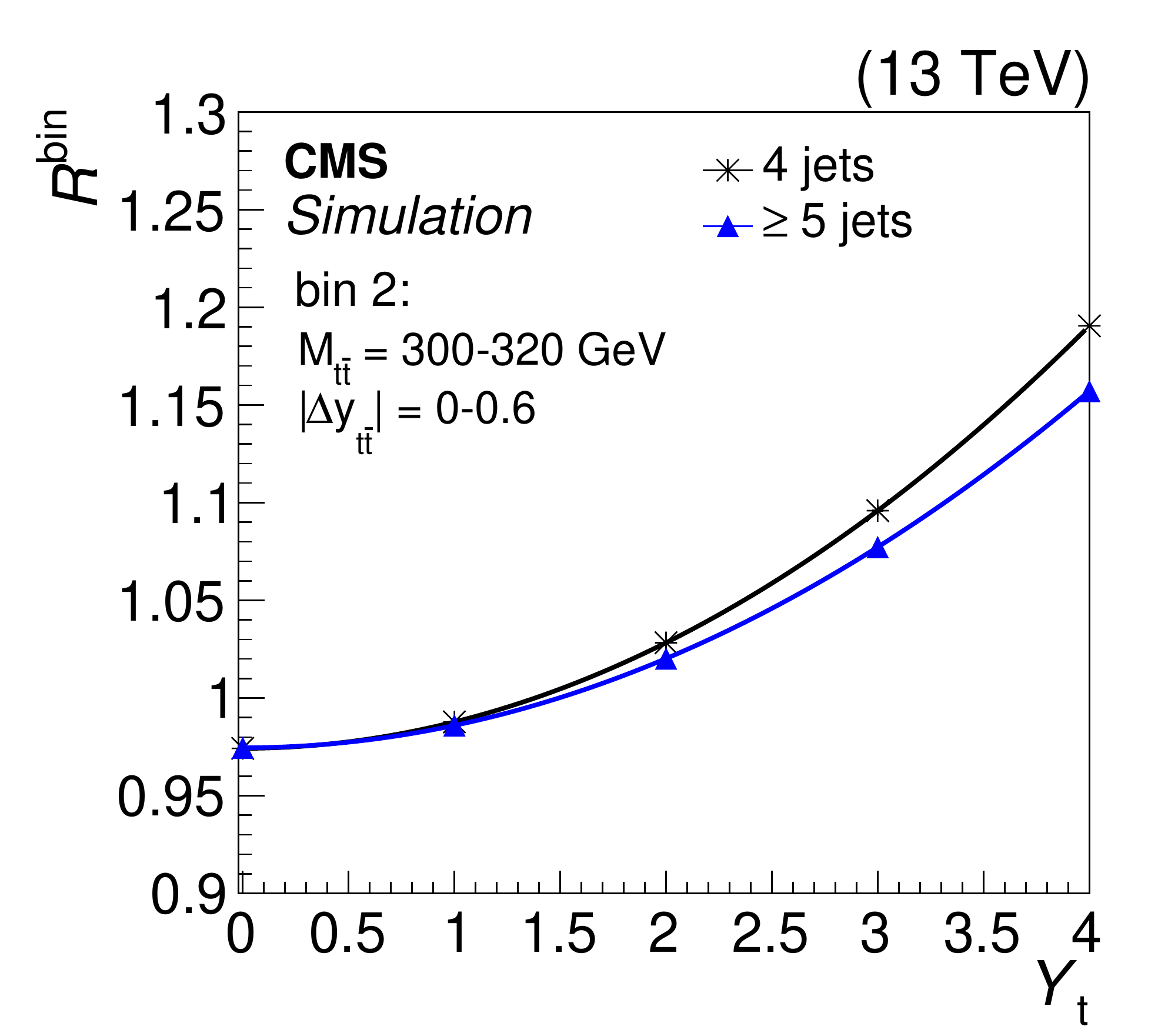}
\includegraphics[width=0.32\textwidth]{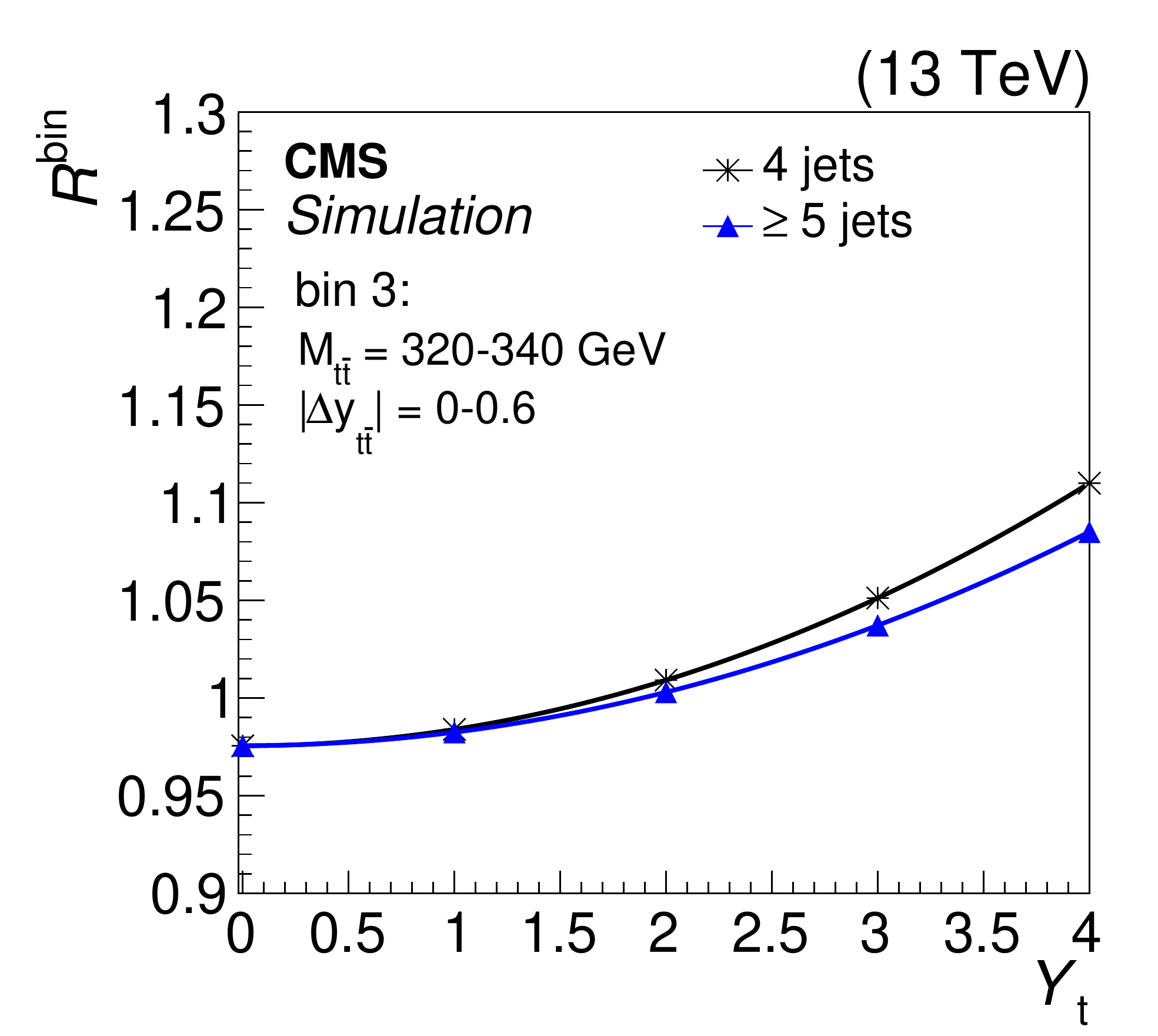}
\includegraphics[width=0.32\textwidth]{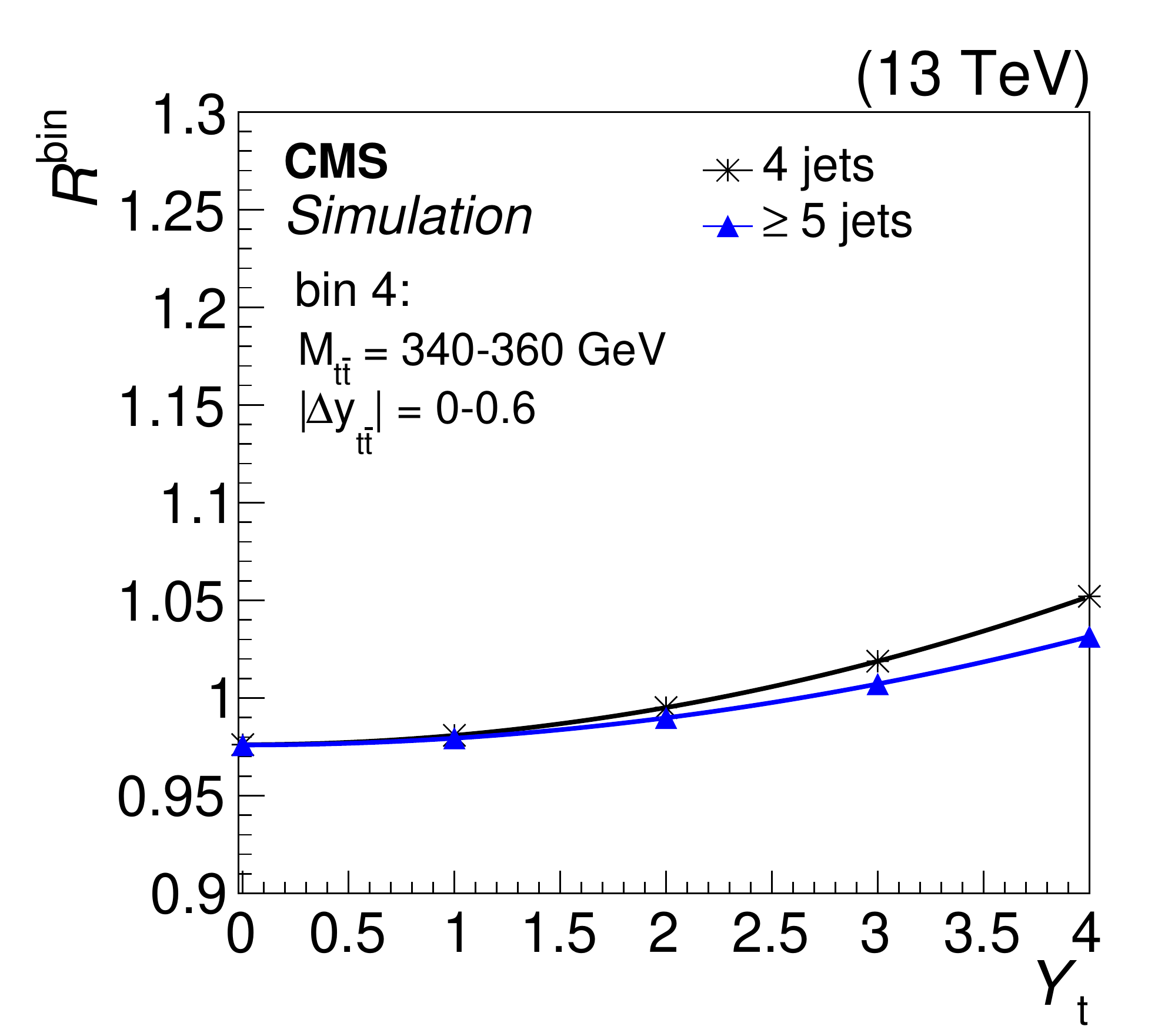}
\includegraphics[width=0.32\textwidth]{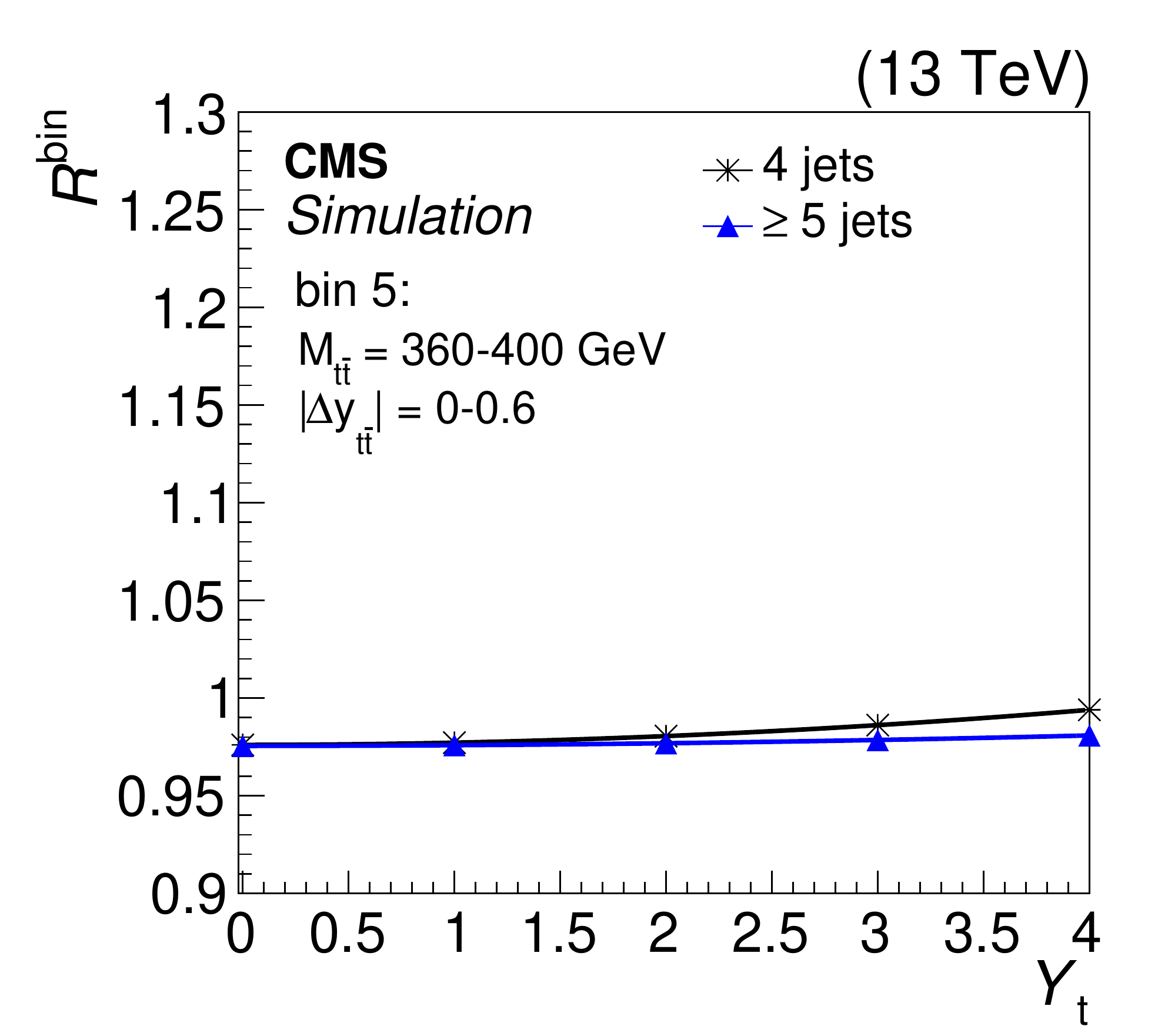}
\includegraphics[width=0.32\textwidth]{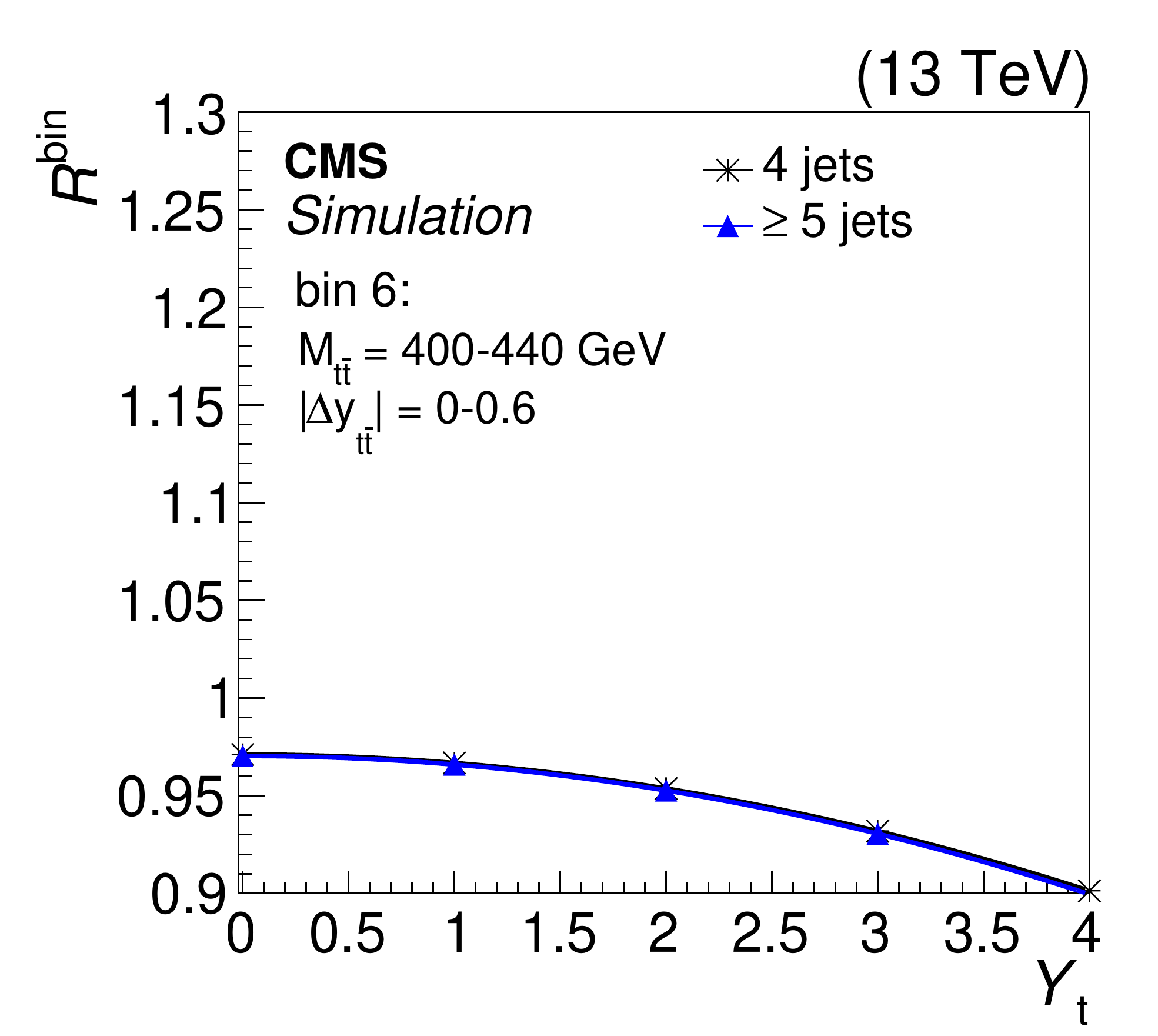}
\caption{The strength of the weak interaction correction, relative to the predicted \POWHEG signal, $R^\mathrm{bin}$, as a function of \yukawa in the categories
with four and five or more jets. The plots correspond to the first six \Mttb bins for $\abs{\deltaY}<0.6$ (as shown in Fig.~\ref{p:results:2dshape:comb}). A quadratic fit is performed in each bin.}
\label{p:sys:model:45j:dely1}
\end{figure*}

The likelihood function is constructed as a product of
Poisson distributions for the observed number of events,
$n^{\mathrm{bin}}_{\mathrm{obs}}$, in each
(\Mttb, $\abs{\deltaY}$) bin~\cite{CMS-NOTE-2011-005}:
\begin{linenomath}
\ifthenelse{\boolean{cms@external}}
{ 
\begin{multline}
\label{eqlikelihood}
\mathcal{L}(\yukawa,\theta) = \prod_{\mathrm{bin}~\in(\Mttb,\abs{\deltaY})} \mathcal{L}_{\mathrm{bin}} = \\
\prod_{\mathrm{bin}} \mathrm{Pois} \left ( n^{\mathrm{bin}}_{\mathrm{obs}} | s^{\mathrm{bin}}(\theta) \, R^{\mathrm{bin}}(\yukawa,\theta) + b^{\mathrm{bin}}(\theta) \right ) \, \rho(\theta),
\end{multline}
} 
{ 
\begin{equation}
\label{eqlikelihood}
\mathcal{L}(\yukawa,\theta) = \prod_{\mathrm{bin}~\in(\Mttb,\abs{\deltaY})} \mathcal{L}_{\mathrm{bin}} = \prod_{\mathrm{bin}} \mathrm{Pois} \left ( n^{\mathrm{bin}}_{\mathrm{obs}} | s^{\mathrm{bin}}(\theta) \, R^{\mathrm{bin}}(\yukawa,\theta) + b^{\mathrm{bin}}(\theta) \right ) \, \rho(\theta),
\end{equation}
} 
\end{linenomath}
where $s^{\mathrm{bin}}$ is the \POWHEG prediction for the number of
signal \ttbar events; $b^{\mathrm{bin}}$ is the  prediction for the
number of events from all background process (single top quark,
V+jets, and QCD multijet production);
$R^{\mathrm{bin}}(\yukawa,\theta)=s^{\mathrm{bin}}(\yukawa)/s^{\mathrm{bin}}(\POWHEG)$
encodes the effect of different \yukawa coupling scenarios,
parametrized with a quadratic dependence on $\yukawa$ in each bin
(shown in Figs.~\ref{p:sys:model:3j:dely1}
and~\ref{p:sys:model:45j:dely1} for the first $\abs{\deltaY}$ bin); and
$\theta$ represents the full suite of nuisance parameters with
$\rho(\theta)$ described by lognormal distributions parametrizing the
uncertainty on each source. The different sources of systematic
uncertainties are described in detail in Section~\ref{sec:sys}. The
quantity $R^{\mathrm{bin}}(\yukawa,\theta)$ is the main parameter of interest
in the fit, as it represents the strength of the weak correction over
the uncorrected \POWHEG yields.

\section{Systematic uncertainties}
\label{sec:sys}

We describe here the different sources of experimental and theoretical
uncertainties and their effect on determining \yukawa.

Systematic uncertainties that do not alter the shape of the
distributions of \Mttb and \deltaY are treated as normalization
uncertainties, while the others are treated as shape uncertainties. The
latter are evaluated bin-by-bin in the likelihood function
Eq.~(\ref{eqlikelihood}). Table~\ref{tab:sys:summary} lists all the
systematic uncertainties.

The uncertainty in the integrated luminosity is 2.5\%~\cite{LUMI}. The
simulated samples are reweighted to match the measured data
distribution in the number of pileup events. The uncertainty in the
total inelastic \Pp{}\Pp{} cross section, which affects the pileup
estimate, is accounted for by varying the average number of pileup
events per bunch crossing by 5\%~\cite{Sirunyan:2018nqx}.

The lepton efficiency scale factors, which account for the differences
in the trigger, reconstruction, and identification efficiencies between
data and simulation, are measured using a tag-and-probe method in $\PZ
\to \ell^+\ell^-$ events~\cite{Khachatryan:2015hwa,TaP13TeV}. These
scale factors, measured in bins of lepton \pt, lepton $\eta$, and jet
multiplicity, are applied to the simulated events The overall
uncertainty in the final measurement from these lepton scale factors is
approximately 2\%.

The uncertainties in the jet energy calibration (JEC) are evaluated by
shifting the energies of jets in simulation up and down by one standard
deviation in bins of \pt and $\eta$. Accounting for different sources
of JEC uncertainties and jet flavors, a total of 19 shape variations
are considered. The uncertainty in the jet energy resolution (JER) is
calculated by broadening the resolution in simulation and recomputing
the acceptances~\cite{Khachatryan:2016kdb}, for which the resulting
effect is a change of less than 1\% in event yields. The {\cPqb}
tagging efficiency in the simulation is corrected using scale factors
in bins of jet \pt and $\eta$ determined from efficiencies measured in
data and simulation~\cite{Sirunyan:2017ezt}. The uncertainty in the
measured scale factors ranges between 1 and 20\% per jet, leading to an
overall effect on the final measurement of 2--3\%.

The single top quark background estimate is affected by a 15\%
normalization uncertainty, evaluated from the combined results of
$t$-channel and $\PW\PQt$
productions~\cite{Sirunyan:2018rlu,Sirunyan:2018bsr}. The systematic
uncertainty in the V+jets background prediction is 30\%, derived from
the leading contribution in the signal region: $\PW$+heavy flavor
production~\cite{Khachatryan:2016ipq}. The systematic uncertainties
described above for the signal are also derived for these background
estimates. The QCD multijet background estimates from the data CR
include a 30\% normalization uncertainty from
Eq.~(\ref{eq:bck:qcdnorm}), and a shape difference observed between
samples with different lepton isolation (as described in
Section~\ref{sec:bck}). The uncertainty from the determination of \ptmiss
due to the electron, muon, and unclustered energy uncertainties,
results in a negligible effect on the acceptance. All the major
experimental uncertainties described above are evaluated for each
process in all reconstruction channels.

In the following, we describe the theoretical uncertainties.
The uncertainties in the factorization and renormalization scales affect
the number of events expected in simulated samples. These
are evaluated by varying each scale independently up and down by a
factor of two. We consider separate variations of the renormalization
and factorization scales by taking the envelope of the observed
variations as the quoted uncertainty. To account for possible
correlation between the two sources of uncertainty, we also add an
additional shape nuisance parameter that corresponds to the
simultaneous variation of both parameters. The different replicas in
the NNPDF3.0 PDF set~\cite{Ball:2014uwa} are used to estimate the
corresponding uncertainty in the shape from the changed acceptance in
each bin, which amounts to a combined variation as large as 5\%. The
different replicas due to the variation of strong coupling constant
$\alpS$ result in changes of the acceptance of around 1\%.

The effect of the top quark mass experimental uncertainty is estimated
by the difference in simulations generated with $\Mtop$ varied by
$\pm$1\GeV~\cite{Khachatryan:2015hba,Aaboud:2018zbu}, and it results in
a shape variation as large as 7\%. The dependence of \Mttb and \deltaY
on the correct description of the top quark $\pt$ in the simulation is
taken into account by checking the difference in the acceptance when
the nominal \POWHEG NLO samples are scaled to match the average top
quark and antiquark $\pt$ distributions calculated at NNLO in $\alpS$ in
Ref.~\cite{Czakon:2017dip}. This uncertainty is treated as a shape nuisance parameter in
the likelihood function for the \ttbar samples.

There are several sources of uncertainties arising from the parton
shower modeling. The uncertainty in matching the matrix element
calculation to the parton shower is estimated by changing the parameter
that regulates the damping of real emissions in the NLO
calculation~\cite{hdamp_underlying}, resulting in an effect of 1--5\%.
The scales, which determine initial- (ISR) and final-state radiation
(FSR) are also varied~\cite{ISR_FSR}, resulting in a maximum change of
4\% in the acceptance and shape variations as large as 10\%. The
uncertainty resulting from the modeling of the amount of multiple
parton interactions is derived following the studies of
Ref.~\cite{hdamp_underlying} and is found to have a negligible effect
on the result. Color reconnection reconfigures color strings after the
parton shower, affecting the hadronic \PW~boson
decays~\cite{hdamp_underlying}. This source of uncertainty typically
results in shape differences smaller than 1\%. The uncertainty in
{\cPqb} quark fragmentation, the momentum transfer from the {\cPqb}
quark to the {\cPqb} hadron, is estimated by varying the parametrized
function in the \PYTHIA simulation. It can produce a shape variation
as large as 3\%. As the {\cPqb} hadron semileptonic branching fractions
may change the {\cPqb} jet energy response, the acceptance is
recalculated after varying the $\PBp$, $\PBz$, $\PBs$, and $\PGLb$
semileptonic branching fractions up and down by their respective
experimental uncertainties~\cite{Tanabashi:2018oca}. The resulting
systematic uncertainty is around 3\%.

Finally, the weak interaction correction is implemented by reweighting
the nominal \POWHEG samples with the ratio of the weak correction over
the LO cross section calculated by \Hathor. As recommended by the
\Hathor authors~\cite{Kuhn:2013zoa}, the associated systematic
uncertainty for this procedure can be estimated from the difference
between the multiplicative and additive treatments, \ie,
$(1+\delta_{\mathrm{QCD}})(1+\delta_{\mathrm{W}})$ and
$(1+\delta_{\mathrm{QCD}} + \delta_{\mathrm{W}})$, where
$\delta_{\mathrm{QCD}}$ is estimated from the effect of varying the
factorization and renormalization scale up and down by a factor of two
on the NLO cross section, and $\delta_{\mathrm{W}}$ is the ratio of the
weak correction over the LO cross section obtained from \Hathor. The
difference is $\delta_{\mathrm{QCD}} \delta_{\mathrm{W}}$, which
is also a function of \yukawa.
This uncertainty is accounted for as a shape nuisance in the likelihood fit.

The experimental uncertainties are treated as 100\% correlated among signal
and background processes and across the jet multiplicity channels.

\begin{table*}[h!tbp]
\centering
\topcaption{Summary of the sources of systematic uncertainty, their effects and magnitudes on signal and backgrounds. If the uncertainty shows a shape dependence in the \Mttb and \deltaY distributions, it is treated as such in the likelihood. Only the luminosity, background normalization, and ISR uncertainties are not considered as shape uncertainties.}
\renewcommand{\arraystretch}{1.1}
\begin{scotch}{lcccc}
Uncertainty & \ttbar & Single \PQt & V+jets & QCD multijet \\
\hline
Integrated luminosity & 2.5\% & 2.5\% & 2.5\% & 2.5\% \\
Pileup & 0--1\%  & 0--1\%  & \NA & \NA \\
Lepton identification/trigger & 1.9\% & 1.9\% & 1.9\% & \NA \\
JEC  & 0--5\%  & 0--5\%  & \NA & \NA \\
JER & 0--0.6\%  & \NA & \NA & \NA \\
{\cPqb} tag scale factor & 3\% & 3\% & 2--3\%  & \NA \\
{\cPqb} mistag scale factor & 0.5\% & 1\% & 3--6\%  & \NA \\
Background normalization & \NA & 15\% & 30\% & 30\% \\
QCD multijet CR definition & \NA & \NA & \NA & 0--60\%  \\[\cmsTabSkip]
Factorization and renormalization scales & 0--6\%  & 2--5\%  & 0--15\%  & \NA \\
PDF & 0.5--1.5\%  & 0.5--1.5\%  & \NA & \NA \\
$\alpS(m_\PZ)$ in PDFs & 1\% & 1.5\% & \NA & \NA \\
Top quark mass & 1--5\%  & \NA & \NA & \NA \\
Top quark \pt modeling & 0--0.5\%  & \NA & \NA & \NA \\
Parton shower & & & & \\
~-NLO shower matching & 1.5--5\%  & \NA & \NA & \NA \\
~-ISR & 2--3\% & \NA & \NA & \NA \\
~-FSR & 0--9\%  & 0--12\%  & \NA & \NA \\
~-Color reconnection & 0--3\%  & \NA & \NA & \NA \\
~-{\cPqb} jet fragmentation & 0--3\%  & 0--5\%  & \NA & \NA \\
~-{\cPqb} hadron branching fraction & 3\% & 2.5--3\%  & \NA & \NA \\
Weak correction $\delta_{\mathrm{QCD}}\delta_{\mathrm{W}}$ & 0--0.2\%  (\yukawa=2) & \NA & \NA & \NA \\
\end{scotch}
\label{tab:sys:summary}
\end{table*}

\section{Results}
\label{sec:limit}

The data events are analyzed in
three exclusive channels, according to the number of jets in the final
state. The expected signal and background estimation shown in
Table~\ref{tab:expyields}, and the systematic uncertainties described in
Section~\ref{sec:sys} are used to construct a binned likelihood
(Eq.~(\ref{eqlikelihood})) as a product of the Poisson probabilities
from all bins in (\Mttb,$\abs{\deltaY}$). From this, we construct a profile
likelihood ratio test statistic
$q(\yukawa) = -2\ln \left [
\mathcal{L}(\yukawa, \hat{\hat{\theta}}) / \mathcal{L}(\hat{\yukawa}, \hat{\theta}) \right ]$,
where $\hat{\hat{\theta}}$ in the numerator
denotes the value of the estimator $\hat{\theta}$ that maximizes the
likelihood for a specific \yukawa, \ie, it is the conditional
maximum-likelihood estimator of $\theta$ (and thus is a function of
\yukawa). The denominator is the maximized (unconditional) likelihood
function, \ie, $\hat{\yukawa}$ and $\hat{\theta}$ are the values of the
estimators that simultaneously maximize the likelihood.
The statistical procedure to extract the parameter of interest is
detailed in Ref.~\cite{Conway:2011in}.

\begin{figure*}[htbp]
\centering
\includegraphics[width=1.0\textwidth]{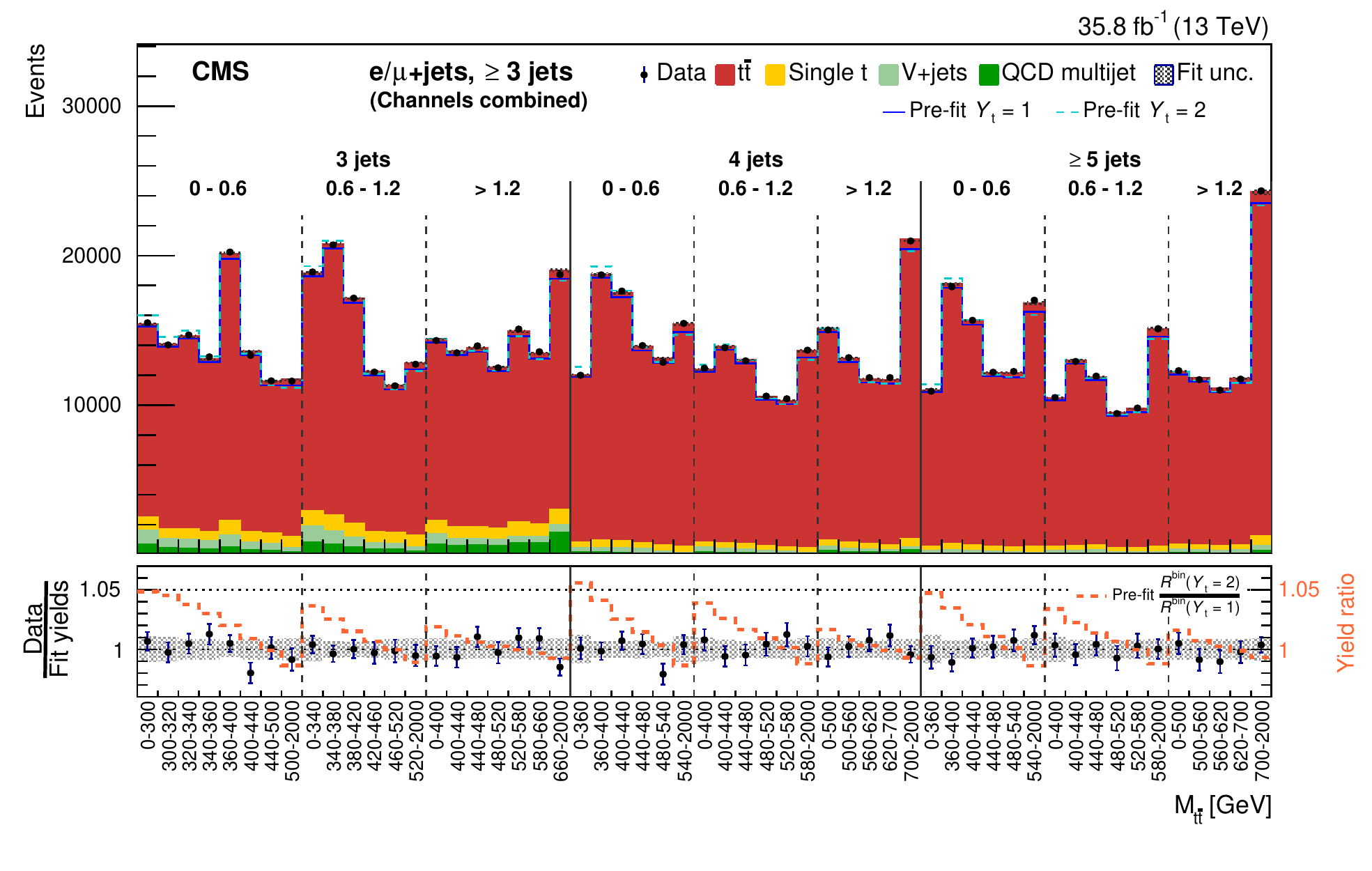}
\caption{The $\Mttb$ distribution in $\abs{\deltaY}$ bins for all events combined, after the simultaneous likelihood fit in all jet channels. The hatched bands show the total post-fit uncertainty. The ratios
of data to the sum of the predicted yields are provided in the lower panel. To show the sensitivity of the data to \yukawa = 1 and \yukawa = 2, the pre-fit yields are shown in the upper panel, and the yield ratio $R^{\mathrm{bin}}(\yukawa=2)/R^{\mathrm{bin}}(\yukawa=1)$ in the lower panel.}
\label{p:results:2dshape:comb}
\end{figure*}

The distributions of \Mttb and $\abs{\deltaY}$ after performing the
combined likelihood fit are shown in Fig.~\ref{p:results:2dshape:comb}.
The analysis covers the phase space from the production
threshold in \Mttb (which is $\approx$200\GeV at the detector level
for events with three reconstructed jets) up to 2\TeV.

\begin{figure*}[tbp]
\centering
\includegraphics[width=0.49\textwidth]{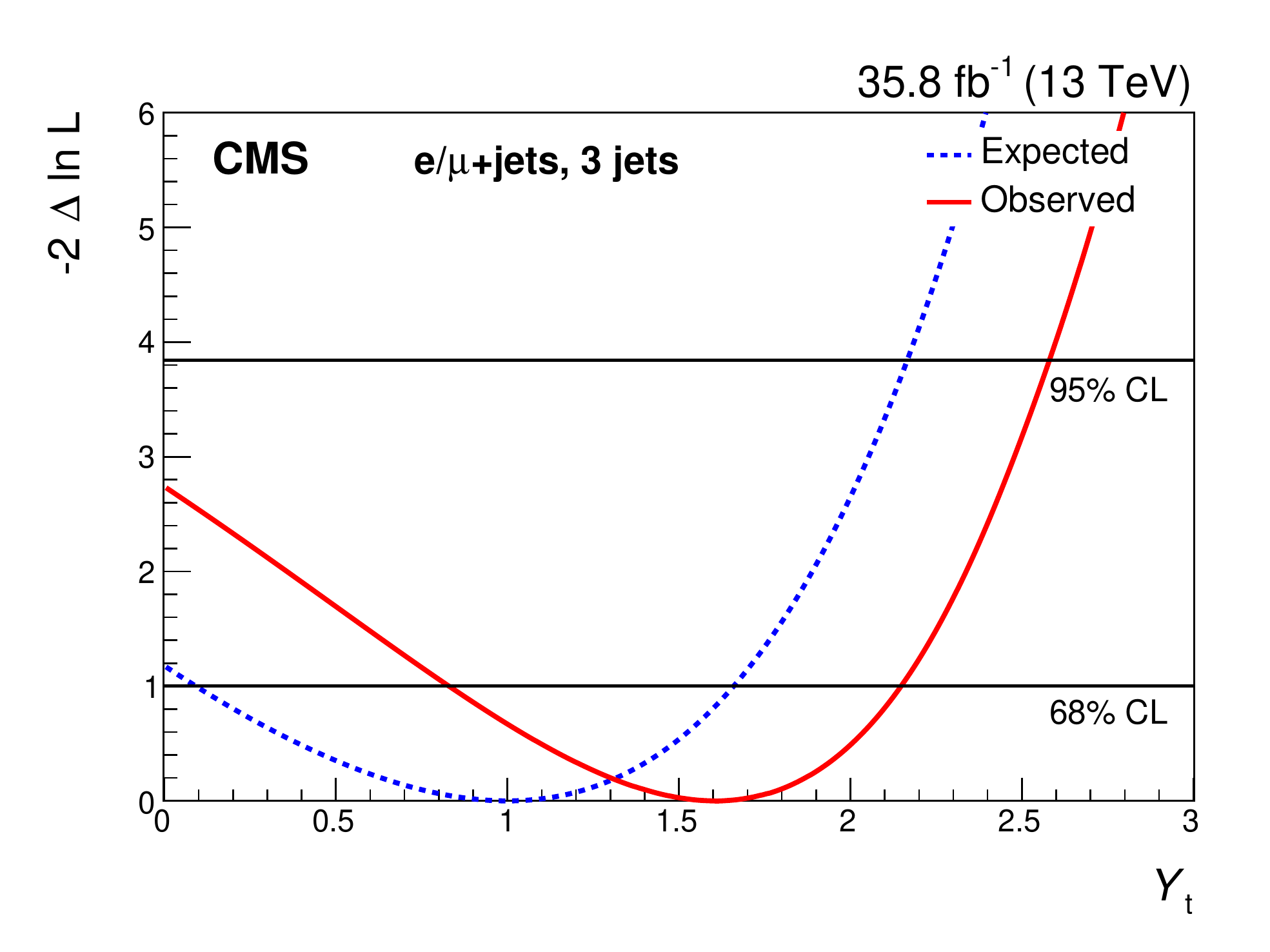}
\includegraphics[width=0.49\textwidth]{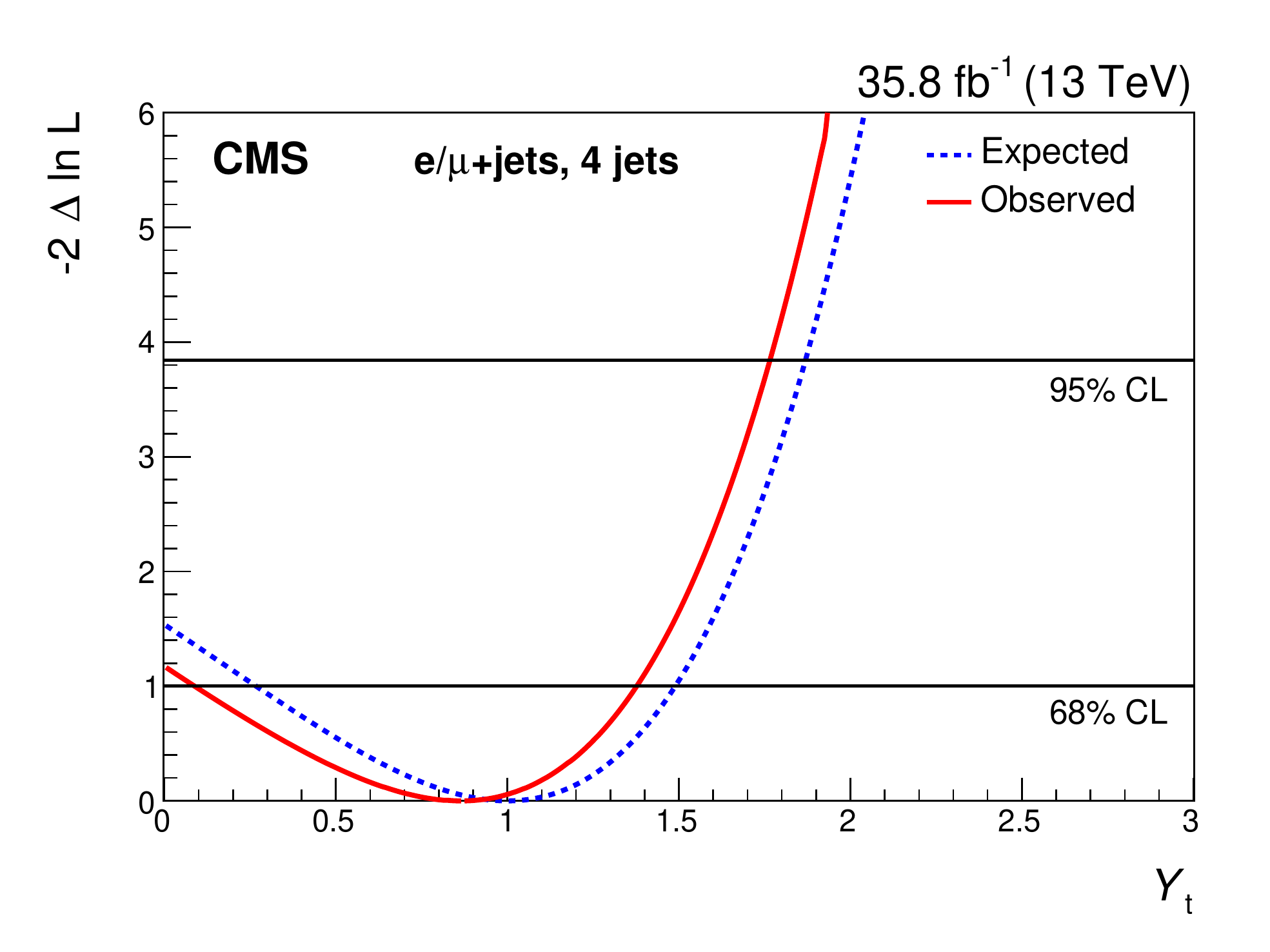}
\includegraphics[width=0.49\textwidth]{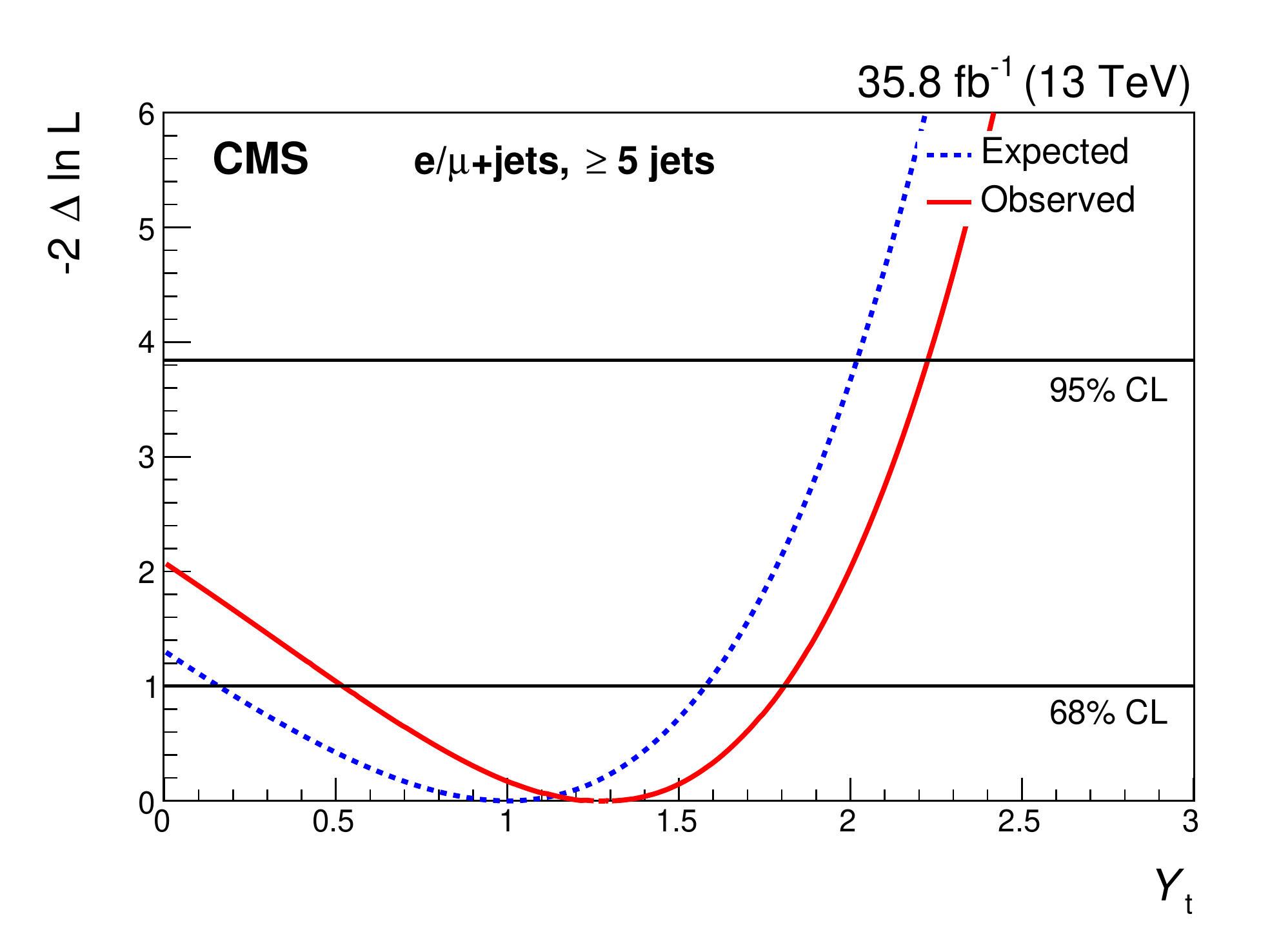}
\includegraphics[width=0.49\textwidth]{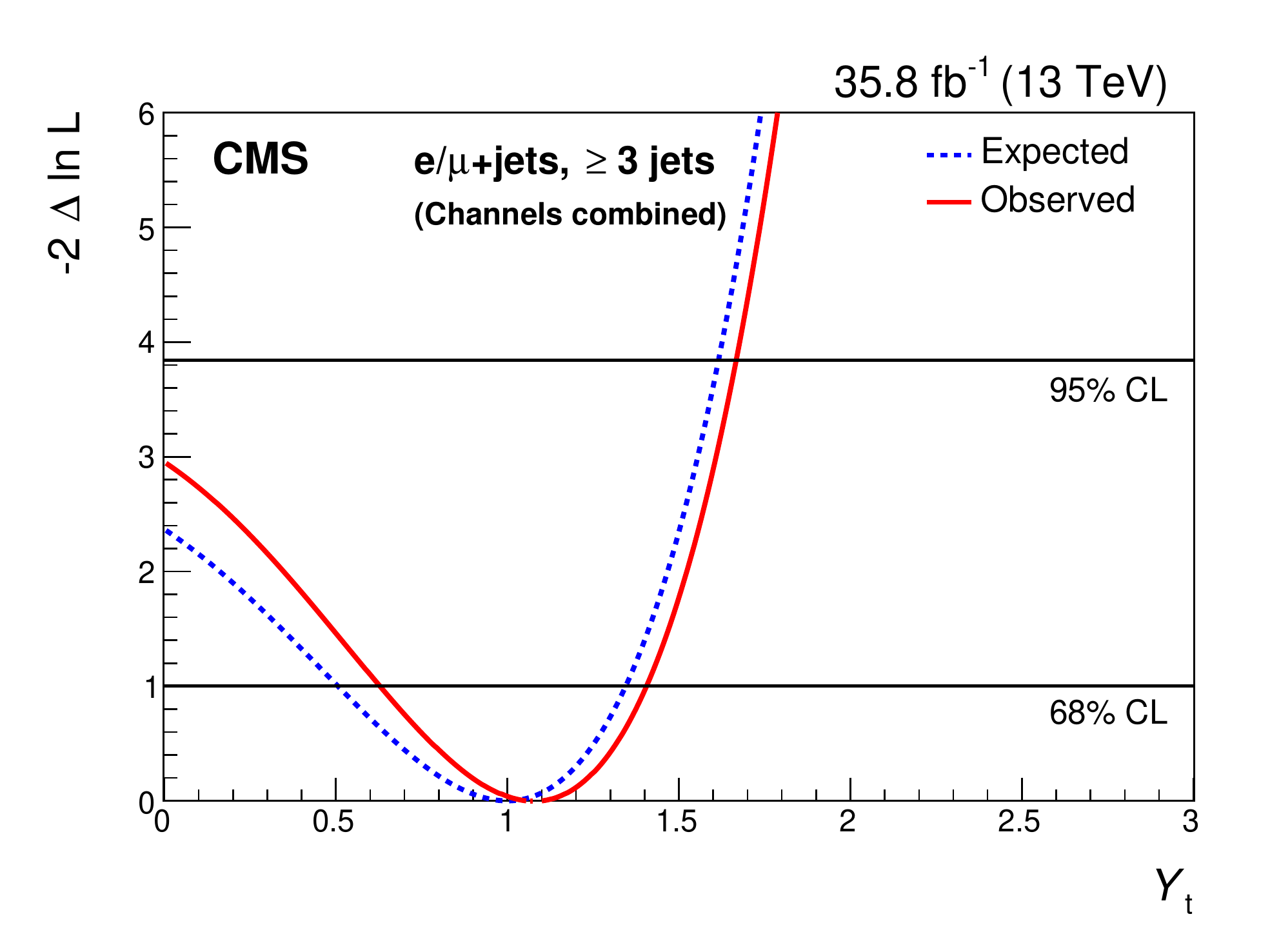}
\caption{The test statistic scan versus \yukawa for each channel (three, four, and five or more jets), and all channels combined. The test statistic minimum indicates the best fit of \yukawa. The horizontal lines indicate 68 and 95\% \CL intervals.}
\label{p:results:likelihoodscan}
\end{figure*}

We measure the top quark Yukawa coupling by scanning the likelihood
function with respect to \yukawa. The likelihood scan distributions can
be found in Fig.~\ref{p:results:likelihoodscan}. The expected and
observed results are presented in Table~\ref{tab:results:limits}.
An upper limit on \yukawa is also determined, using a modified frequentist \CLs
procedure~\cite{Junk:1999kv,Read:2002hq} with the asymptotic method~\cite{Cowan:2010js}.

\setlength\extrarowheight{5pt}
\begin{table}[htbp]
\centering
\topcaption{The expected and observed best fit values and 95\% \CL upper limits on \yukawa.}
\renewcommand{\arraystretch}{1.2}
\begin{scotch}{lcccc}
Channel & \multicolumn{2}{c}{Best fit \yukawa} & \multicolumn{2}{c}{95\% \CL upper limit} \\
        & Expected & Observed & Expected & Observed \\
\hline
3 jets & $1.00^{+0.66}_{-0.90}$ & $1.62^{+0.53}_{-0.78}$ & ${<}2.17$ & ${<}2.59$ \\
4 jets & $1.00^{+0.50}_{-0.72}$ & $0.87^{+0.51}_{-0.77}$ & ${<}1.88$ & ${<}1.77$ \\
$\geq$5 jets & $1.00^{+0.59}_{-0.83}$ & $1.27^{+0.55}_{-0.74}$ & ${<}2.03$ & ${<}2.23$ \\[\cmsTabSkip]
Combined & $1.00^{+0.35}_{-0.48}$ & $1.07^{+0.34}_{-0.43}$ & ${<}1.62$ & ${<}1.67$ \\
\end{scotch}
\label{tab:results:limits}
\end{table}

\section{Summary}
\label{sec:summary}

A measurement of the top quark Yukawa coupling is presented, extracted
by investigating \ttbar pair production in final states with an
electron or muon and several jets, using proton--proton data collected
by the CMS experiment at $\sqrt{s} = 13\TeV$, corresponding to an
integrated luminosity of 35.8\fbinv. The \ttbar production cross
section is sensitive to the top quark Yukawa coupling through weak
force corrections that can modify the distributions of the mass of top
quark--antiquark pairs, \Mttb, and the rapidity difference between top
quark and antiquark, \deltaY. The kinematic properties of these final
states are reconstructed in events with  at least three jets, two of
which are identified as originating from bottom quarks. The inclusion
of events with only three reconstructed jets using a dedicated
algorithm improves the sensitivity of the analysis by increasing the
signal from events in the low-\Mttb region, which is most sensitive to
the Yukawa coupling. The ratio of the top quark Yukawa coupling to its
expected SM value, \yukawa, is extracted by comparing the data with the
expected $\ttbar$ signal for different values of \yukawa in a total of
55 bins in \Mttb, $\abs{\deltaY}$, and the number of reconstructed
jets. The measured value of \yukawa is $1.07^{+0.34}_{-0.43}$,
compared to an expected value of $1.00^{+0.35}_{-0.48}$. The observed upper
limit on \yukawa is 1.67 at 95\% confidence level (\CL), with an expected value of
1.62.

Although the method presented in this paper is not as sensitive as the combined
CMS measurement of \yukawa performed using Higgs boson production and
decays in multiple channels~\cite{Sirunyan:2018koj}, it has the
advantage that it does not depend on any assumptions about the
couplings of the Higgs boson to particles other than the top quark. The
result presented here is more sensitive than the only other result from
CMS exclusively dependent on \yukawa, namely the limit on the ${\ttbar}{\ttbar}$ cross section,
which constrains \yukawa to be less than 2.1 at 95\%
\CL~\cite{Sirunyan:2017roi}.

\begin{acknowledgments}
\label{ack}

We congratulate our colleagues in the CERN accelerator departments for the excellent performance of the LHC and thank the technical and administrative staffs at CERN and at other CMS institutes for their contributions to the success of the CMS effort. In addition, we gratefully acknowledge the computing centers and personnel of the Worldwide LHC Computing Grid for delivering so effectively the computing infrastructure essential to our analyses. Finally, we acknowledge the enduring support for the construction and operation of the LHC and the CMS detector provided by the following funding agencies: BMBWF and FWF (Austria); FNRS and FWO (Belgium); CNPq, CAPES, FAPERJ, FAPERGS, and FAPESP (Brazil); MES (Bulgaria); CERN; CAS, MoST, and NSFC (China); COLCIENCIAS (Colombia); MSES and CSF (Croatia); RPF (Cyprus); SENESCYT (Ecuador); MoER, ERC IUT, PUT and ERDF (Estonia); Academy of Finland, MEC, and HIP (Finland); CEA and CNRS/IN2P3 (France); BMBF, DFG, and HGF (Germany); GSRT (Greece); NKFIA (Hungary); DAE and DST (India); IPM (Iran); SFI (Ireland); INFN (Italy); MSIP and NRF (Republic of Korea); MES (Latvia); LAS (Lithuania); MOE and UM (Malaysia); BUAP, CINVESTAV, CONACYT, LNS, SEP, and UASLP-FAI (Mexico); MOS (Montenegro); MBIE (New Zealand); PAEC (Pakistan); MSHE and NSC (Poland); FCT (Portugal); JINR (Dubna); MON, RosAtom, RAS, RFBR, and NRC KI (Russia); MESTD (Serbia); SEIDI, CPAN, PCTI, and FEDER (Spain); MOSTR (Sri Lanka); Swiss Funding Agencies (Switzerland); MST (Taipei); ThEPCenter, IPST, STAR, and NSTDA (Thailand); TUBITAK and TAEK (Turkey); NASU and SFFR (Ukraine); STFC (United Kingdom); DOE and NSF (USA).

\hyphenation{Rachada-pisek} Individuals have received support from the Marie-Curie program and the European Research Council and Horizon 2020 Grant, contract Nos.\ 675440 and 765710 (European Union); the Leventis Foundation; the A.P.\ Sloan Foundation; the Alexander von Humboldt Foundation; the Belgian Federal Science Policy Office; the Fonds pour la Formation \`a la Recherche dans l'Industrie et dans l'Agriculture (FRIA-Belgium); the Agentschap voor Innovatie door Wetenschap en Technologie (IWT-Belgium); the F.R.S.-FNRS and FWO (Belgium) under the ``Excellence of Science -- EOS" -- be.h project n.\ 30820817; the Beijing Municipal Science \& Technology Commission, No. Z181100004218003; the Ministry of Education, Youth and Sports (MEYS) of the Czech Republic; the Lend\"ulet (``Momentum") Program and the J\'anos Bolyai Research Scholarship of the Hungarian Academy of Sciences, the New National Excellence Program \'UNKP, the NKFIA research grants 123842, 123959, 124845, 124850, 125105, 128713, 128786, and 129058 (Hungary); the Council of Science and Industrial Research, India; the HOMING PLUS program of the Foundation for Polish Science, cofinanced from European Union, Regional Development Fund, the Mobility Plus program of the Ministry of Science and Higher Education, the National Science Center (Poland), contracts Harmonia 2014/14/M/ST2/00428, Opus 2014/13/B/ST2/02543, 2014/15/B/ST2/03998, and 2015/19/B/ST2/02861, Sonata-bis 2012/07/E/ST2/01406; the National Priorities Research Program by Qatar National Research Fund; the Programa Estatal de Fomento de la Investigaci{\'o}n Cient{\'i}fica y T{\'e}cnica de Excelencia Mar\'{\i}a de Maeztu, grant MDM-2015-0509 and the Programa Severo Ochoa del Principado de Asturias; the Thalis and Aristeia programs cofinanced by EU-ESF and the Greek NSRF; the Rachadapisek Sompot Fund for Postdoctoral Fellowship, Chulalongkorn University and the Chulalongkorn Academic into Its 2nd Century Project Advancement Project (Thailand); the Welch Foundation, contract C-1845; and the Weston Havens Foundation (USA).
\end{acknowledgments}

\bibliography{auto_generated}   
\cleardoublepage \appendix\section{The CMS Collaboration \label{app:collab}}\begin{sloppypar}\hyphenpenalty=5000\widowpenalty=500\clubpenalty=5000\vskip\cmsinstskip
\textbf{Yerevan Physics Institute, Yerevan, Armenia}\\*[0pt]
A.M.~Sirunyan$^{\textrm{\dag}}$, A.~Tumasyan
\vskip\cmsinstskip
\textbf{Institut für Hochenergiephysik, Wien, Austria}\\*[0pt]
W.~Adam, F.~Ambrogi, T.~Bergauer, J.~Brandstetter, M.~Dragicevic, J.~Erö, A.~Escalante~Del~Valle, M.~Flechl, R.~Frühwirth\cmsAuthorMark{1}, M.~Jeitler\cmsAuthorMark{1}, N.~Krammer, I.~Krätschmer, D.~Liko, T.~Madlener, I.~Mikulec, N.~Rad, J.~Schieck\cmsAuthorMark{1}, R.~Schöfbeck, M.~Spanring, D.~Spitzbart, W.~Waltenberger, J.~Wittmann, C.-E.~Wulz\cmsAuthorMark{1}, M.~Zarucki
% ...
\vskip\cmsinstskip
\dag: Deceased\\
1:  Also at Vienna University of Technology, Vienna, Austria\\
% ...
\end{sloppypar}
\end{document}